\documentclass[prb, aps, 10pt, twocolumn, floatfix, superscriptaddress]{revtex4-1}
\usepackage{amssymb,amsfonts,amsmath,amsthm}
\usepackage{graphicx}%esto va sin el DVIPS!
\usepackage[english]{babel}
\usepackage[latin1]{inputenc}
\usepackage[pdftex,colorlinks=true,pdfstartview=FitH,linkcolor=blue,citecolor=blue,urlcolor=blue]{hyperref}

\usepackage{bm}
\usepackage{float}
\usepackage{mathtools}
\usepackage{color,soul}
\usepackage{times}
\usepackage{upgreek}
\usepackage{MnSymbol}
\usepackage{verbatim}
\usepackage[bottom]{footmisc}
\usepackage{bbold}

\usepackage[dvipsnames,svgnames,table]{xcolor}
\usepackage{fancyhdr}

\usepackage{subfigure}

\hypersetup{
pdfstartview={FitH},% fits the width of the page to the window
colorlinks=true,    % false: boxed links; true: colored links
linkcolor=NavyBlue, % color of internal links
citecolor=Blue,   % color of links to bibliography
filecolor=NavyBlue, % color of file links
urlcolor=NavyBlue   % color of external links
}

\newcommand{\ve}{\varepsilon}

\newcommand{\bea}{\begin{eqnarray}}
\newcommand{\eea}{\end{eqnarray}}
\newcommand{\be}{\begin{equation}}
\newcommand{\ee}{\end{equation}}

\begin{document}
%%%%%%%%%%%%%%%%%%%%%%%%%%%%%%%%%%%%%%%%%%%%%%%%%%%%%%%%%%%%%%
\title{Asynchronous Locking in Metamaterials of Fluids of Light and Sound}
%%%%%%%%%%%%%%%%%%%%%%%%%%%%%%%%%%%%%%%%%%%%%%%%%%%%%%%%%%%%%

\author{D. L. Chafatinos}
\affiliation{Centro At{\'{o}}mico Bariloche and Instituto Balseiro,
Comisi\'on Nacional de Energ\'{\i}a At\'omica (CNEA)- Universidad Nacional de Cuyo (UNCUYO), 8400 Bariloche, Argentina.}
\affiliation{Instituto de Nanociencia y Nanotecnolog\'{i}a (INN-Bariloche), Consejo Nacional de Investigaciones Cient\'{\i}ficas y T\'ecnicas (CONICET), Argentina.}

\author{A.~S. Kuznetsov}
\affiliation{Paul-Drude-Institut f\"{u}r Festk\"{o}rperelektronik, Leibniz-Institut im Forschungsverbund Berlin e.V., Hausvogteiplatz 5-7,\\ 10117 Berlin, Germany.}

\author{A.~A. Reynoso}
\affiliation{Centro At{\'{o}}mico Bariloche and Instituto Balseiro,
Comisi\'on Nacional de Energ\'{\i}a At\'omica (CNEA)- Universidad Nacional de Cuyo (UNCUYO), 8400 Bariloche, Argentina.}
\affiliation{Instituto de Nanociencia y Nanotecnolog\'{i}a (INN-Bariloche), Consejo Nacional de Investigaciones Cient\'{\i}ficas y T\'ecnicas (CONICET), Argentina.}
\affiliation{Departamento de F\'isica Aplicada II, Universidad de Sevilla, E-41012 Sevilla, Spain}

\author{G. Usaj}
\affiliation{Centro At{\'{o}}mico Bariloche and Instituto Balseiro,
Comisi\'on Nacional de Energ\'{\i}a At\'omica (CNEA)- Universidad Nacional de Cuyo (UNCUYO), 8400 Bariloche, Argentina.}
\affiliation{Instituto de Nanociencia y Nanotecnolog\'{i}a (INN-Bariloche), Consejo Nacional de Investigaciones Cient\'{\i}ficas y T\'ecnicas (CONICET), Argentina.}

\author{P. Sesin}
\affiliation{Centro At{\'{o}}mico Bariloche and Instituto Balseiro,
Comisi\'on Nacional de Energ\'{\i}a At\'omica (CNEA)- Universidad Nacional de Cuyo (UNCUYO), 8400 Bariloche, Argentina.}
\affiliation{Instituto de Nanociencia y Nanotecnolog\'{i}a (INN-Bariloche), Consejo Nacional de Investigaciones Cient\'{\i}ficas y T\'ecnicas (CONICET), Argentina.}

\author{I. Papuccio}
\affiliation{Centro At{\'{o}}mico Bariloche and Instituto Balseiro,
Comisi\'on Nacional de Energ\'{\i}a At\'omica (CNEA)- Universidad Nacional de Cuyo (UNCUYO), 8400 Bariloche, Argentina.}
\affiliation{Instituto de Nanociencia y Nanotecnolog\'{i}a (INN-Bariloche), Consejo Nacional de Investigaciones Cient\'{\i}ficas y T\'ecnicas (CONICET), Argentina.}

\author{A.~E. Bruchhausen}
\affiliation{Centro At{\'{o}}mico Bariloche and Instituto Balseiro,
Comisi\'on Nacional de Energ\'{\i}a At\'omica (CNEA)- Universidad Nacional de Cuyo (UNCUYO), 8400 Bariloche, Argentina.}
\affiliation{Instituto de Nanociencia y Nanotecnolog\'{i}a (INN-Bariloche), Consejo Nacional de Investigaciones Cient\'{\i}ficas y T\'ecnicas (CONICET), Argentina.}

\author{K. Biermann}
\affiliation{Paul-Drude-Institut f\"{u}r Festk\"{o}rperelektronik, Leibniz-Institut im Forschungsverbund Berlin e.V., Hausvogteiplatz 5-7,\\ 10117 Berlin, Germany.}

\author{P.~V. Santos}
\email[Corresponding author, e-mail: ]{santos@pdi-berlin.de}
\affiliation{Paul-Drude-Institut f\"{u}r Festk\"{o}rperelektronik, Leibniz-Institut im Forschungsverbund Berlin e.V., Hausvogteiplatz 5-7,\\ 10117 Berlin, Germany.}

\author{A. Fainstein}
\email[Corresponding author, e-mail: ]{afains@cab.cnea.gov.ar}
\affiliation{Centro At{\'{o}}mico Bariloche and Instituto Balseiro,
Comisi\'on Nacional de Energ\'{\i}a At\'omica (CNEA)- Universidad Nacional de Cuyo (UNCUYO), 8400 Bariloche, Argentina.}
\affiliation{Instituto de Nanociencia y Nanotecnolog\'{i}a (INN-Bariloche), Consejo Nacional de Investigaciones Cient\'{\i}ficas y T\'ecnicas (CONICET), Argentina.}

\date{\today}

%%%%%%%%%%%%%%%%%%%%%%%%%%%%%%%%%%%%%%%%%%%%%%%%%%%%%%%%%%
\begin{abstract}
{Phonons, the quanta of vibrations, are very important for the equilibrium and dynamical properties of matter. GHz coherent phonons can also interact with and act as interconnects in a wide range of quantum systems. Harnessing and tailoring their coupling to opto-electronic excitations thus becomes highly relevant for engineered materials for quantum technologies. With this perspective we introduce polaromechanical metamaterials, two-dimensional arrays of $\mu$m-size zero-dimensional traps confining light-matter polariton fluids and GHz phonons. 
A strong exciton-mediated polariton-phonon interaction determines the inter-site polariton coupling with remarkable consequences for the dynamics. When locally perturbed by optical excitation, polaritons respond by locking the energy detuning between neighbor sites at integer multiples of the phonon energy, evidencing synchronization involving the polariton and phonon fields. These results open the path for the coherent control of quantum light fluids with hypersound in a scalable platform.
}
\end{abstract}
%%%%%%%%%%%%%%%%%%%%%%%%%%%%%%%%%%%%%%%%%%%
%\pacs{63.22.+m,78.30.Fs,78.30.-j,78.67.Pt}
\maketitle
%%%%%%%%%%%%%%%%%%%%%%%
\section{Introduction}
%%%%%%%%%%%%%%%%%%%%%%

Microcavity exciton-polariton fluids of light (the quantum states formed by strongly coupled excitons and photons in microcavities) constitute a hybrid system~\cite{Kurizki2015} that displays a plethora of striking properties. These include Bose-Einstein condensation (BEC)~\cite{Kasprzak2006}, superfluidity~\cite{Amo2009}, and Josephson-like oscillations~\cite{Lagoudakis2010,Abbarchi2013}, with peculiarities stemming from the involved exciton mediated repulsive Coulomb interactions and the driven-dissipative nature of the fluid~\cite{CarusottoRMP2013}. The light-matter strong-coupling coherently connects the optical domain (100's of THz) with inter-band optoelectronic excitations~\cite{Kurizki2015}.   The engineering of coupled pairs of polariton traps~\cite{Lagoudakis2010,Abbarchi2013} and arrays~\cite{Hartmann2006,Winkler2015,Alyatkin2020} with controllable interactions~\cite{Kalinin2020} has attained a degree of maturity that enables the implementation of quantum simulators~\cite{Kim2017,Kalinin2020b,Boulier2020,Gosh2020} and topological photonics~\cite{Solnyshkov2021}.  Another emerging area is that of optomechanical crystals (OMXs)~\cite{Thomas2006,PainterOMX,RMP,Ren2020}, hybrid structures that bridge the optical domain with acoustics (MHz-GHz-range). OMXs exploit the Bragg co-localisation of mechanical and optical modes to greatly enhance their interaction. Interestingly, cavity optomechanics has also been exploited to induce gauge fields as a resource for effectively breaking the time-reversal symmetry in topological photonics~\cite{Schmidt2015,Shen2018,Ruesink2018} and phononics~\cite{Brendel2017,Mathew2020}.  
%In this domain one active trend is the search for the ultra-strong coupling regime, characterized by an optomechanical coupling rate exceeding the decay rates of phonons and light, as well as the phonon frequency.~\cite{Forn-Diaz2019,Frisk2019,Hughes2021}
The question then naturally arises: can the powerful developments of cavity optomechanics be used in polariton systems relevant for optoelectronics and quantum technologies? Moreover: can the behaviour of driven-dissipative light fluids be intertwined with coherent vibrations in a lattice to yield a collective behaviour qualitatively different from that of its individual components? We address these questions and answer them positively in this work.

%%%%%%%%%%%%%%%%%%%%%%%%%%%%%%%%%%%%%%%%%%%%%
\begin{figure*}[!hht]
 \begin{center}
    \includegraphics[trim = 0mm 0mm 0mm 0mm,scale=0.4,angle=0]{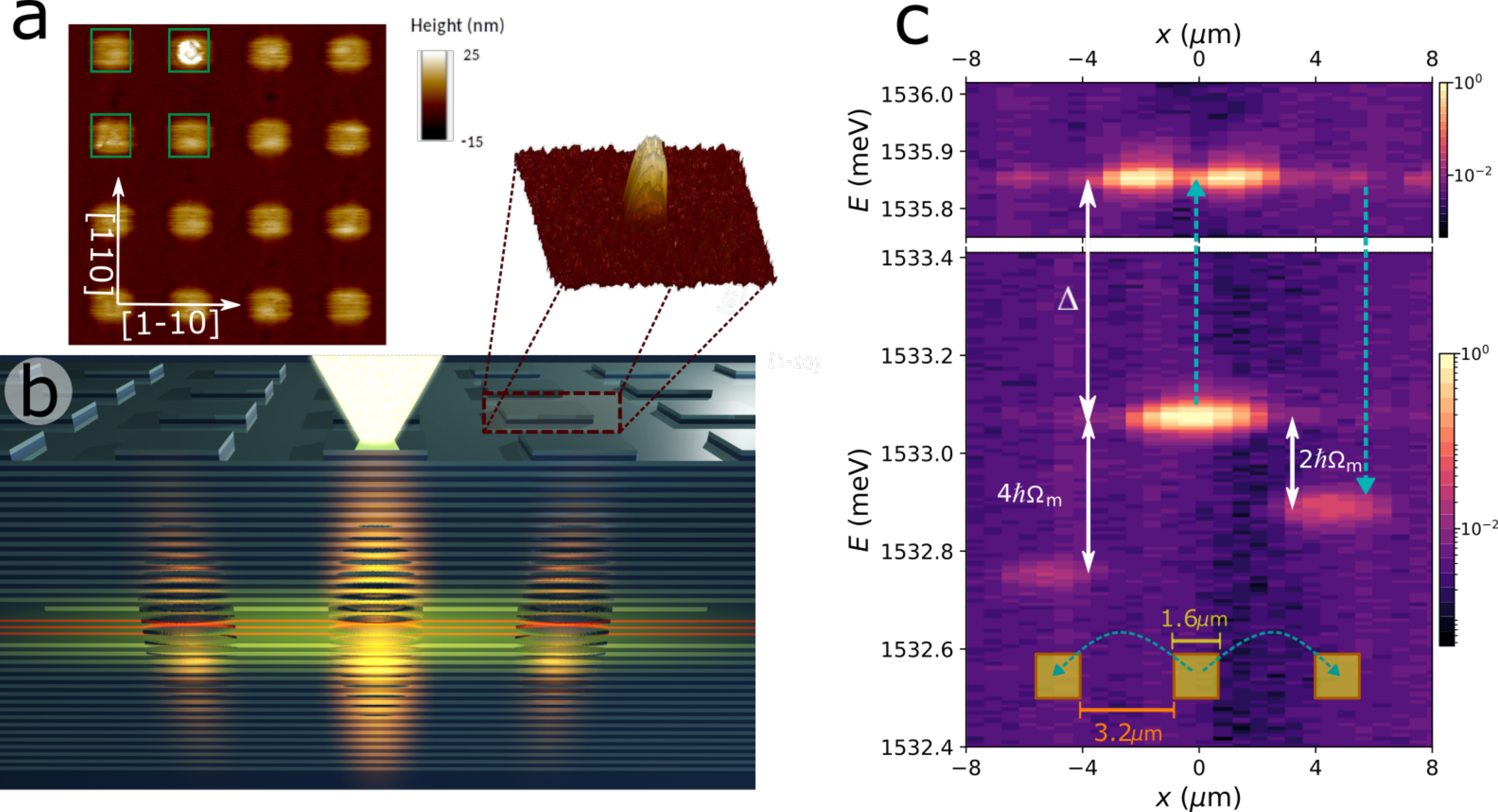}
\end{center}
\vspace{-0.8 cm}
%\hspace{-1.6 cm}
\caption{
\textbf{Polaromechanical metamaterials.}
Panel \textbf{a} is a 2D (top-view) atomic-force microscope (AFM) image of an array of $1 \mu$m square intra-cavity traps separated by $1 \mu$m-wide barriers. The observed rounding of the structures results from the overgrowth of the top-DBR. A scheme of the microcavity patterned with square-shape micrometer-size traps is presented in panel \textbf{b}. The three red lines in the spacer represent the quantum wells where excitons are hosted. The spacer thickness is larger at the position of the traps.
The laser focuses on one of the traps, leading to coupled polaritons and vibrations in the individual unit cells of the array. The inset shows an AFM image of an individual $2 \mu\mathrm{m} \times 2 \mu\mathrm{m}$ square trap. 
Panel \textbf{c} shows the spatially resolved emission spectra obtained for such focused non-resonant excitation with powers above the polariton condensation threshold, for a square array of $1.6 \mu$m square traps separated by $3.2 \mu$m barriers. The central trap ($x=0$) and the two closest neighbors ($x \pm 4.8 \mu$m) can be identified. Both s-like symmetry ground ($\sim 1533$~meV) and p-like symmetry excited ($\sim 1535.9$~meV) states of polaritons in the traps are observed. Note that the neighbor trap ground state energies asynchronously lock at relative detunings corresponding to $\delta E \sim -2 \hbar \Omega_\mathrm{m}$ and $\delta E = -4 \hbar \Omega_ \mathrm{m}$, where $\Omega_ \mathrm{m}/2\pi \sim 20$~GHz is the confined phonon frequency ($\hbar \Omega_\mathrm{m} \sim 80 \mu$eV). The dashed blue-arrows represent the optomechanically-induced inter-trap coupling mediated by virtual transitions to the excited-state.
}
\label{Fig1}
\end{figure*}
%%%%%%%%%%%%%%%%%%%%%%%%%%%%%%%%%%%%%%%%%%%%%

Polariton condensates are of great relevance in cavity optomechanical phenomena due to their long coherence times and the resonantly-enhanced exciton-mediated optomechanical coupling~\cite{Fainstein2013,Restrepo2014,Jusserand2015,Kuznetsov2021}. Both parameters, coherence time and coupling strength,  are critical to boost the optomechanical cooperativity~\cite{RMP}. 
Photons and phonons follow the same wave equation in isotropic materials. Consequently, the same GaAs/AlAs-based distributed Bragg reflector (DBR) planar microcavities leading to cavity polaritons also confine hypersound in the 20~GHz range~\cite{Fainstein2013},  and this allowed the recent demonstration of a polariton-driven GHz phonon laser~\cite{Chafatinos2020}. Inspired then by the idea of OMXs~\cite{Thomas2006,PainterOMX,RMP,Ren2020}, we go beyond the concept of Bragg structures and propose metamaterials based on resonant unit cells~\cite{Lemoult2013}. These consist of  micrometer-size 0D intra-cavity polariton traps~\cite{Winkler2015,Kuznetsov2018}, that confine and co-localize polaritons and acoustic vibrations (`polaromechanical'' individual resonators), arranged into periodic arrays (see the AFM image of an actual array in Fig.~\ref{Fig1}{\bf a} and the scheme in Fig.~\ref{Fig1}{\bf b}). The concept of polaromechanical metamaterials becomes particularly powerful in the regime of negligible Josephson-like inter-site coupling (relatively large inter-trap separations). In this regime, the on-site optomechanical interactions lead to a phonon-mediated strong inter-trap tunnelling mechanism of the polariton condensates~\cite{Reynoso2022}. Striking signatures of the coherent polariton-phonon coupling emerge in this case when the structures are optically driven with a continuous  non-resonant and spatially localized optical excitation close to and above the threshold for condensation. Namely, the ground-states of polariton condensates at neighbor traps asynchronously lock with energies differing by integer numbers of the confined phonon energy, as illustrated in Fig.~\ref{Fig1}{\bf c}. In order to explain this notable result we begin by describing the polariton and phonon bands of polaromechanical arrays. We then focus on the behavior of single traps and coupled double traps to show how such bands emerge from the resonant unit cells. Finally, the characteristics of the novel synchronization phenomenon are analyzed and theoretically modeled, followed by a discussion and outlook.

%%%%%%%%%%%%%%%%%%%%
\section{Results}
%%%%%%%%%%%%%%%%%%%%

%%%%%%%%%%%%%%%%%%%%%%%%%%%%%%%%%%%%%%%%%%%%%
\begin{figure*}[!hht]
 \begin{center}
    \includegraphics[trim = 0mm 0mm 0mm 0mm,scale=0.6,angle=0]{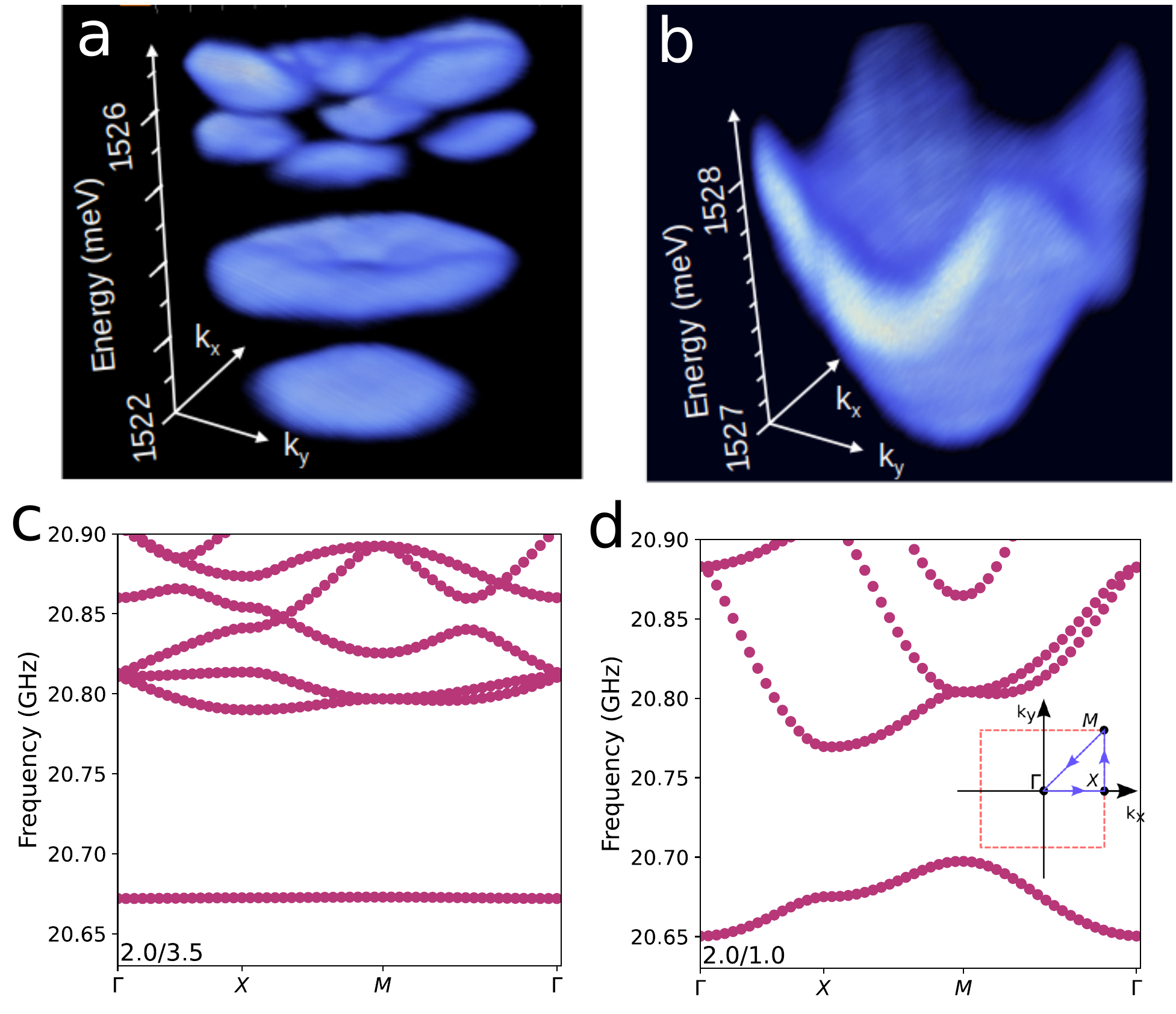}
\end{center}
\vspace{-0.8 cm}
%\hspace{-1.6 cm}
\caption{
\textbf{Polariton and phonon bands in lattices of 0D resonators.}
Panels \textbf{a} and \textbf{b} present the spectrally and wavevector resolved polariton energies obtained by photoluminescence at low excitation powers in lattices of $4 \mu\mathrm{m} \times 4 \mu\mathrm{m}$ and $1 \mu\mathrm{m} \times 1 \mu\mathrm{m}$ square traps, respectively, both with $1 \mu\mathrm{m}$-wide barriers. The larger $4 \mu\mathrm{m} \times 4 \mu\mathrm{m}$ traps display several confined modes, while only the ground state is confined for the smaller $1 \mu\mathrm{m} \times 1 \mu\mathrm{m}$ structures. Panels \textbf{c} and \textbf{d} display the calculated in-plane dispersion of the acoustic modes of lattices of $2 \mu\mathrm{m} \times 2 \mu\mathrm{m}$ square traps separated by $3.5 \mu\mathrm{m}$ and $1 \mu\mathrm{m}$-wide barriers, respectively. 
%These two bands arise from the confined phonon mode at $\sim 20$~GHz, respectively from the s and p-like symmetry states laterally localised in the individual traps. 
%The red circle indicates the $\Gamma$ point mode accessible through our pump and probe technique.
}
\label{Fig2}
\end{figure*}
%%%%%%%%%%%%%%%%%%%%%%%%%%%%%%%%%%%%%%%%%%%%%

\paragraph*{\textbf{Polaromechanical metamaterials: co-localised polariton and phonon bands in lattices of 0D resonators.}} The proposed system is based on $\mu$m-sized traps  previously studied in the context of polariton phenomena~\cite{Winkler2015,Kuznetsov2018,Chafatinos2020}, and created by micro-structuring the spacer of an (Al,Ga)As microcavity in-between growth steps by molecular beam epitaxy (MBE). 
Etching of the microcavity spacer prior to the growth of the top DBR into regions of narrow and wider thickness gives rise to optical cavity modes of higher or lower energy, respectively (see details in the Supplementary Note 1). These provide the means to define traps and barriers in spatially tailored effective potentials. The magnitude of the effective potential modulation acting on the polaritons (typically some meV) is determined by the spacer etching thickness (typically around 10-15~nm), and by the cavity-exciton detuning.  The etching is performed far from the quantum wells (QWs) so that the quality of the resulting excitonic system remains conserved. Arrays of polariton traps can also be fabricated with alternative technologies, as  investigated by other groups (see e.g., Ref.~\onlinecite{CarusottoRMP2013} and references therein). We note, however, that in order to display strong optomechanical phenomena as reported here, the embedded QWs need to be displaced away from the position of the maximum cavity optical field. This is opposed to what is usually done to maximize the photon-exciton strong-coupling. At the position of the maximum of the optical field the strain associated with the confined phonon field is zero, and thus the exciton-mediated polariton-phonon coupling vanishes (see details in the Supplementary Note 1B).

Panels {\bf a} and {\bf b} in Fig.~\ref{Fig2} present the measured 2D polariton in-plane energy dispersion (i.e., along the $k_x-k_y$ plane) for a $7 \times 7$ array of $4 \mu\mathrm{m} \times 4 \mu\mathrm{m}$ square traps and a $10 \times 10$ array of $1 \mu\mathrm{m} \times 1 \mu\mathrm{m}$ square traps, respectively, in both cases separated by $1 \mu$m barriers. 
These dispersions were measured by angular-resolved photoluminescence (PL) with low optical powers (i.e., well below the BEC threshold). The full array was homogeneously illuminated with a large spot of $\sim 50 \mu$m diameter. The larger $4 \mu\mathrm{m} \times 4 \mu\mathrm{m}$ traps confine several states with energies below the finite barriers~\cite{Kuznetsov2018}. When organized in arrays this results in the formation of several bands, as displayed in Fig.~\ref{Fig2}{\bf a}. The lower energy band derives from trap states of s-like symmetry, and is comparatively flat due to the larger degree of confinement (lower hybridization with neighbor traps) of these states. In contrast, the smaller $1 \mu\mathrm{m} \times 1 \mu\mathrm{m}$ traps only confine the ground state (Fig.~\ref{Fig2}{\bf b}) . Note that due to the smaller size, in this latter case the ground state is closer to the barrier edge. Consequently, the hybridization is larger resulting in a broader band (when compared to the $4 \mu\mathrm{m} \times 4 \mu\mathrm{m}$ case). The reciprocal space dispersion in this case resembles the one for dispersive electron bands in a tight-binding model, where the role of the atomic electron level is played by the discrete polariton ground state of the 0D traps.

As mentioned above, in planar microcavities vibrations are also confined~\cite{Fainstein2013}, with fundamental frequency  $\Omega_\mathrm{m}/2\pi \sim 20$~GHz corresponding to a breathing of the cavity spacer along the growth direction. The wavelength of these confined phonons is the same as the one for the confined photons (determined by the cavity spacer thickness). The frequency difference (from tens of GHz for the phonons to hundreds of THz for the photons) just bears the relation between the respective wave speeds. The local etching of the spacer thickness blue-shifts the phonon mode energy in the same proportion as for the confined photons (see Supplementary Note 4) and, consequently, an effective lateral potential develops for the confined acoustic phonons as it does for the polaritons. The calculated {\em phonon} dispersion around the $\Gamma$ point of the Brillouin zone ($k_x=k_y=0$) for arrays of
$2 \mu\mathrm{m} \times 2 \mu\mathrm{m}$ traps with $\sim 3.5 \mu$m and $\sim 1 \mu$m-wide barriers are presented in Figs.~\ref{Fig2}{\bf c} and d, respectively (the model used is described in the Methods section). Phonon bands arise for the 2D lattice of traps and, mimicking what was observed for the polaritons, the lower energy band is flatter for the case of more isolated ground states (Fig.~\ref{Fig2}{\bf c}). Note that for the polariton dispersions presented in Figs.~\ref{Fig2}{\bf a} and {\bf b} the width of the bands is tuned by the size of the traps, while for the calculated phonon bands in Figs.~\ref{Fig2}{\bf c} and {\bf d} this same objective is accomplished by changing the barrier width.

%%%%%%%%%%%%%%%%%%%%%%%%%%%%%%%%%%%%%%%%%%%%%
\begin{figure*}[!hht]
    \begin{center}
    \includegraphics[trim = 0mm 0mm 0mm 0mm, clip=true, keepaspectratio=true, width=1.8\columnwidth, angle=0]{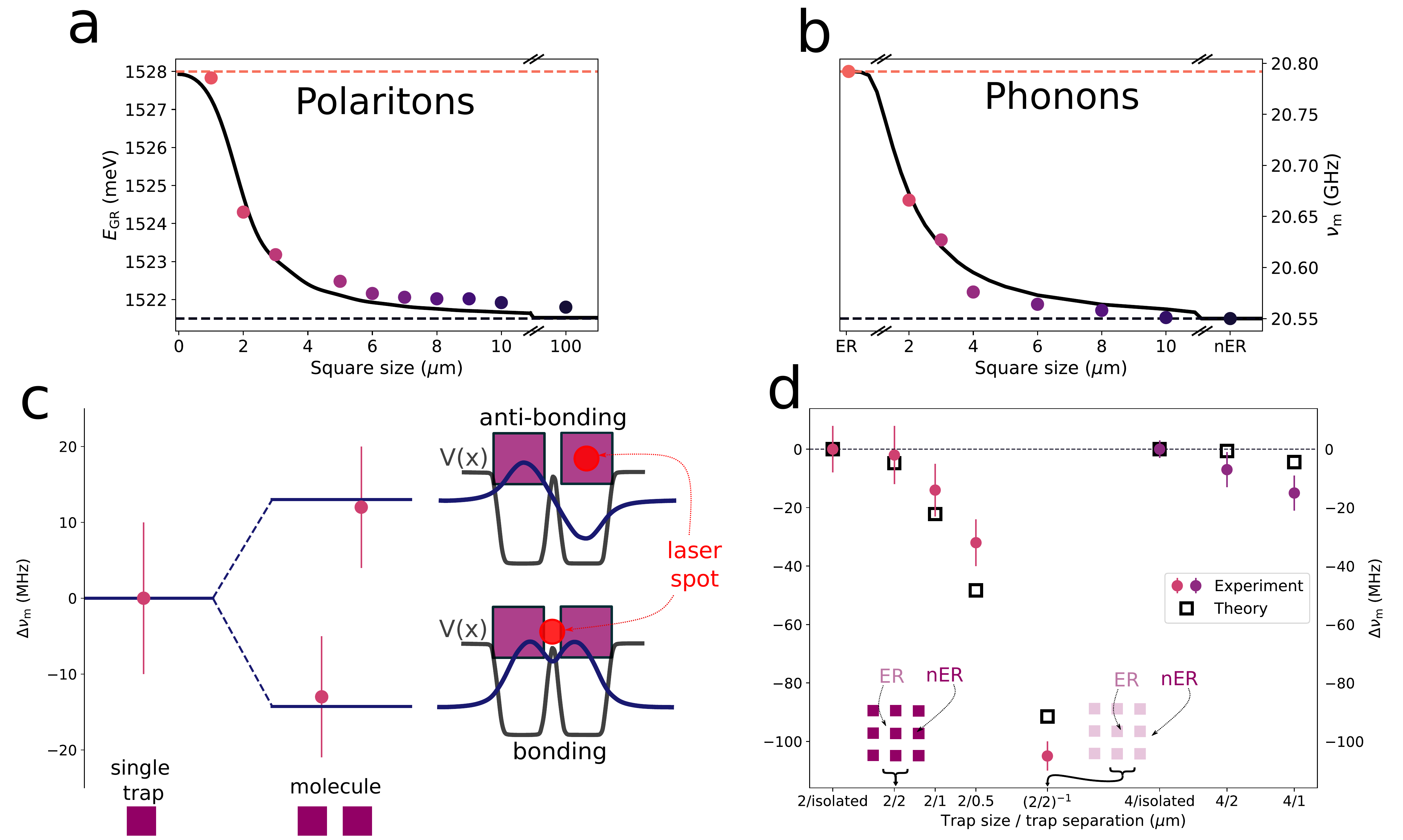}
    \end{center}
    \vspace{-0.8 cm}
\caption{
{\bf Experimental study of confined phonons in traps and lattices of traps.}
%Panel \textbf{a} shows the finite-element 3D modelling of the spatial distribution of the volumetric strain associated to the fundamental confined acoustic mode of an individual circular trap of $5 \mu$m diameter. The cylinder is the full modelled region, much larger than the trap (the non-etched region is indicated with the darker color in the top). A scheme of the experimental pump and probe method used to study the confined phonons is included (see text for details). 
Panel \textbf{a} (\textbf{b}) presents the measured size dependence of the polariton (phonon) ground confined mode in isolated square traps. The limits of the induced potential, corresponding to planar etched (ER, barrier) and non-etched (nER, well) regions are indicated by the dashed horizontal lines. The continuous black curve is the effective potential model. The case for a molecular-like phonon double-trap made of two $2 \mu\mathrm{m} \times 2 \mu\mathrm{m}$ traps separated by $0.5 \mu$m, is illustrated in panel \textbf{c}. The phonon potential, and the calculated bonding and anti-bonding-like levels are shown, together with a scheme of the respective excitation condition at different sites. Symbols correspond to the experiments, while the solid lines are the theoretical predictions.
%Because different positions in the wafer correspond to different absolute energies, we have slightly shifted rigidly in the figure the energy of the two molecule states so that its centre of mass coincides with the single trap frequency.
Panel \textbf{d} summarises the phonon experiments on arrays of traps of 4 and 2 $\mu$m lateral size, with different inter-trap separation, and their comparison with the effective potential model. Again the energies are given with respect to that of the corresponding single trap (identified as ``2/isolated'' and ``4/isolated'' in this panel). The arrays are labeled as $a/b$, with $a$ the trap size, and $b$ the inter-trap separation. The case of an inverted array of $2 \mu$m etched regions separated by $2 \mu$m non-etched regions (labeled as $(2/2)^{-1}$ is also included. Circular full (square empty) symbols are the experimental (theoretical) values. 
}
\label{Fig3}
\end{figure*}
%%%%%%%%%%%%%%%%%%%%%%%%%%%%%%%%%%%%%%%%%%%%%

\paragraph*{\textbf{Experimental study of phonon confinement in traps.}} 

Figure~\ref{Fig3} expands on an experimental study of the phonons confined in individual square traps, molecule-like double-traps, and in arrays of traps.  
%The spatial distribution of the mechanical fundamental eigenmode ($\sim 20$~GHz) calculated using finite elements is shown in panel a (see the Supplementary Material IVB). This was done for the simplest geometry consisting of a circular trap at the center of a cylindrical pillar limited by vacuum. It is conceptually similar to the square trap, but computationally simpler because the rotation symmetry requires solving only a 2D problem. Full consideration of the sample layer structure is included as prescribed by the MBE growth. The trap is modeled as a central $5 \mu$m diameter non edged region with larger spacer thickness (indicated in the top of the figure with a darker color). The color coded acoustic amplitudes clearly show that the traps indeed confine phonons in 0D, in much the same way as they confine light: i.e., the Bragg reflectors strongly confine the phonons along the growth direction, while the microstructuring of the spacer thickness creates an effective lateral potential that confines the phonons in the remaining two dimensions. 
Panels {\bf a} and {\bf b} in Fig.~\ref{Fig3} present a comparative study of the polariton and phonon ground-state energy for {\em isolated} traps of varying square size. The polariton energies in panel {\bf a} were obtained from PL data recorded at 5~K under low excitation power (i.e., well below BEC)~\cite{Kuznetsov2018}. The trap spectra correspond to discrete levels, the number of confined states depending on the size of the traps (only data for the ground state are presented in Fig.~\ref{Fig3}{\bf a}). The limits of the induced potential, corresponding to measurements performed on planar etched (barrier) and non-etched (well) regions are indicated by the dashed horizontal lines. The expected size-dependence for a trapping potential with finite barrier height is observed. The experimental data are compared in Fig.~\ref{Fig3}{\bf a} with a theoretical calculation based on an effective potential model with a realistic description of the trap potential~\cite{Kuznetsov2018} (black solid curve, see Supplementary Note 4 for details), showing excellent agreement.

To experimentally study the high frequency vibrations in the same traps, we used a picosecond coherent phonon pump and probe technique~\cite{Thomsen1986,Matsuda2005,Kimura2011}. A ps-laser pulse is used to resonantly excite the optical cavity mode, generating coherent phonons by a displacive mechanism.  These mechanical oscillations in turn modulate the cavity energy, thus allowing their detection using a delayed probe pulse. Traps were individually addressed using a microscope set-up with a $\sim 3~\mu$m-wide Gaussian spot (more details are provided in Supplementary Note 2). 
%The concept is illustrated by the scheme in Fig.~\ref{Fig3}a and in more details in the Methods section. 
The dashed horizontal lines in Fig.~\ref{Fig3}{\bf b} represent the measurements in extended planar non edged and edged regions, respectively. These define, as for the polariton case discussed above, the limits of the induced lateral phonon effective potential (traps and barriers, respectively). Within these limits the phonon trap energies increase with decreasing size, as expected for confined states in a trap with finite barriers (examples of experimental spectra are provided in the Supplementary Note 3). The observed shift is very well described by a phonon effective potential model (black solid curve) based precisely on the same parameters for the traps as used to describe the polariton energies (see the Methods section).  The similarity with the behavior of polaritons in Fig.~\ref{Fig3}{\bf a} emphasises the concept of polaromechanical traps in which both polaritons and phonons are confined in 0D resonators. We have detected confined polaritons and phonons in traps with dimensions down to $1\mu\mathrm{m} \times 1 \mu\mathrm{m}$, exhibiting record coherence times for polariton condensates (ns-long) and for confined phonons (100's ns) with no observable reduction with decreasing trap size. We note, for comparison, that a study of cavity confined phonon dynamics in etched micropillars~\cite{Anguiano2017} showed a significant decrease of the mechanical mode lifetime for pillar diameters below $7 \mu$m. It was argued there that phonon losses induced by lateral surfaces become more relevant for smaller devices because of an increased surface to volume ratio.

We now turn to architectures of coupled 0D resonators. Experiments demonstrating the formation of polariton bands were described in Figs.~\ref{Fig2}{\bf a} and {\bf b}, we concentrate here on the phonon properties. The case of a molecule-like structure made of two $2 \mu\mathrm{m} \times 2 \mu\mathrm{m}$ traps separated by $0.5 \mu$m is presented in Fig.~\ref{Fig3}{\bf c}. The symbols correspond to the experimental values, and the horizontal lines are the calculated energies (see the Supplementary Note 4).  To selectively excite each state we positioned the laser spot either symmetrically between the traps, or on top of one of them (as shown in the scheme of Fig.~\ref{Fig3}{\bf c}).
Two states of anti-bonding and bonding character arise, split relative to the individual trap by the interaction energy $\pm J$. The modelled effective phonon potential corresponding to the double-trap structure is also shown, together with the calculated spatial shape of the resulting bonding and anti-bonding states.  The measured splitting of the modes, and their ordering (bonding state at lower energy), are in excellent agreement with the theory.

Figure~\ref{Fig3}{\bf d} summarizes experiments on a series of arrays of square traps of $2 \mu$m and $4 \mu$m lateral size, and different inter-trap distance. These are labeled as $a/b$, where $a(b)$ identifies the trap size(separation) in $\mu$m. An ``inverted'' array is also included, corresponding to etched squares of $2 \mu\mathrm{m} \times 2 \mu\mathrm{m}$ size, separated by $2 \mu$m-wide non-etched channels [labeled as $(2/2)^{-1}$].  All frequencies are given with respect to that of the respective isolated single trap. The measured frequencies are also compared to those calculated for the $\Gamma$ point ``s''-like vibrations (i.e., the lower energy mode at $k_y=0$ in Figs.~\ref{Fig2}{\bf c} and {\bf d}), shown with open squares in Fig.~\ref{Fig3}{\bf d}.  Due to the experimental geometry with light incident within a small cone around the normal direction, the probe pulse couples with $k \sim 0$ vibrations. Since, in addition, the spatial distribution of the Gaussian pulse is uniform at the scale of an individual trap, it is sensitive to the more symmetric ``s''-like ground state. Note that the observed red-shift of the detected modes with decreasing inter-trap separation (e.g, from $2/2$ to $2/0.5$) is a measure of the array phonon half band-width ($J$). For inter-trap separations $\gtrsim 2~\mu$m the $\Gamma$ point mode is almost coincident in frequency with that of the individual trap, signalling the flat-band limit. Note also the weaker red-shift of the $\Gamma$ point mode of the $4 \mu\mathrm{m} \times 4 \mu\mathrm{m}$ trap lattices, when compared to those constructed from $2 \mu\mathrm{m} \times 2 \mu\mathrm{m}$ traps,  reflecting the relatively flatter-band nature of the former. Indeed, as for the polariton bands in Fig.~\ref{Fig2}{\bf a}, for larger traps the on-site energies are smaller and thus red-shifted farther down from the barrier edge. Consequently, the states become less delocalised when traps are coupled in an array. Several additional examples of the dispersion of phonon modes in arrays of different trap sizes and separations can be seen in the Supplementary Note 4.

%%%%%%%%%%%%%%%%%%%%%%%%%%%%%%%%%%%%%%%%%%%%%
\begin{figure*}[!hht]
    \begin{center}
    \includegraphics[trim = 0mm 0mm 0mm 0mm, clip=true, keepaspectratio=true, width=1.2\columnwidth, angle=0]{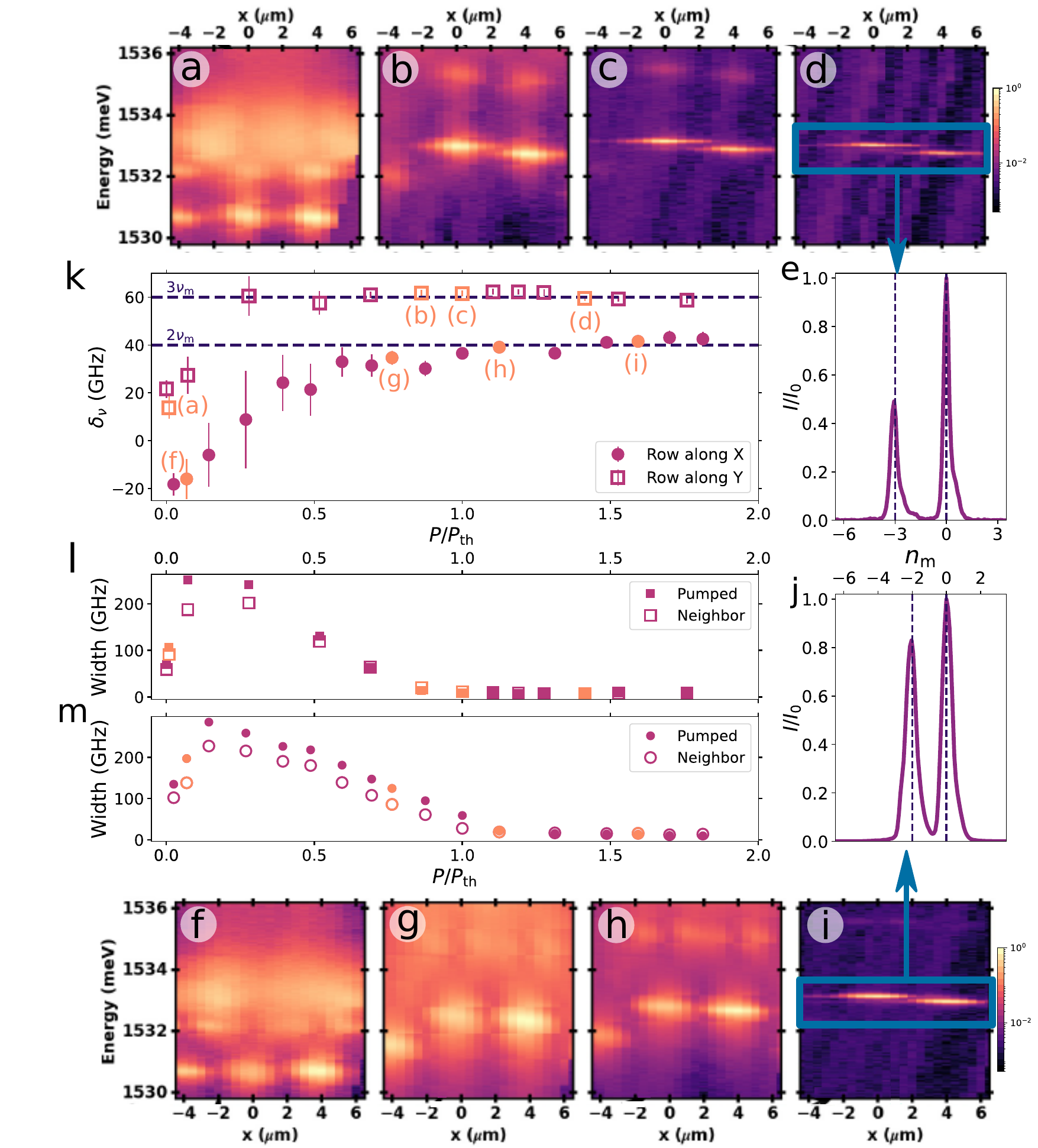}
    \end{center}
    \vspace{-0.8 cm}
\caption{
{\bf Polaromechanics: asynchronous locking.}
Excitation power dependence of the polariton emission spectra for an array of square traps of $1.3 \mu$m lateral size separated by $2.6~\mu$m- etched regions. Examples of energy-resolved spatial images for some selected pump powers are shown in panels \textbf{a-d} for a row of traps along the crystal direction Y [-1 -1 0] (see the AFM image in Fig.~\ref{Fig1}{\bf a}). The spectrum corresponding to high-powers is displayed in panel \textbf{e} (the vertical dashed line indicates an integer number of phonon quanta).
A similar sequence of spatial images and high-power spectrum corresponding to a row of traps along the orthogonal direction X [-1 1 0], are shown in panels \textbf{f-j}. Panel \textbf{k} summarizes the frequency detuning between the pumped and a neighbor trap as a function of pump power (given in terms of $P_\mathrm{Th}$, the condensation threshold power), for the two studied cases. The symbols corresponding to the shown spatial images are differentiated with orange color. Panel \textbf{l}(\textbf{m}) is the power dependence of the emission linewdith corresponding to the pumped (solid symbol) and most intense neighbor trap (empty symbol) for the data taken along Y (X). In panels \textbf{k-m} circles represents measurements along X, and squares along Y.
}
\label{Fig4}
\end{figure*}
%%%%%%%%%%%%%%%%%%%%%%%%%%%%%%%%%%%%%%%%%%%%%

\paragraph*{\textbf{Driven polaromechanical crystals: asynchronous energy locking.}} 
We turn now to the main result of our work, i.e. the observation of a locking of the inter-trap polariton energy detuning 
at fixed differences that correspond to integer numbers of the phonon energy (asynchronous locking).
This occurs when the in-plane polariton dynamics in these polaromechanical metamaterials is induced by optical driving the traps with excitation power close to and above the threshold for BEC. This is done through non-resonant continuous wave optical excitation with the $\sim 3 \mu$m diameter laser spot. A Gaussian shape exciton reservoir is formed, with a size that can be somewhat larger but of the order of the spot size. The related Coulomb repulsion, together with residual disorder, and the inter-trap coupling $J$, are expected to determine the local energies (and thus the dynamics) of polaritons in the traps. When polariton traps are coupled (large $J$), the Josephson flow is responsible for the phase locking between the condensates~\cite{Wouters2008,Eastham2008,Eastham2021}. If the potential difference between the traps (determined by local disorder~\cite{Baas2008,Ohadi2018} and eventually by an inhomogeneous Coulomb interaction~\cite{Wertz2010},  as is the case for our Gaussian shape excitation) exceeds a certain critical value (limit of small $J$), the Josephson flow cannot reach a steady state and the condensates cannot synchronize. 
%Synchronisation between distant polariton condensates has been theoretically proposed,~\cite{Wouters2008,Eastham2008} and also experimentally observed.\cite{Baas2008,Ohadi2018} 
%Indeed synchronisation is a quite general behaviour intimate to dynamical non-linear systems, and pervasive to very different domains of nature.~\cite{Synchro}
Quite notably, however, as shown in Fig.~\ref{Fig1}{\bf c} for an array of $1.6 \mu\mathrm{m} \times 1.6 \mu\mathrm{m}$ traps separated by $3.2 \mu$m barriers, the neighbour site polariton condensate ground state energies in our experiments neither follow a simple Gaussian distribution (as would be expected for uncoupled polariton condensates) nor share the same frequency (as would happen if they synchronize), but lock at detunings that correspond to integer multiples of the mechanical phonon energy $\hbar \Omega_m$ - the asynchronous locking.

Indeed this striking result is not an exception but the rule, as further shown in Fig.~\ref{Fig4} with a similar experiment in a different square array of $1.3 \mu\mathrm{m} \times 1.3 \mu\mathrm{m}$ square traps separated by $2.6~\mu$m barriers (see also Supplementary Note 7 for a larger set of data). Figures~\ref{Fig4}{\bf a-d} present energy-resolved spatial images for increasing non-resonant excitation power, obtained from one row of traps aligned along the crystal direction Y [-1 -1 0].  At very small powers, three traps characterized by relatively broad lines are observed approximately at the same ground state energy (the central trap is at $y=0~\mu$m, the closest neighbours appear at $\sim \pm 4~\mu$m). As the power increases two of the traps blue-shift (panels {\bf a-c}), as expected from the Coulomb repulsion with the exciton reservoir, the third remaining farther away in energy. This asymmetry between left and right traps arises from an involuntary misalignment of the laser spot towards one of the neighbours. As the modes shift with increasing pump power, a clear line-narrowing and non-linear increase of intensity signals the transition to the condensation of the light fluid above a threshold power ($P_\mathrm{th}$). At the highest powers, the blue-shift saturates due to higher order non-linearities. Note that the energy of the neighbor trap that also blue-shifts gets rapidly locked with increasing power at a detuning that corresponds to three times the phonon frequency (see the high-power spectra displayed in Fig.~\ref{Fig3}{\bf e}). Figures~\ref{Fig4}{\bf f-i} display the corresponding spatial images from the same array, but now for a row of traps aligned along the orthogonal crystal direction X [-1 1 0] (square traps grow with a slight rectangular asymmetry, so that the coupling $J$ of the ground states along Y is somewhat larger than along X~\cite{Kuznetsov2018}). The blue-shift and saturation of the modes with increasing excitation power is observed, again particularly for two of the traps that evolve in energy close one to the other. A similar line narrowing and intensity increase indicates polariton condensation and, as evidenced by the high-power spectrum in Fig.~\ref{Fig4}{\bf j},  the neighbor trap detuning locks again at an integer number of the phonon energy quanta (two in this experiment).

The full power dependence of the detuning between the most intense neighbor trap and the central one for the two reported cases (along X displayed with circles, and along Y shown with squares) is shown in Fig.~\ref{Fig4}{\bf k} further illustrating the locking at integer numbers of the phonon energy quanta. Figures~\ref{Fig4}{\bf l} and \ref{Fig4}{\bf m} present the corresponding linewidth as a function of optical excitation power.  Solid (open) symbols correspond to the central (neighbor) trap. For both experiments, the linewidth of the central trap emission at low powers is larger than that of the neighbor trap. This arises due to the larger fluctuations of the reservoir, that is more populated at the position of the central trap. Notably, when polaritons in the two traps asynchronously lock their linewidths become identical. Moreover, for these experiments polaritons in the two traps condense at the same applied power.  For decoupled traps $P_\mathrm{th}$ is expected to be determined by the local exciton reservoir population, which is different for the two traps. The identical linewidth and  $P_\mathrm{th}$ values thus provide strong evidence that the polaritons localized within neighbor traps couple in a single coherent state when they asynchronously lock mediated by the phonon induced coupling.

Our results evidence a very strong interaction between polaritons and phonons in these resonant polaromechanical metamaterials.  We have observed that the confined vibrations play a particularly relevant role in the described collective dynamics when the direct Josephson-like coupling between the ground states of neighbour traps is negligible, i.e. in the flat-band limit discussed in the previous sections. This is indeed the case for the data displayed in Figs.~\ref{Fig1}{\bf c} and \ref{Fig4}, corresponding to arrays of relatively distant traps (see the analysis in Supplementary Note 6).  It can be shown~\cite{Reynoso2022} via second-order perturbation theory that in this case a phonon-induced inter-site {\em quadratic} optomechanical coupling $g_\mathrm{2}$ arises, due to virtual transitions  between the isolated polariton ground states and an extended p-like excited level at energy $\Delta$ above the ground state, induced by the on-site linear optomechanical coupling $g_\mathrm{0}$ (see the scheme in Fig.~\ref{Fig1}{\bf c}  and the Supplementary Note 6). For small traps, such as those leading to the locking behavior shown in Fig.~\ref{Fig1}{\bf c} and Fig.~\ref{Fig4}, the involved p-like state at $\Delta \sim 3$~meV is indeed shared by both sites. For this case it follows that $g_{\mathrm{2}}=g_\mathrm{0}^2/\Delta$~\cite{Reynoso2022,Paraiso2015}. The concept of a polaromechanical metamaterial thus becomes central in this regime, because the polariton inter-trap coupling is determined by the on-site optomechanical interactions of the hybrid metamaterial resonant unit-cell.

Our observations are related to the physics of synchronization, which has been observed in polariton systems~\cite{Wouters2008,Eastham2008,Eastham2021,Baas2008,Ohadi2018}. We will base in this previous developments, extending them to the case of a dynamical mechanically-mediated inter-trap coupling. To model the observed physics we follow Wouters~\cite{Wouters2008}, and describe two coupled polariton modes and the corresponding reservoir as (we take $\hbar=1$):
\bea
\nonumber
i \dot{\psi_j}&=&(\ve_j+U_j |\psi_j|^2+U_j^R n_j)\psi_j-J\psi_{3-j}+\frac{i}{2}(R n_j-\gamma)\psi_j \\
\dot{n_j}&=& P_j-\gamma_R n_j-R|\psi_j|^2 n_j.
\label{GP}
\eea
Here $\ve_j$ is the bare energy of the $j$-mode ($j=1,2$), $U_j$ and $U_j^R$ are the polariton-polariton and polariton-reservoir interaction couplings, $J$ describes a hopping term between modes, $\gamma$ the polariton decay and $R$ the stimulated loading from the reservoir. The dynamic of the later is controlled by the pump power $P_j$, the excitonic decay rate $\gamma_R$ and the stimulated decay to the condensate.

One seeks for a solution of the form~\cite{Wouters2008}: $\psi_j=\sqrt{\rho_j}e^{-i\omega t\pm \theta/2}$, that is, two states with a synchronized identical frequency $\omega$ (here the $+$ sign corresponds to $j=1$) and $\dot{n_j}=0$. After some algebra, and taking $R|\psi_j|^2 >>\gamma_R$, one finds that synchronization exists whenever there is a solution for $\theta$ that satisfies $\Delta\ve = f(\theta)$, with:
%\begin{widetext}
\be
f(\theta)=J\cos\theta\left(\frac{1}{\alpha}-\alpha\right)
+\frac{U_1^R \xi _1^0}{\xi _1}n_0-\frac{U_2^R \xi _2^0}{\xi _2}n_0
+\xi_1 U_1 \rho_0
-\xi_2 U_2  \rho_0.
\label{Theta}
\ee
%\end{widetext}
In the above  $\alpha=\sqrt{\rho_2/\rho_1}$, $\Delta\ve=\ve_2-\ve_1$ and we have introduced some dimensionless  parameters so that $\xi_j=\rho_j/\rho_0$, $\rho_0=\gamma_R/R$, and $n_0=\gamma/R$.  
%Once solved, the locking frequency is given by
%\be
%\omega=\frac{\bar{\ve}_1+\bar{\ve}_2}{2}-\sqrt{\left(\frac{\bar{\ve}_1+\bar{\ve}_2}{2}\right)^2+J^2\cos^2(\theta})
%\ee
%where $\bar{\ve}_j=\ve_j+U_j \rho_0 \xi_j+U_j^Rn_0 \frac{\xi_j^0}{\xi _j}$ is the dressed energy. 
A solution for Eq.~\ref{Theta} exists even in the absence of polariton-polariton and polariton-reservoir interactions when $\Delta\ve \leq \frac{2J^2}{\gamma}$~\cite{Wouters2008}. It turns out, however, that the presence of $U_j$ and $U_j^R$ strongly favors the appearance of a synchronized phase (see e.g., Fig.~1 of Ref.~\onlinecite{Wouters2008}).

In Ref.~\onlinecite{Wouters2008}, $J$ describes a time-independent direct Josephson coupling which, as argued above, in the case described here can be considered null for the ground state. For an inter-trap coupling that is quadratic in the phonon displacement, we have instead a time dependent hopping $J(t)=J_m (e^{i 2 \Omega_m t}+e^{-i 2 \Omega_m t}+2)$ with $J_m=g_2 n_b$.  This follows from assuming the presence of a coherent population of $n_b$ phonons, i.e., $b(t)+b(t)^* =2\sqrt{n_b} \cos(\Omega_m t)$, and an optomechanical interaction having a hopping term between the two polariton modes with a prefactor containing the second-order power of the phonon displacement, namely, $-\hbar g_2 (\hat{b}+\hat{b}^\dagger)^2$ (see Supplementary Note 8). 
In the rotating wave approximation, keeping only the resonant terms, we have (assuming $\ve_1>\ve_2$):
\begin{widetext}
\bea
\nonumber
i \dot{\psi_1}&=&(\ve_1+U_1 |\psi_1|^2+U_1^R n_1)\psi_1-J_me^{-i2\Omega_m t}\psi_{2}+\frac{i}{2}(R n_1-\gamma)\psi_1 \\
\nonumber
i \dot{\psi_2}&=&(\ve_2+U_2 |\psi_2|^2+U_2^R n_2)\psi_2-J_me^{i2\Omega_m t}\psi_{1}+\frac{i}{2}(R n_2-\gamma)\psi_2. \\
\label{GP_ph}
\eea
\end{widetext}
By comparison with Eq.~\ref{GP} above it follows that now solutions can be found proposing $\psi_1=\sqrt{\rho_1}e^{-i\omega t+\theta/2}$ and $\psi_2=\sqrt{\rho_2}e^{-i(\omega-2\Omega_m) t-\theta/2}$, i.e. two condensates with frequencies locked at a fixed detuning given by $2\Omega_m$. The resulting synchronization conditions are the same as above (Eq.~\ref{Theta}) except that now $\ve_2 \rightarrow  \ve_2+2\Omega_m$, and hence $\ve_2=\ve_1-2\Omega_m+f(\theta)$. Consequently all the conclusions derived for synchronized polariton condensates also apply here, except that synchronization is now represented by the mechanically induced {\em asynchronous} locking of the polariton condensate energies. It can be shown that our system satisfies the conditions to be in this pseudo-synchronized phase (see the numerical simulations in the Supplementary Note 8).  Note that the observation that the number of exchanged quanta in our experiments is $n>1$ is compatible with a higher-order non-linear phonon-mediated coupling between the polariton condensates of neighbour traps, as proposed here. 
%A mechanically induced inter-trap coupling linear in the mechanical displacement (as is more standard in cavity optomechanical systems~\cite{RMP}) would lead to asynchronous locking detuned by the energy of a single phonon.
%Summarizing, we have shown that non-linear interactions, manifested for example by the polariton blue shift that in our samples vary between 1 to 2~meV, contribute to the formation of a quite peculiar synchronized phase delocalized over potential traps. Indeed, nonlinear interactions allow the modification of the initial detuning between polariton states corresponding to different potential minima (which may be different due to fabrication uncertainties and the local Coulomb interaction with the reservoir), up to the point where the mechanically-induced inter-trap coupling can operate to induce the asynchronous locking (at values that corresponding to integer numbers of the phonon energy $\sim 20$GHz ($\sim 80 \mu$eV). 
Interestingly, a quite similar asynchronous locking behavior has been reported recently in a completely different setting. Indeed, it was observed that {\it Pitangus sulphuratus}, a bird from the Americas, manages to lock the frequency difference between its two vocal cords.
The mechanics of this bird-singing responds to non-linear dynamical equations very similar to those used above to describe the coupled condensates~\cite{Doppler2020}.
% highlighting the universality of the observed behavior. 
Polaromechanical metamaterials thus express universal physics emerging from interacting optomechanics in nonlinear polaritonic systems~\cite{Bobrovska2017,Zambon2022}.

\section{Discussion and outlook}
%%%%%%%%%%%%%%%%%%%%%%%%%%%%%%%%

%Synchronisation between distant polariton condensates has been theoretically proposed,\cite{Wouters2008} and also experimentally observed.\cite{Ohadi2016,Ohadi2018,Topfer2020} Indeed synchronisation is a quite general behaviour intimate to dynamical non-linear systems, and pervasive to very different domains of nature.~\cite{Synchro} Interestingly, when the deep quantum limit is attained, energy quantisation hinders synchronisation of identical {\em non-linear} oscillators.~\cite{Lorch2017} Interaction requires an exchange of energy, and in the quantum regime the possible quanta of energy are discrete. If the extractable energy of one oscillator does not exactly match the amount that the second oscillator may absorb, interaction, and thereby synchronisation, is blocked.  Polariton condensates are single quantum coherent states, but they are not in the deep quantum regime in the sense that they contain a large number of particles. However, a similar phenomenon emerges because of i) the intrinsic non-linearity of the system required for the asynchronous locking, ii) the inter-trap coupling that is mediated by vibrations, and iii) the BEC linewidth that is much smaller than the phonon frequency. In the polaromechanical metamaterials the polariton condensates and the phonon fields become intimately intertwined in a collective mode that involves the exchange of energy in finite amounts given by the confined phonon energy quanta. 

The polariton trap arrays studied here are of the same kind as the one investigated in Ref.~\onlinecite{Chafatinos2020} and in which polariton-driven phonon lasing was reported.  The condition for the observation of such mechanical self-oscillation is that two polariton states have to satisfy the proper resonance condition, i.e., to be detuned by an integer number of phonon quanta, $n>1$. The evidence for such phonon lasing was the observation of clear and intense sidebands separated by $\hbar \Omega_m$ from the emission of both ground and excited states of the traps. It follows from the extensive experiments reported here (see Supplementary Note 7) that this is not necessarily always the case: sometimes we do indeed observe very clear and strong optomechanically-induced sidebands, in other cases we find that the levels asynchronously lock as reported here but there are no clear sidebands, and sometimes the two effects happen together (asynchronous locking plus weaker but discernible sidebands). Notably, from our systematical studies we find that the locking without intense sidebands is more frequent than the case of large amplitude sidebands. It seems also to be the case that a larger excitonic component favors the appearance of the sidebands. For the model in Eq.~\ref{GP_ph}, inclusion in the rotating wave approximation of only the resonant terms leads to locking without sidebands. Consideration of the counter-rotating term is accompanied in the simulated spectra by sidebands. Moreover, it turns out that mechanical coherent oscillations of amplitude of just a few percent of the phonon energy are enough to establish the asynchronous locking, but would lead to very weak and thus hardly observable sidebands (see the Supplementary Note 8). It is our understanding that both, the emergence of sidebands and the asynchronous locking, are signatures of the existence of a mechanical coherent oscillation.

In summary, we have demonstrated a new concept for polaromechanical metamaterials, based on planar arrays of intra-cavity traps that confine, co-localise, and strongly couple vibrations and polariton light fluids. The building blocks are micron-size high-Q resonators for polaritons and sound, that can be arranged and inter-linked in arbitrary tailored architectures. The involvement of polariton condensates assures very long coherence times, exceeding the mechanical oscillation period, and leads through an exciton-mediated resonant interaction to hugely enhanced optomechanical couplings. Novel phenomena arise in these polaromechanical crystals, particularly the asynchronous locking of condensed light fluids from neighbour traps at detunings that scale with integer numbers of the phonon energy. This evidences a coherent collective dynamics of the polariton and phonon fields, opening the path to novel hybrid scalable platforms applicable for the bi-directional coherent conversion of light-to-microwaves in the 20~GHz frequency range, and for the ultrafast coherent mechanical control of light fluids in quantum technologies.

\paragraph*{Methods:} 
%(to be completed)
\paragraph*{\textbf{Photoluminescence spectroscopy.}}  
For the polariton PL experiments in the traps at 5-10~K, an external cavity-stabilized $cw$ Spectra Physics Ti-Sapphire Matisse laser was used for the non-resonant excitation at $\sim 760$~nm. The wavevector dependent energy dispersions in Fig.~\ref{Fig2} were obtained by exciting the trap-arrays with a large $\sim 50 \mu$m laser spot, using small optical powers well below the condensation threshold. The wavevector dependence of the spectra was determined with standard methods based on angle-resolved light collection. The optical driving of the condensates to observe the reported asynchronous locking phenomena (Figs.~\ref{Fig1}{\bf c} and \ref{Fig4}) was performed through focused excitation purposely positioned in one of the central traps of the array using microcope optics to reduce the spot size down to $\sim 3 \mu$m. The same microscope objective (NA=0.3) was used to collect the emitted light. In this latter case a triple additive Jobin-Yvon T64000 spectrometer was used to obtain the required high spectral resolution ($\sim 5$~GHz$^{-1} \sim 20~\mu$eV). \\ 

\paragraph*{\textbf{Pump and probe phonon spectroscopy.}}  

A ps-laser pulse is used to resonantly excite the optical cavity mode. A rapid change of index of refraction is induced by the pump through carrier excitation. In addition to this electronic response, the pump pulse launches coherent phonons by a displacive mechanism~\cite{Winter2012,Ruello2015}. These mechanical oscillations modulate the cavity energy through two mechanisms, interface displacement and photoelastic interaction, which are detected using a delayed probe pulse that samples the cavity's reflectivity. A typical spectrum displays characteristic lines corresponding to the $\sim 20$~GHz fundamental confined breathing mode of the structures, and weaker contributions at the higher energy overtones at $\sim 60$~GHz and  $\sim 100$~GHz. More details are provided in Supplementary Note 2.

\paragraph*{\textbf{Effective potential phonon modelling.}}  

We assume the non-etched effective quadratic dispersion relation arising when $k_z$ is quantized, i.e., $E(k_x,k_y)=E_{\mathrm{cav,ne}}+\hbar^2(k_x^2+k_y^2)/(2 m_{\mathrm{eff}})$, with homogenous in-plane speed of sound $v_s$ defining an effective mass $ m_{\mathrm{eff}}=E_{\mathrm{cav,ne}}/v_s^2$.   This is incorporated in a 2D Schroedinger-like equation, \[\left(-\hbar^2/(2 m_{\mathrm{eff}})\nabla^2+E_{\mathrm{cav,ne}}+V_e(x,y)\right)\Psi(x,y)=E\Psi(x,y)\], 
that adds the potential $V_e(x,y)$ to effectively describe the trapping induced by the etching. The full height of the potential in an etched region is $V_{\mathrm{max}}=(E_{\mathrm{cav,e}}-E_{\mathrm{cav,ne}})$, with 
$E_{\mathrm{cav,e}}$  the energy of the phonon mode in a large etched region. Each square trap $i$, centered in $(x_i,y_i)$ contributes to $V_e(x,y)$ the potential $V_i(x,y)=V_{\mathrm{max}}(1-v_i(x-x_i)v_i(y-y_i))$ where the trap profile along each direction is given by $v_i(\alpha)=\frac{1}{2}\left(\mathrm{erfc}\left(\frac{\alpha-\frac{w_i}{2}}{0.55\delta_i}\right)-\mathrm{erfc}\left(\frac{\alpha+\frac{w_i}{2}}{0.55\delta_i}\right)\right)$ with $w_i$ the trap width and $\delta_i$ the 10\% to 90\% transition length. The eigenvalue problem is solved using finite differences by the customary approach of imposing periodic conditions fulfilling the Bloch theorem.  For the width of the nER to ER transition regions, we take $0.35 \mu$m, consistent with both the modelling of the polariton properties and STM studies in similar structures. More details and examples of the calculated phonon dispersion in trap arrays can be found in Supplementary Note 4. 

\paragraph*{\textbf{Estimation of the optomechanical coupling $g_\mathrm{0}$.}}

Electrically generated mechanical waves have been used in individual similar traps to obtain $g_\mathrm{om}/2\pi = \Delta E_\mathrm{p}/\Delta u \sim 50$~THz/nm~\cite{Kuznetsov2021} (change of polariton energy $\Delta E_\mathrm{p}$ per unit displacement $\Delta u$). This parameter is related to the on-site linear optomechanical coupling by $g_\mathrm{0}=g_\mathrm{om} x_\mathrm{zpf}$ ($x_\mathrm{zpf}$ is the displacement due to zero point fluctuations)~\cite{RMP}. The effective mass associated to the oscillator can be estimated as $m_\mathrm{eff} \sim 0.5$~pg for a structure of $2 \mu$m lateral size (see the Supplementary Note 5) and from this we obtain $x_\mathrm{zpf} \sim 1$~fm. It follows that $g_\mathrm{0}/2\pi \sim 50$~MHz, quite a huge value when compared with other reported optomechanical systems~\cite{RMP}.  By involving the deformation potential interaction associated to the exciton component of polaritons, $g_\mathrm{0}/2\pi$ is thus amplified by three orders of magnitude from the $\sim 50$kHz calculated for purely optical radiation pressure interaction~\cite{g_0}.  
To enhance cavity optomechanical phenomena materials with electronic resonances are usually avoided due to the related absorption that reduces the optical Q-factor~\cite{RMP}. Counter intuitively, the opposite occurs in our polaromechanical crystals. Indeed, the measured BEC coherence time is $\sim 1-2$~ns, implying $Q \sim 7 \times 10^5$, while the bare cavity photon lifetime is only $\sim 10$~ps. This means that, by involving the exciton-mediated optomechanical interaction, $g_\mathrm{0}$ is amplified by three orders of magnitude, while at the same time because the system is in the (polariton) strong coupling regime the Q-factor is {\it enhanced} by a factor $\sim 100$ .

%\paragraph*{Online content:}
%Any methods, additional references, Nature Research reporting
%summaries, source data, extended data, supplementary information, acknowledgements, peer review information; details of author
%contributions and competing interests; and statements of data and
%code availability are available at https://doi.org/.\\

\section*{Data availability}
The source data that support the findings of this study are available from the corresponding author upon reasonable request. All these data are directly shown in the corresponding figures without further processing.

%\bibliography{refe}

%Corresponding author: A.F., afains@cab.cnea.gov.ar

%\begin{references}

%\section*{Acknowledgements} 

\begin{acknowledgments}
We acknowledge partial financial support from the ANPCyT-FONCyT (Argentina) under grants PICT-2015-1063, PICT-2018-03255, PICT-2016-0791 and PICT 2018-1509, CONICET grant PIP 11220150100506, SeCyT-UNCuyo grant 06/C603, from the  German Research Foundation (DFG) under grant 359162958, and the joint Bilateral Cooperation Program between the German Research Foundation (DFG) and the Argentinian Ministry of Science and Technology (MINCyT) and CONICET. AF thanks the Alexander von Humboldt Foundation for the support, and the Paul Drude Institute for the hospitality during which this work was finished. 
A.A.R. aknowledges support by PAIDI 2020 Project No. P20-00548 with FEDER funds.
We thank Ignacio Carraro and Facundo Fainstein for enlightening discussions, and FF for pointing to us Ref.~\onlinecite{Doppler2020}.
\end{acknowledgments}

\section*{Author contributions}
D.L.C. and A.S.K. have contributed equally. 
D.L.C, I.P. and A.E.B. contributed to the pump and probe and locking experiments, and A.S.K to the sample characterization and measurement of the polariton dispersions. 
A.S.K., K.B., and P.V.S. designed and fabricated the structured microcavity sample. 
P.S., A.A.R., G.U, A.E.B., and A.F. outlined theoretical aspects, 
with A.A.R and G.U developing the synchronization theory and performing the related numerical simulations. 
All authors contributed to the discussion and analysis of the results. 
P.V.S. and A.F. conceived and directed the project. 
A.F. prepared the manuscript with inputs from all co-authors.

\section*{Competing interests}
The authors declare no competing interests.

\section*{Additional information}

\paragraph*{\textbf{Supplementary information}} is available for this paper at https://doi.org/... \\

\paragraph*{\textbf{Correspondence}} and requests for materials should be addressed to A.F.

%%%%%%%%%%%%%%%%%%%%%%%%%%%%%%%%%%%%%%%%%%%%%%%%%%%%%%%%%%%%%%%%%%%%%%%
%%%%%%%%%%%%%%%%%%%%%%%%%%%%%%%%%%%%%%%%%%%%%%%%%%%%%%%%%%%%%%%%%%%%%%%
%        Beginning of SM
%%%%%%%%%%%%%%%%%%%%%%%%%%%%%%%%%%%%%%%%%%%%%%%%%%%%%%%%%%%%%%%%%%%%%%%
%%%%%%%%%%%%%%%%%%%%%%%%%%%%%%%%%%%%%%%%%%%%%%%%%%%%%%%%%%%%%%%%%%%%%%%

\onecolumngrid

\pagebreak

%\clearpage
\setcounter{section}{0}
\setcounter{page}{1}
\setcounter{figure}{0}
\renewcommand{\thesection}{Supplementary Note \arabic{section}}
\renewcommand{\figurename}{Supplementary Figure \!\!}
\renewcommand{\thefigure}{S\arabic{figure}}

\begin{center}
\textbf{\large Supplementary Material:\\Asynchronous Locking in Metamaterials of Fluids of Light and Sound}
\end{center}

\section*{Supplementary Note 1: Sample details and description}\label{sec: Sample details and description}
\setcounter{subsection}{0}
\subsection{The sample-fabrication}
%%%%%%%%%%%%%%%%%%%%%%%%%%%%%%%%%%%%%%%%%%%%%
\begin{figure}[!hht]
    \begin{center}
	\subfigure[]{\includegraphics[trim = 5mm 10mm 5mm 25mm, clip=true, keepaspectratio=true, width=0.4\columnwidth, angle=0]{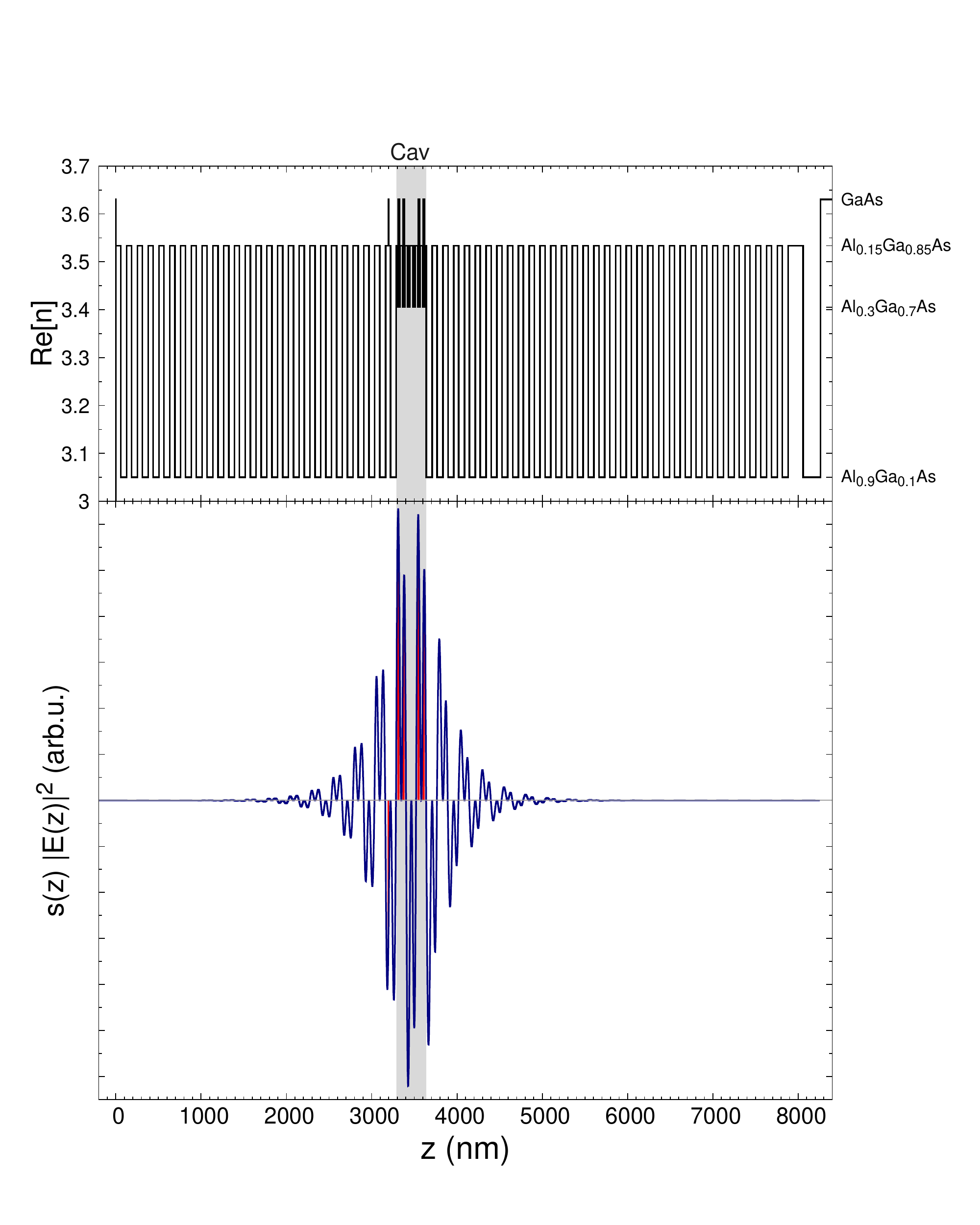}}
	\subfigure[]{\includegraphics[trim = 5mm 10mm 5mm 25mm, clip=true, keepaspectratio=true, width=0.4\columnwidth, angle=0]{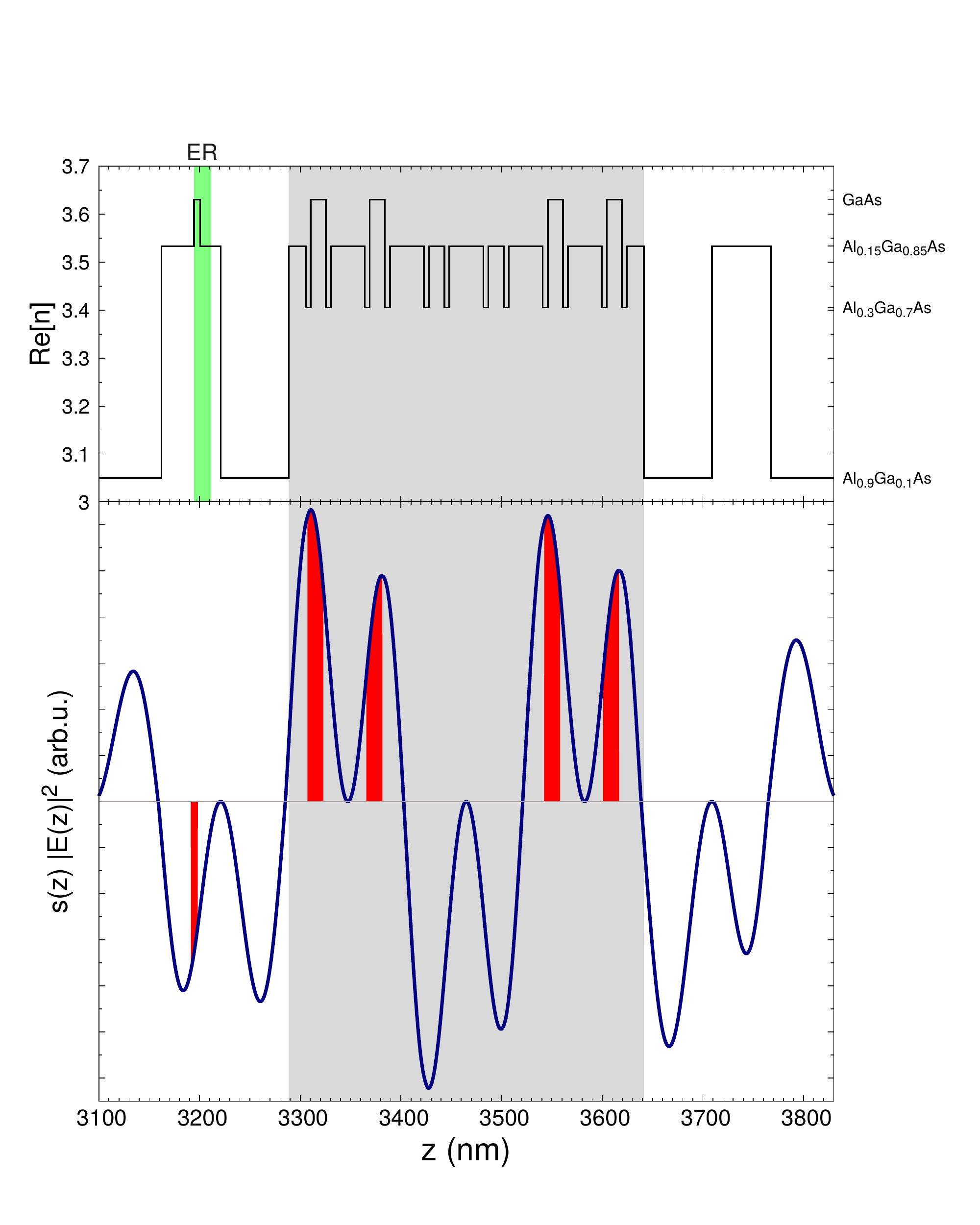}}
    \end{center}
    \vspace{-0.4 cm}
\caption{\textbf{Sample's structure and strain-electric field overlap.} \textbf{(a)} The top panel displays the profile of the real part of the index of refraction ($n$) along the MBE growth direction $z$. $z$=0\,nm corresponds to the air-sample surface, and the substrate begins at $\sim$8.2\,$\mu$m. Top and bottom DBRs are separated by the cavity spacer (gray shaded region, \textrm{Cav}) that embeds the four GaAs QWs. The bottom panel shows the overlap between the strain field $s(z)$ and the squared electric field $E(z)$. \textbf{(b)} corresponds to the detail of the cavity region in \textbf{(a)}. The green area (ER) corresponds to the part that is removed during the etching process. The optimized overlap of $s(z)\,|E(z)|^2$ with the GaAs cavity QWs is indicated in red.}
\label{Fig-index-n-Efield}
\end{figure}
%%%%%%%%%%%%%%%%%%%%%%%%%%%%%%%%%%%%%%%%%%%%%

Figure \ref{Fig-index-n-Efield}(a) (top panel) shows a scheme of the profile of the planar sample. Here the real part of the refractive index corresponding to each layer is plotted as a function of the MBE growth direction ($z$). Figure \ref{Fig-index-n-Efield}(b) details the region of the hybrid photon/phonon cavity with the embedded QW's. The region of the cavity is highlighted in gray, and the region (ER) highlighted in green in the top panel of Fig.\ref{Fig-index-n-Efield}(b) corresponds to the portion of the sample ($\sim$12\,nm) that is removed during the etching process to attain the in-plane patterning required to define the traps [see Fig.1(a) of the main text].\\

The microcavities reported in this work were designed to optimize the \emph{electrostrictive polariton-phonon interaction}. To achieve this situation, the quantum wells (QWs) were displaced from the anti-nodes of the cavity electric field [$E(z)$], which turn out to be nodes of the strain field [$s(z)$], to the position were the product $s(z)\,\left|E(z) \right|^2$ is maximized. The bottom panels of the respective Figs.\,\ref{Fig-index-n-Efield}(a) and \ref{Fig-index-n-Efield}(b) show this optimized magnitude along the $z$-direction for the planar sample. As can be noted, in panel (a) the strain-electric field product is peaked at the cavity spacer, and the overlap is optimal at the QWs position. This region of QWs and strain-electric field's superposition is indicated by the red-shaded area in Fig.\ref{Fig-index-n-Efield}(b).

The fabrication method and micrometer size lateral patterning of the traps, in the context of cavity polariton, is well described in Refs.[\onlinecite{SM_Daif2006, SM_Winkler2015, SM_Kuznetsov2018}]. It consists of an MBE growth of the bottom DBR and the cavity spacer with the embedded QWs. The processes continues with the patterning and wet-etching of the traps outside the MBE chamber, and is concluded with a subsequent MBE overgrowth of the top DBR. The sample consisted of a top(bottom) DBR formed by 25(33) $\frac{\lambda}{4}/\frac{\lambda}{4}$ periods of Al$_{0.15}$Ga$_{0.85}$As/Al$_{0.90}$Ga$_{0.10}$As layers, embedding a $\frac{3}{2}\lambda$ cavity with four 15\,nm GaAs QWs as schematized in Fig.\ref{Fig-index-n-Efield}(b) (top panel), grown on a 350$\mu$m thick GaAs(001) substrate. A scheme of the resulting structure is shown in Fig.1(b) of the main text. One point worth mentioning is that, because the etching is performed at the limits of the cavity spacer, far from the QWs, the quality of the latter is not affected by the etching and overgrowth processes

\subsection{Optimization of the electrostrictive polariton-phonon interaction}

%%%%%%%%%%%%%%%%%%%%%%%%%%%%%%%%%%%%%%%%%%%%%
\begin{figure}[!!!ttt]
    \begin{center}
    \subfigure[]{\includegraphics[trim = 0mm 0mm 0mm 0mm, clip=true, keepaspectratio=true, width=0.4\columnwidth, angle=0]{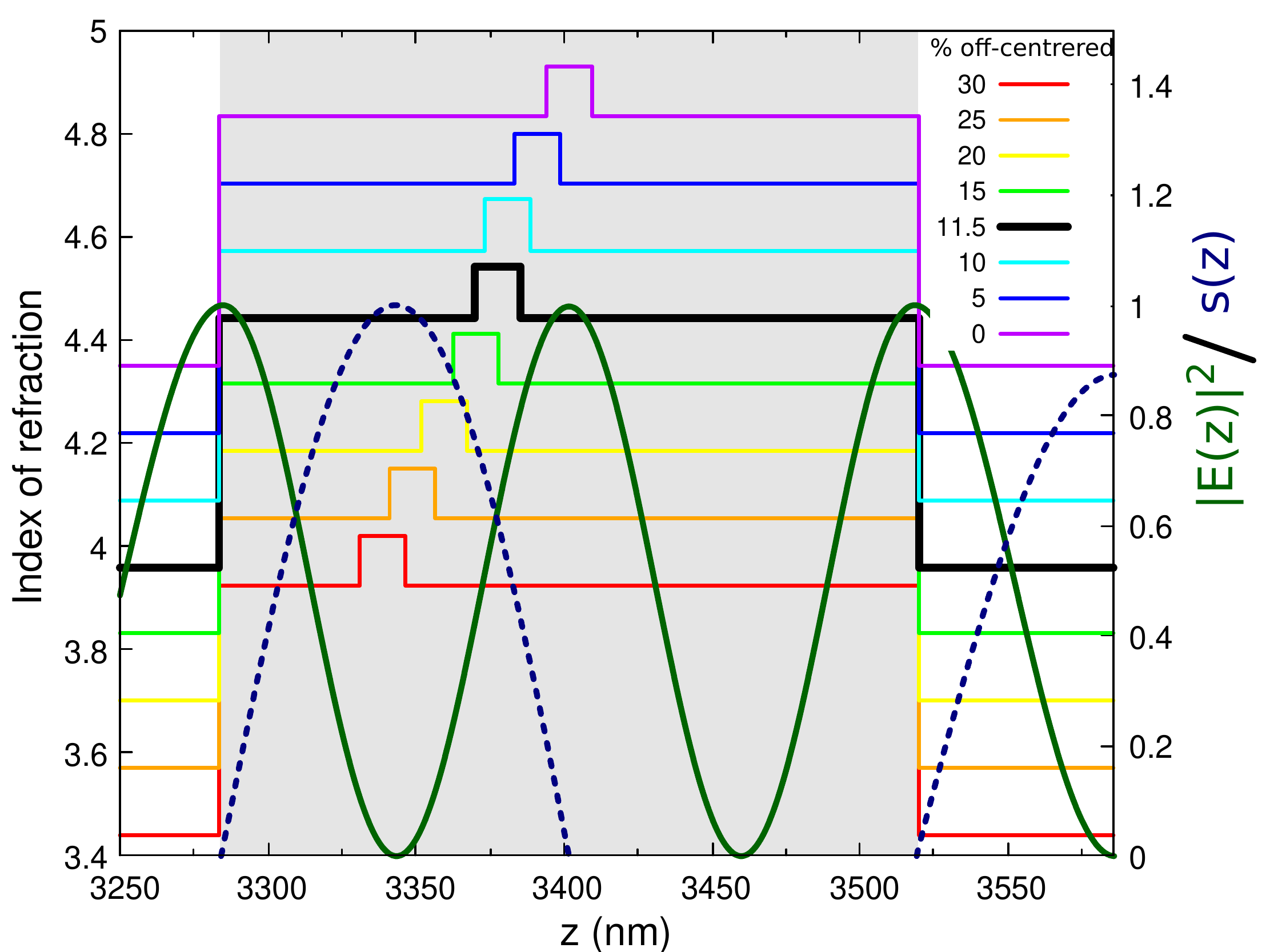}}
    \subfigure[]{\includegraphics[trim = 0mm 0mm 0mm 0mm, clip=true, keepaspectratio=true, width=0.4\columnwidth, angle=0]{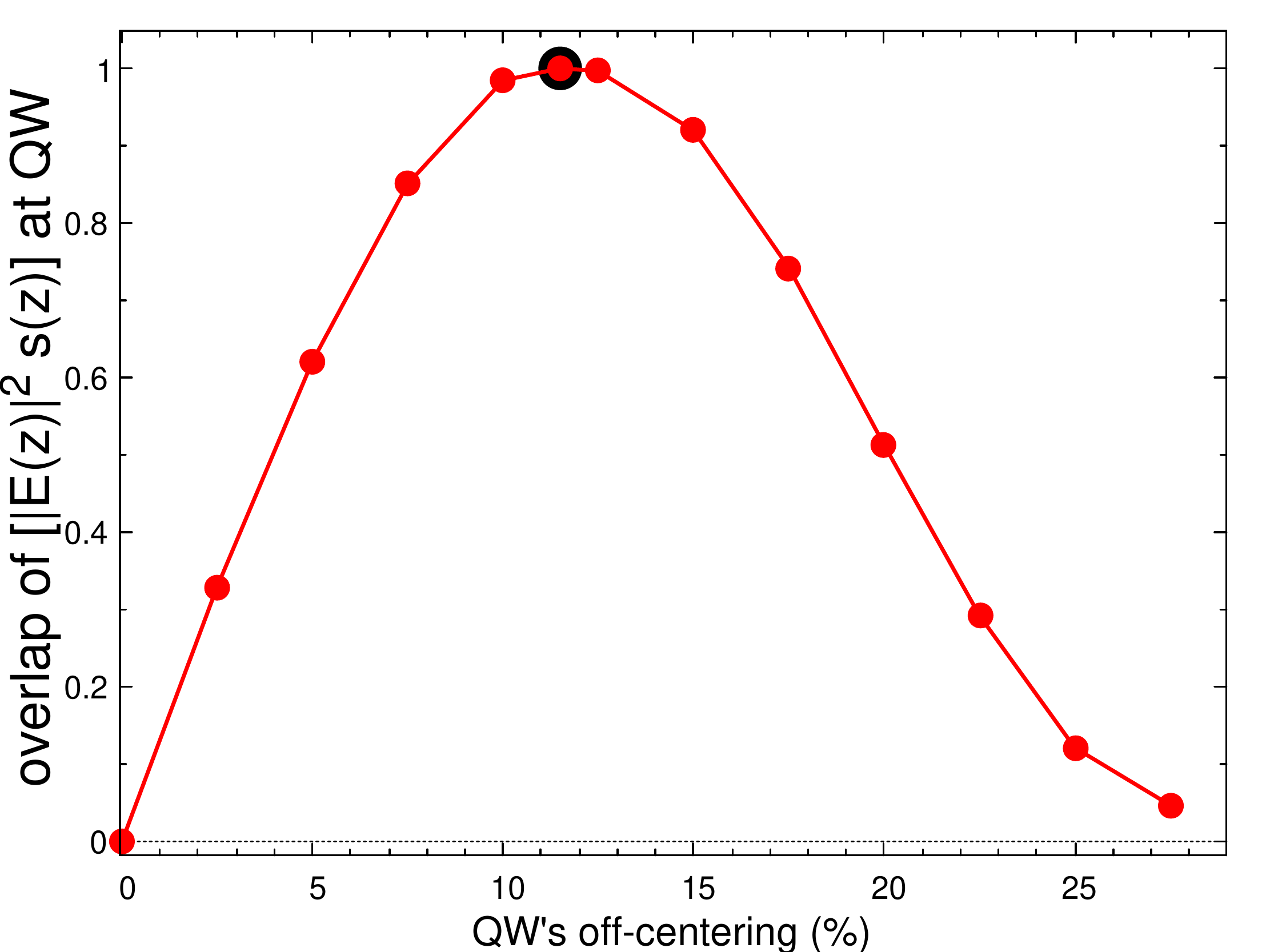}}
    \end{center}
    \vspace{-0.4 cm}
\caption{\textbf{Optimization of the polariton-phonon coupling within a SQW cavity}. \textbf{(a)} Detail of a $\lambda$ cavity spacer with an embedded SQW. The left y-axis corresponds to the index of refraction profile. Initially, the SQW is centered (0\%) at the cavity spacer (gray-shaded region), and it is gradually off-centered. Superimposed (right y-axis), corresponds to the normalized electric field's intensity $|E(z)|^2$ (full dark-green) and acoustic strain $s(z)$ (dashed-blue). \textbf{(b)} Overlap integral of $|E(z)|^2$ and $s(z)$ for the different QW's positions depicted in panel (a). The optimized situation (maximum) is indicated by the black circle, which corresponds to the thicker black index of refraction profile in panel (a).}
\label{Fig_gDP_vs_QWpos}
\end{figure}
%%%%%%%%%%%%%%%%%%%%%%%%%%%%%%%%%%%%%%%%%%%%%
%Variation of $g_0^{\textrm{DP}}$ with the cavity embedded quantum well's position
In Fig.\,\ref{Fig_gDP_vs_QWpos} we schematically show how the values for the single-photon electrostrictive coupling rate $g_0^{\textrm{DP}}$ (deformation potential interaction) evolves when the relative position within the cavity spacer is varied. 
As mentioned in the preceding paragraphs, and well observed in this figure, the polariton-phonon coupling strength is optimized when the quantum wells (QWs) are displaced in such way that the overlap between the product of $s(z)$, the acoustic strain field, and $\left|E(z) \right|^2$, the amplitude of the cavity's electric field, is maximized.

Figure \ref{Fig_gDP_vs_QWpos}(a) exemplifies the case of a simple $\lambda$ cavity spacer (indicated by the gray-shaded area), for which the embedded single quantum well (SQW) is displaced from the cavity center. Superimposed (right y-axis), $\left|E(z) \right|^2$ and $s(z)$ are indicated by the full dark-green and dashed curves, respectively. Initially (violet curve, 0\% off-centred), the SQW is centred at the electric field's anti-node, a situation that is usually chosen for enhancing the exciton-photon interaction. However, we notice that at that point the strain is zero so the electrostrictive polariton-phonon coupling vanishes. The SQW is gradually off-centred towards the front cavity spacer limit, departing from the optimal exciton-photon overlap but increasing its superposition with the acoustic confined strain field. Since $g_0^{\textrm{DP}}$ is proportional to the overlap integral of $s(z)\,\left|E(z) \right|^2$ within the SQW \cite{SM_Sesin2021}, it will gradually change when the off-centring percentage is increased. In Fig. \ref{Fig_gDP_vs_QWpos}(b) we show $\int_{\textrm{SQW}}s(z)\,\left|E(z) \right|^2 dz$, where it is evident that the cavity centred QW is the worse situation. The optimal strain--e-field interaction, equivalently the optimal phonon-polariton interaction, is achieved at $\sim$11.5\% off-centring [black thicker curve in Fig.\,\ref{Fig_gDP_vs_QWpos}(a)], and decreases afterwards.

\subsection{List of the studied samples}

In the Supplementary Table \ref{tab:sample} we summarize the different samples studied in this work. The main characteristics that differentiate them are stated: cavity-exciton detuning of non-etched region ($\Delta_{\textrm{nER}}$) and the depth of the polaritonic  barrier well ($\Delta_{\textrm{LP}}$). The experimental technique used in each case is indicated: for phonon properties \textit{Pump and Probe phonon spectroscopy} (P$\&$P), and for polariton properties \textit{Photoluminescence spectroscopy} (PL). The figures of the main text and supplementary notes, where the corresponding data appears, are additionally included in column ``Figure ID''. All of the samples are either single traps, two couple traps or array of traps, and were designed and fabricated as explained earlier in this Supplementary Note 1.

%The samples used to characterize the phonon properties (with P$\&$P phonon technique) result to be more photonic than those used to study the polariton properties (see column $\Delta_{nER}$). This means that the samples studied with the PL technique have a higher optomechanical factor than the other. In these optomechanical samples the asynchronous locking and optomechanical parametric oscillations were analyzed \cite{Reynoso2022}.

\begin{table*}[]
\begin{tabular}{|c|cc|c|c|c|c|}
\hline
Type of sample        & \multicolumn{2}{c|}{Structure parameters ($\mu$m)} & \begin{tabular}[c]{@{}c@{}}Experimental \\ technique\end{tabular} & $\Delta_{\textrm{nER}}$ (meV) & $\Delta_{\textrm{LP}}$ (meV) & Figure ID                \\ \hline
                      & \multicolumn{1}{c|}{Trap size}    & Separation     &                                                                   &                               &                               &                          \\ \hline
Array of square traps & \multicolumn{1}{c|}{1.3}          & 2.6            & PL                                                   & -5.4                          & -7.1                          & 4, S14                   \\ \hline
Array of square traps & \multicolumn{1}{c|}{1.6}          & 3.2            & PL                                                    & -5.4                          & -7.1                          & 1(c), S10, S11, S12, S13, S14 \\ \hline
Array of square traps & \multicolumn{1}{c|}{2.0}          & 4.0            & PL                                                    & -5.4                          & -7.1                          & S14                      \\ \hline
Array of square traps & \multicolumn{1}{c|}{2.5}          & 5.0            & PL                                                    & -5.4                          & -7.1                          & S14                      \\ \hline
Array of square traps & \multicolumn{1}{c|}{4.0}          & 4.0            & PL                                                    & -10.6                         & -6.7                          & 2(a)                     \\ \hline
Array of square traps & \multicolumn{1}{c|}{1.0}          & 1.0            & PL                                                    & -10.6                         & -6.7                          & 2(b)                     \\ \hline
Array of square traps & \multicolumn{1}{c|}{4.0}          & 2.0, 1.0       & P$\&$P & -12.5                         & -10.8                         & 3(d)                     \\ \hline
Array of square traps & \multicolumn{1}{c|}{2.0}          & 2.0, 1.0, 0.5  & P$\&$P & -12.5                         & -10.8                         & 3(d)                     \\ \hline
Double coupled traps   & \multicolumn{1}{c|}{2.0}          & 0.5            & P$\&$P & -12.5                         & -10.8                         & 3(c)                     \\ \hline
Single square trap           & \multicolumn{1}{c|}{10.0 - 1.0}   & --              & P$\&$P  & -12.5                         & -10.8                         & 3(b)                     \\ \hline
Single square trap           & \multicolumn{1}{c|}{10.0 - 1.0}   & --              & PL                                                   & -10.6                         & -6.7                          & 3(a)                     \\ \hline
\end{tabular}
\caption{\textbf{Parameters of the different analysed samples.} The first column presents the different types of geometries studied. The second, the size and separation of the traps (except for isolated traps). The next column indicates the main experimental technique used in each experiment: Photoluminescence (PL) spectroscopy and Pump and Probe (P$\&$P) phonon spectroscopy. $\Delta_{\textrm{nER}}$ is the detuning between cavity-mode and exciton energies from the non-etched region (nER), and $\Delta_{\textrm{LP}}$ is the energy difference between the lower-polariton in the extended non-etched and etched regions. The latter defines the modulation of the polariton effective potential (as in Ref.\cite{SM_Kuznetsov2018}). The last column associates each sample with the data presented in the different figures of this work. The letter ``S'' describes a figure in the Supplementary Material.}
\label{tab:sample}
\end{table*}

\section*{Supplementary Note 2: Time-resolved pump-probe phonon spectroscopy}\label{sec: pump-probe spec}

%%%%%%%%%%%%%%%%%%%%%%%%%%%%%%%%%%%%%%%%%%%%%
\begin{figure}[!!!ttt]
    \begin{center}
    \includegraphics[trim = 0mm 0mm 0mm 0mm, clip=true, keepaspectratio=true, width=0.4\columnwidth, angle=0]{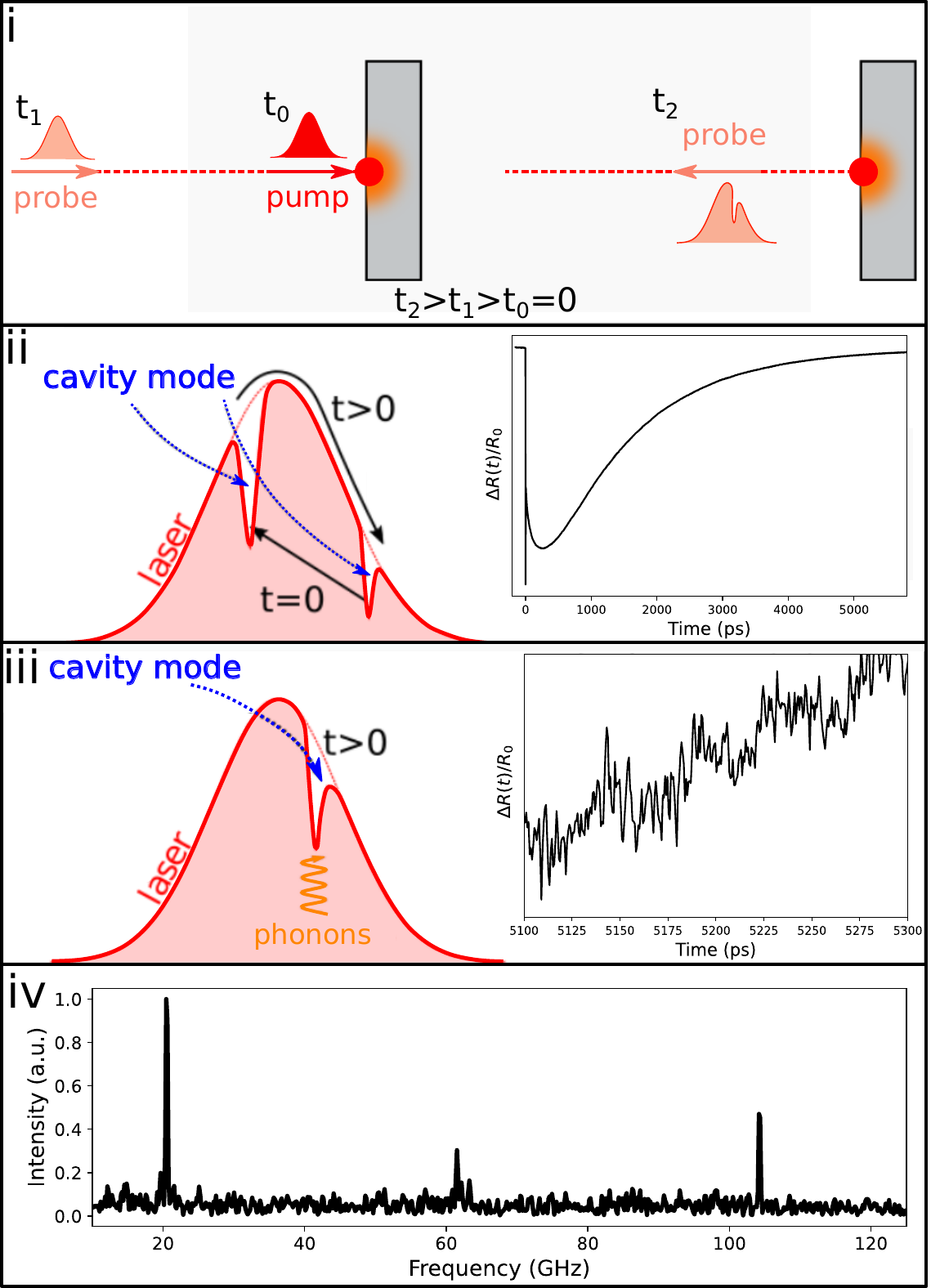}
    \end{center}
%    \vspace{-0.8 cm}
\caption{\textbf{Picosecond coherent phonon pump-probe method.} A scheme of this technique is presented in panel (i). At $t_0$ the pump pulse strikes the sample inducing a change of the refractive index, and launching coherent phonons. At a later time $t_1$ a delayed pulse probes the cavity reflectivity. The change in the refractive index induces a shift of the cavity mode, leading to the so-called electronic-signal (ii). The coherent phonons modulate the cavity mode leading to the vibrational signal (iii). Processing of the signal obtained as a function of the probe delay, and its Fourier transform, lead to the vibrational spectra (iv). Typical cavity confined modes are observed around 20, 60, and 100~GHz.}
\label{Fig_PP_method}
\end{figure}
%%%%%%%%%%%%%%%%%%%%%%%%%%%%%%%%%%%%%%%%%%%%%

The method used to investigate the dynamical aspects of the phononic states of the individual traps, as well as the double-coupled trap systems (molecules) and 2D trap arrays, is the ultra-fast laser spectroscopy method called picosecond coherent phonon pump and probe technique \cite{SM_Thomsen1986, Matsuda2005}, particularly adapted for semiconductor optical microcavities \cite{SM_Fainstein2013}, combined with microscopy to enable the addressing of individual traps \cite{SM_Chafatinos2020}.

The concept of this method is illustrated in Fig.~\ref{Fig_PP_method}(i)-(iii). A ps-laser pulse is used to resonantly excite the optical cavity mode. Because of direct (above-gap excitation) or residual (below-gap excitation) carrier generation, a rapid change of index of refraction is induced by the pump, which recovers within a time-scale defined by electron-hole evolution and recombination. As a consequence of this change in the refractive index, the cavity mode strongly blue-shifts in a ps scale, to recover its equilibrium in longer ns-scale times. In addition to this so-called electronic response, the pump pulse through the same carrier excitation launches coherent phonons mediated by a deformation-potential induced displacive mechanism \cite{SM_Winter2012, SM_Ruello2015}. These mechanical oscillations modulate the cavity energy involving mainly two mechanisms, interface displacement and photoelastic interaction \cite{SM_Winter2012, SM_Ruello2015}, and can thus be detected using a delayed probe that samples the cavity's reflectivity \cite{SM_Kimura2007, SM_Kimura2011}.

In Fig.\ref{Fig_PP_method}(ii)(right panel) a typically obtained transient reflectance $\Delta R(t)/R_{\textrm{o}}$ is shown. The reflectivity of the probe pulse is recorded as a function of the time delay with respect to the arrival of the exciting pump pulse that defines $t$=0\,ps. Using a convenient filtering treatment, the slow varying envelope evolution, ascribed to the optical constant's temporal variations due to electronic states and thermal modifications, the faster dynamics defined by the vibrational oscillatory components can be extracted [see right panel in Fig\ref{Fig_PP_method}(iii)]. After extraction a spectral analysis is performed by applying a numerical Fourier transformation (nFT) of the measured window.\\ 
A typical spectrum is shown in Fig.\,\ref{Fig_PP_method}(iv), with the characteristic stronger signal corresponding to the $\sim$20\,GHz fundamental confined breathing mode of the structures, and weaker contributions at the higher energy overtones at $\sim$60\,GHz and  $\sim$100\,GHz. These acoustic cavity modes have been previously studied in planar microcavities \cite{SM_Fainstein2013},  and also in individual micrometer-size pillars obtained by deep etching planar structures \cite{SM_Anguiano2017, SM_Lamberti2017}. This previous knowledge is exploited here to experimentally identify the phonon confinement in \emph{individual} polariton traps generated by weak distant modulation of the spacer thickness, and is extended to study more complex structures with different dimensionality (molecules, and 2D arrays).\\

\section*{Supplementary Note 3: Phonon confinement in traps: experimental results} \label{sec: phonon confinement exp results}

\begin{figure}[!!!ttt]
    \begin{center}
%	\subfigure[]{\includegraphics[trim = 0mm 0mm 0mm 0mm, clip=true, keepaspectratio=true, width=0.8\columnwidth, angle=0]{figs-Chafa/smooth/smooth_10.0_120.0.png}}
%	\subfigure[]{\includegraphics[trim = 0mm 0mm 0mm 0mm, clip=true, keepaspectratio=true, width=0.45\columnwidth, angle=0]{figs-Chafa/smooth/smooth_zoom_10.0_120.0.png}}
%	\subfigure[]{\includegraphics[trim = 0mm 0mm 0mm 0mm, clip=true, keepaspectratio=true, width=0.45\columnwidth, angle=0]{figs-Chafa/smooth/smooth_zoom_17.0_23.0.png}}
	\includegraphics[trim = 0mm 0mm 0mm 0mm, clip=true, keepaspectratio=true, width=0.6\columnwidth, angle=0]{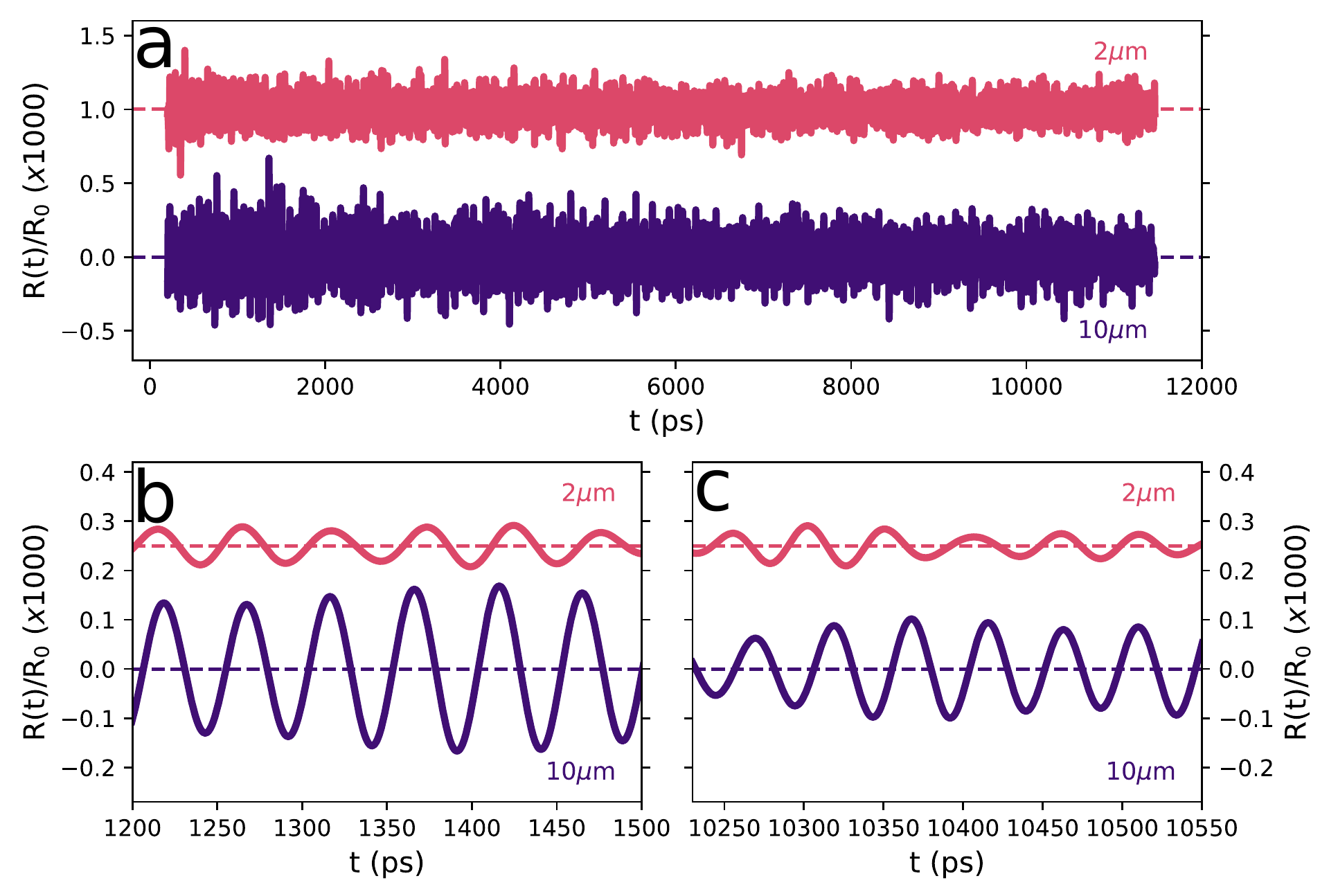}
    \end{center}
    \vspace{-0.4 cm}
\caption{\textbf{Extracted phonon oscillations for two different isolated traps.} \textbf{(a)} The bottom curve of this panel corresponds to the transient measured on a 10\,$\mu$m\,$\times$\,10\,$\mu$m trap, whereas the upper curve correspond to the equivalent results on a 2\,$\mu$m\,m$\times$\,2\,$\mu$m trap. \textbf{(b)} and \textbf{(c)} correspond to the upper transients in \textbf{(a)} with a band-pass filter centered at 20.55\,GHz and 20.68\,GHz, respectively. \textbf{(b)} corresponds to a detail for short delay times, whereas \textbf{(c)} displays the situation near the end of the temporal measured window.
}
\label{Fig:Extracted Osc transients nER 2x2}
\end{figure}

Examples of measurements using this technique on a single micrometer size polariton trap are depicted in Fig.\ref{Fig:Extracted Osc transients nER 2x2}. In panel (a), the transient curves corresponding to the vibrational contributions (the electronic part was removed as explained above), extracted for an isolated 2\,$\mu$m$\,\times$\,2\,$\mu$m trap (pink curve), and for an isolated 10\,$\mu$m\,$\times$\,10\,$\mu$m$^2$ trap (violet curve) are shown. Both transients show clear oscillations, which result from the superposition of the spectral components [see Fig.\,\ref{Fig_PP_method}(iv)] of $\sim$20, $\sim$60, $\sim$100\,GHz (periods respectively of $\sim$50, $\sim$17, $\sim$10\,ps). For the extended temporal scale chosen in Fig.\ref{Fig:Extracted Osc transients nER 2x2}(a) the individual oscillations, which are coherent with respect the impulsive excitation of the pump pulse, cannot be distinguished. However, it is clearly noticeable that the signal remains basically undamped up to the observed $\sim$11\,ns (the maximum scan time is limited by the $\sim 80$~MHz repetition rate of the pulsed laser). It actually well exceeds this wide acquired time frame. To better see this prevalescence, in Fig.\ref{Fig:Extracted Osc transients nER 2x2}(b) and (c) we show the transient oscillations of the fundamental mode of the traps, filtered using a band-pass $\pm$3\,GHz centered around the fundamental modes (at $\sim$20.68\,GHz and $\sim$20.55\,GHz, respectively). Figure \ref{Fig:Extracted Osc transients nER 2x2}(b) corresponds to a time window at early delay times, and Fig.\ref{Fig:Extracted Osc transients nER 2x2}(c) at almost the end of the measured window.  
The observation of the vibrational modes of these micrometer size polariton traps, which exhibit lifetimes that well exceed 12\,ns, is in contrast to what has been previously observed in optomechanical semiconductor micropillar resonators \cite{SM_Anguiano2017, SM_Anguiano2018a, SM_Anguiano2018b}.

%%%%%%%%%%%%%%%%%%%%%%%%%%%%%%%%%%%%%%%%%%%%%
%%%%%%%%%%%%%%%%%%%%%%%%%%%%%%%%%%%%%%%%%%%%%

\begin{figure}[!!!ttt]
    \begin{center}
        \includegraphics[trim = 0mm 0mm 0mm 0mm, clip=true, keepaspectratio=true, width=0.5\columnwidth, angle=0]{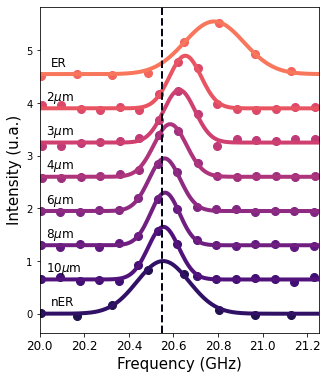}
    \end{center}
    \vspace{-0.8 cm}
\caption{\textbf{Variation with the traps' size of the confined fundamental phonon mode.} Spectra showing the size dependence of the acoustic confined ground state in square traps. Symbols are the experimental results obtained from a numerical Fourier transformation (nFT) of the extracted picosecond coherent time-resolved pump-probe measurements. The continuous curves are the corresponding Gaussian fits. The clear shift is evidenced, due to phonon confinement in the traps, with respect to the planar (non-structured) situation (nER) marked by the vertical dotted line.}
\label{Fig_PP_size-dependence}
\end{figure}

Several single square polariton traps of varying lateral sizes have been analyzed using picosecond coherent time-resolved pump-probe phonon spectroscopy in combination with microscopy. In Fig.\ref{Fig_PP_size-dependence} we show a detail of the spectral region of the fundamental 3D confined phononic mode around 20\,GHz [see Fig.\,\ref{Fig_PP_method}(iv)], when varying the trap size as indicated. As a reference, the results for the planar samples on the non-edged (nER) and edged (ER) regions are included. Due to the measured temporal window (same for all traps, see Figs.\ref{Fig:Extracted Osc transients nER 2x2}), the spectra are resolution limited. However, the blue-shift for decreasing trap size is clearly observed. The determination of the peaks' position, reported in Fig.2(c) of the main text, was performed by fitting the data using Gaussian distribution functions. The dotted vertical line indicates the frequency of the non-etched planar sample. The spectra for nER and ER have been measured using a shorter temporal window, with the consequent increase of the Fourier-limited peak's width.

%%%%%%%%%%%%%%%%%%%%%%%%%%%%%%%%%%%%%%%%%%%%%
%%%%%%%%%%%%%%%%%%%%%%%%%%%%%%%%%%%%%%%%%%%%%

\section*{Supplementary Note 4: Theoretical considerations of phonon confinement in traps and arrays of traps} \label{sec: theoretical considerations}
\setcounter{subsection}{0}

\subsection{Phonon confinement in planar structures}\label{subsec - SMsubsec phon conf planar structures}
%%%%%%%%%%%%%%%%%%%%%%%%%%%%%%%%%%%%%%%%%%%%%
\begin{figure}[!!!ttt]
    \begin{center}
    \includegraphics[trim = 0mm 0mm 0mm 0mm, clip=true, keepaspectratio=true, width=0.8\columnwidth, angle=0]{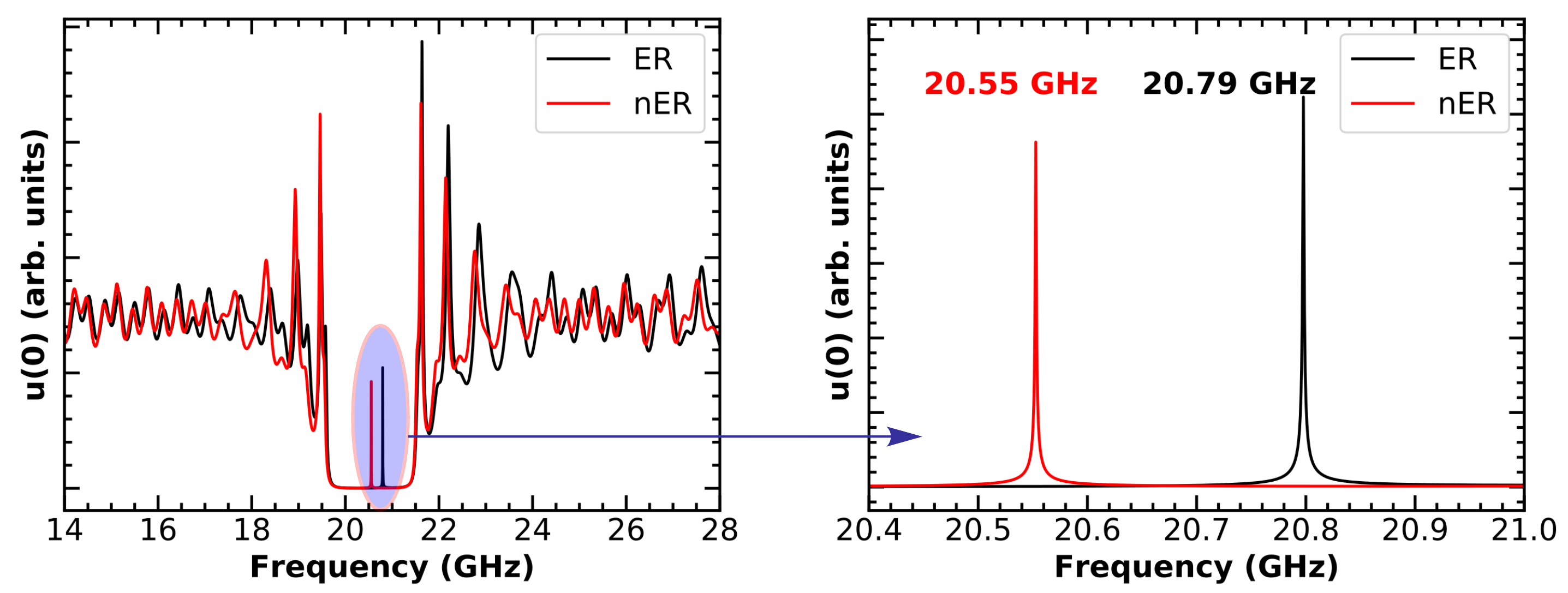}
    \end{center}
%    \vspace{-0.8 cm}
\caption{{\bf Phonon confinement in planar cavities.}
The panel on the left shows the calculated surface displacement (equivalent to the transmission) for white sound propagating from the substrate for both the non-etched (nER) and etched (ER) planar regions. The calculations were performed using a transfer matrix method. The phonon cavity mode for each structure can be seen within the phonon DBR stop-bands (detailed with a zoom, right panel). The frequency difference ($\sim$0.24\,GHz) between both modes is a measure of the modulation amplitude for the effective phonon potentials affecting the in-plane propagation of these cavity-confined phonons when the lateral patterning defines the hybrid micrometer-size phonon-polariton traps.
}
\label{Fig-pl-structures}
\end{figure}
%%%%%%%%%%%%%%%%%%%%%%%%%%%%%%%%%%%%%%%%%%%%%

The vertical confinement (along the $z$ direction) is illustrated in Fig.~\ref{Fig-pl-structures}, where the calculated acoustic transmission is displayed for the cases of the planar non-etched (nER, red curve) and etched (ER, black curve) regions. The acoustic transmission is evaluated as the frequency resolved surface displacement for ``white'' sound incident from the substrate. The typical DBR phonon stop-band~\cite{SM_Trigo2002} can be identified between approximately 19.8 and 21.3\,GHz. 

The simultaneous photon-phonon field's co-localization of this hybrid device (see main text),\cite{SM_Fainstein2013} is evidenced for the non-etched region (nER) by the cavity mode appearing at the acoustic stop-band's center ($\sim$20.55\,GHz). The acoustic cavity mode for the etched region (ER) is blue-shifted out from the stop-band's center approximately by 0.24~GHz. Extending the same concept used for photonic traps to the acoustic phonons, one concludes that the lateral patterning described above should lead to an additional phononic lateral trapping potential, with modulations of $\sim 0.24$~GHz between non-etched and etched regions. This potential should determine the in-plane propagation of the vertically confined acoustic phonons in a way that should be equivalent to the described confinement of photons \cite{SM_Kuznetsov2018}.

\subsection{3D Phonon confinement: Finite elements calculation}\label{subsec - SMsubsec 3D phon conf}

%%%%%%%%%%%%%%%%%%%%%%%%%%%%%%%%%%%%%%%%%%%%%
\begin{figure}[!!!hhh]
    \begin{center}
    \includegraphics[trim = 0mm 0mm 0mm 0mm, clip=true, keepaspectratio=true, width=0.7\columnwidth, angle=0]{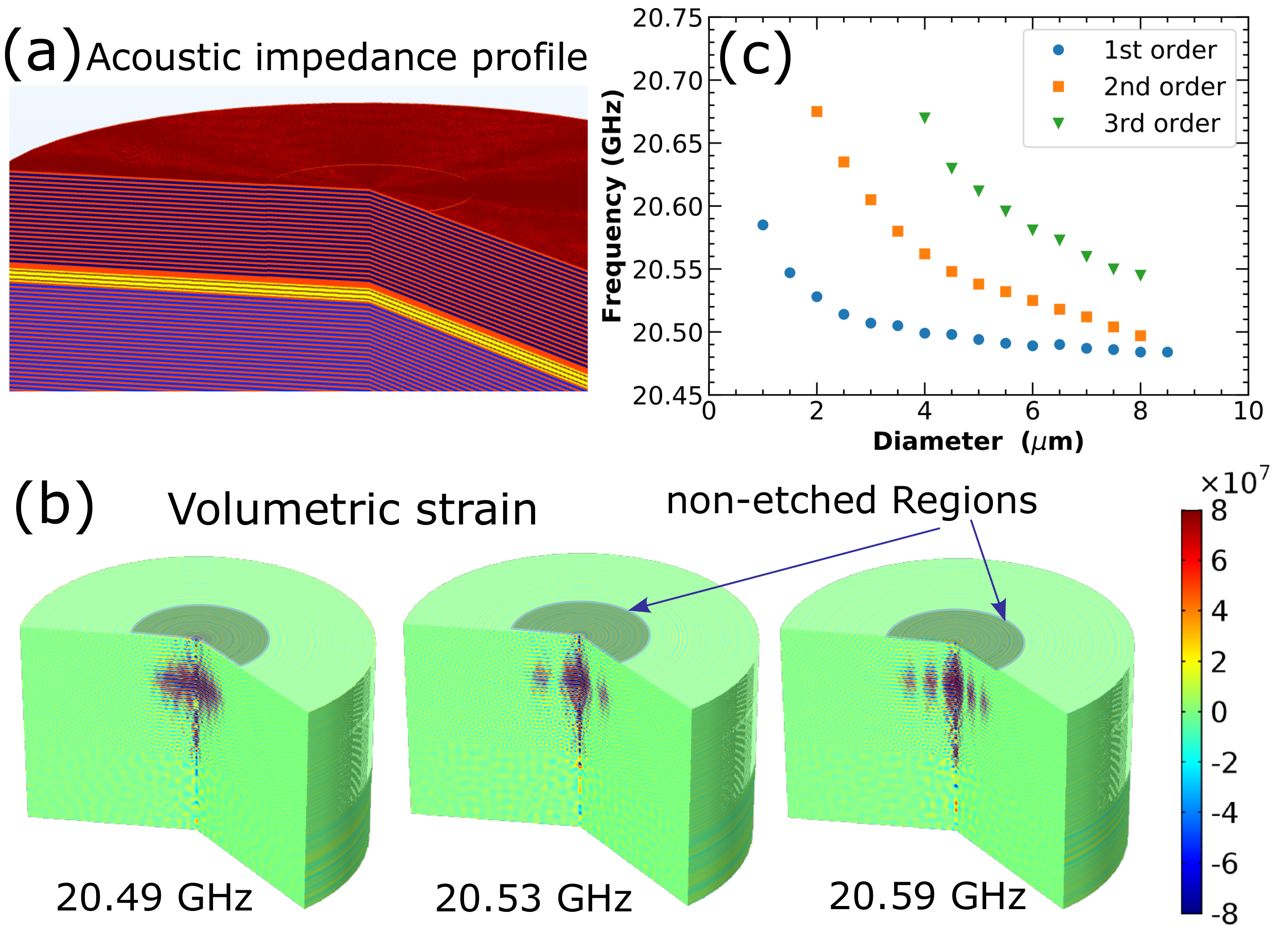}
    \end{center}
%    \vspace{-0.8 cm}
\caption{
{\bf Phonon confinement in microcavity 3D traps.}
Finite-elements 3D modeling of the spatial distribution and lateral size energy dependence, respectively, of the acoustic modes of single circular traps. \textbf{(a)} A detail of the modeled circular traps is presented, where the layered structure consists of: The top and bottom DBRs (indicated with red and violet layers, respectively), the spacer (yellow), and the patterned region (orange). The calculation is performed for a large cylindrical pillar of 20$\mu$m in diameter. At the center of this cylinder, the trap with the larger spacer thickness is patterned. The circular step forming the disk, is visible at the upper air-sample surface of this sketch (see text for details). In \textbf{(b)} the simulated volumetric strain ($\Delta v/V$) for first three confined mechanical modes is shown for a circular trap of a 5.5\,$\mu$m diameter trap. \textbf{(c)} Results for the calculated frequency of the first three orders of mode confinement for corresponding varying trap diameter. 
}
\label{Fig-3D-confinement}
\end{figure}
%%%%%%%%%%%%%%%%%%%%%%%%%%%%%%%%%%%%%%%%%%%%%

To test and demonstrate the concept of lateral in-plane effective confinement in these kind of traps, we calculated the mechanical eigenmodes using a commercial finite elements analysis software (COMSOL Multiphysics \cite{SM_COMSOL}) for the simplest geometry: a circular trap of cylindrical symmetry. GaAs and AlAs are considered as mechanically isotropic materials, and no mechanical absorption processes are considered. The used geometry is presented in Fig.~\ref{Fig-3D-confinement}(a). The material layering shown in this panel corresponds to the nominal values as defined by the MBE growth. The trap is modeled as a central circular region with larger spacer thickness as prescribed by the performed microfabrication process, and consistent with the measured polariton confinement (Fig.\,1(a) of the main text). The full calculated structure is a cylindrical pillar limited by vacuum.   In contrast with the optical response of the structure, phonons are subject to total reflection at the semiconductor/vacuum interfaces. To avoid any artificial effect induced by the pillar geometry, for the calculations we chose a total diameter large enough so that the free boundary condition of the sample-vacuum interfaces do not affect the solutions of modes localized at the smaller trap centered at the pillar. Also, in contrast with the light wave solutions, different mechanical directions of vibration are coupled through the Poisson's ratio (the ratio of transverse to axial elastic strain). In our case, a mode fulfills two simultaneous resonance conditions: the vertical confinement, determined by the DBRs and the spacer thickness, and the radial confinement determined by the effective potential induced by the local variation of the spacer thickness. Fulfilling both conditions results in a coupling between vertical and radial strains described by Poisson's ratio. Traps of diameter from 1-8$\,\mu$m were modeled, choosing 20$\,\mu$m for the external pillar diameter, and all the structure mounted on a 5\,$\mu$m thick GaAs substrate. 

As the material parameters, such as density, Young's modulus, and the Poisson ratio, well established and published values were used for the Al$_x$Ga$_{1-x}$As layered system \cite{SM_Adachi1985}. The volumetric strain ($\Delta v/V$) for the mechanical fundamental eigenmode and the first two excited eigenmodes, for a $5.5 \mu$m diameter trap are shown in Fig.\,\ref{Fig-3D-confinement}(b). These modes correspond to laterally well localized symmetric vibrations with different number of radial nodes. The trap-size dependence of the frequency of these modes is presented in Fig.\,\ref{Fig-3D-confinement}(c), showing the expected $1/D$ increase with decreasing diameter $D$. For the ground state this confinement induced blue-shift amounts to somewhat less than 100~MHz for $D \sim 1 \mu$m. Quite clearly the mechanical modes become strongly localized within the trap by the effective potential and, interestingly, due to the Poisson-ratio the breathing-like character of the modes implies that an expansion(compression) along z is accompanied by an in-plane compression(expansion).

\subsection{Effective potential model and phonon trap lattices}
%%%%%%%%%%%%%%%%%%%%%%%%%%%%%%%%%%%%%%%%%%%%%
\begin{figure}[!!!ttt]
    \begin{center}
	\subfigure[~ 1\,$\mu$m\,$\times$\,1\,$\mu$m\text{ traps separated by }1\,$\mu$m]{\includegraphics[trim = 0mm 0mm 0mm 0mm, clip=true, keepaspectratio=true, width=0.4\columnwidth, angle=0]{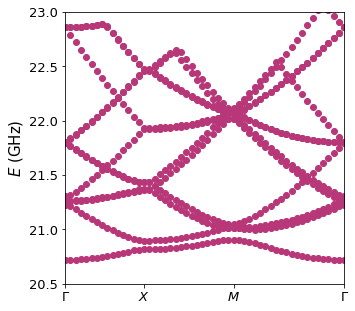}}
	\subfigure[~ 1\,$\mu$m\,$\times$\,1\,$\mu$m\text{ traps separated by }3.5\,$\mu$m]{\includegraphics[trim = 0mm 0mm 0mm 0mm, clip=true, keepaspectratio=true, width=0.4\columnwidth, angle=0]{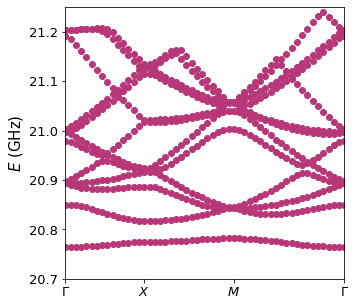}}\\
	\subfigure[~ Path followed in the 1st Brilloulin zone]{\includegraphics[trim = 0mm 0mm 0mm 0mm, clip=true, keepaspectratio=true, width=0.3\columnwidth, angle=0]{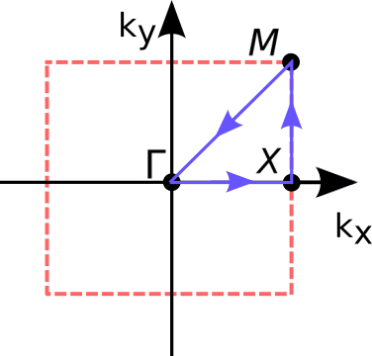}}
    \end{center}
    \vspace{-0.4 cm}
\caption{
{\bf Phonons in polariton-phonon hybrid lattices.}
The in-plane dispersion of the acoustic modes of a square lattice of $1\,\mu$m$\,\times 1\mu\mathrm{m}$ square traps, separated by 1 and 3.5 $\mu$m, are shown respectively in \textbf{(a)} and \textbf{(b)}. \textbf{(c)} depicts the path followed in the first Broullin zone and shown in  {(a)} and {(b)}.}
%\rojo{Yo la dispersión lo pondría con puntos sino con lines}
\label{FigS7}
\end{figure}
%%%%%%%%%%%%%%%%%%%%%%%%%%%%%%%%%%%%%%%%%%%%%

As addressed and demonstrated in the previous subsection \ref{subsec - SMsubsec phon conf planar structures} and \ref{subsec - SMsubsec 3D phon conf}, the traps perform as resonators that confine phonons in all three dimensions. In addition, the finite height of the phonon effective potential barriers allows for the design of lattices, with the concomitant emergence of acoustic 2D phonon bands, in a similar way as what happens to polaritons.% and is illustrated in Fig.\,1(d) of the main text.\\

To model how the cavity phonon mode's energy is affected by the different trap and array configurations (isolated traps, double-traps, and 2D arrays) we start from the non-etched effective quadratic dispersion relation arising when $k_z$ is quantized, i.e.
\begin{eqnarray}
E(k_x,k_y)=E_{\mathrm{cav,ne}}+\frac{\hbar^2(k_x^2+k_y^2)}{2 m_{\mathrm{eff}}}. 
\end{eqnarray}
Notice that for a material with homogeneous speed of sound $v_s$ the effective mass $ m_{\mathrm{eff}}=E_{\mathrm{cav,ne}}/v_s^2$. Such an effective mass description of the unconfined non-etched zone is incorporated in a 2D Schr\"odinger equation, 
\begin{eqnarray}
\left[-\frac{\hbar^2}{2 m_{\mathrm{eff}}}\nabla^2+E_{\mathrm{cav,ne}}+V_{\textrm{e}}(x,y)\right]\,\Psi(x,y)=E\,\Psi(x,y)\,, 
\end{eqnarray}
that adds, via the potential $V_{\textrm{e}}(x,y)$, the effect of the trapping induced by the etching. Taking into account that the energy of the phonon mode in a large etched region is $E_{\mathrm{cav,e}}$ we can assume that the full height of the potential in an etched zone is $V_{\mathrm{max}}=(E_{\mathrm{cav,e}}-E_{\mathrm{cav,ne}})$. Consequently, each square trap with a given index $i$, centered at $(x_i,y_i)$ contributes to $V_{\textrm{e}}(x,y)$ the potential
\begin{eqnarray}
V_i(x,y)=V_{\mathrm{max}}\,\big[1-v_i(x-x_i)v_i(y-y_i)\big]
\end{eqnarray}
where, following Ref.[\onlinecite{Kuznetsov2018}], the profile of the trap ``$i$'' along each direction is given by 
\begin{eqnarray}
v_i(\alpha)=\mbox{$\frac{1}{2}$}\left[\mathrm{erfc}\left(\mbox{$\frac{\alpha-\frac{w_i}{2}}{0.55\,\delta_i}$}\right)-\mathrm{erfc}\left(\mbox{$\frac{\alpha+\frac{w_i}{2}}{0.55\,\delta_i}$}\right)\right] ,
\end{eqnarray}
where $w_i$ is the trap width, $\delta_i$ is the 10\% to 90\% transition length, and $\mathrm{erfc}(...)$ stands for the \textit{complementary error function}. Using finite differences we solve the eigenvalue problem by discretising a large 2D zone containing the traps. For the case of molecules and arrays, all traps have $w_i=w$ and are separated by the distance $\Delta$. The simulated zone restricts to the supercell, where the super-lattice spacing is $a=w+\Delta$. After the numerical calculation, we obtain the eigen-energies for each value of $\bm{k}$ (1D or 2D) by the customary approach of imposing periodic conditions that fulfill the Bloch's theorem. For the width of the transition regions between the nER to ER, we take 0.35\,$\mu$m, consistent with both the modeling of the polariton properties and scanning tunneling microscopy (STM) studies in similar structures\cite{SM_Kuznetsov2018}. \\

\begin{figure}[!!!ttt]
    \begin{center}
   	\subfigure[~ 2\,$\mu$m\,$\times$\,2\,$\mu$m\text{ traps separated by }2\,$\mu$m]{\includegraphics[trim = 0mm 0mm 0mm 0mm, clip=true, keepaspectratio=true, width=0.4\columnwidth, angle=0]{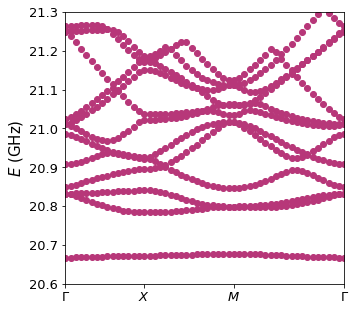}}
    \subfigure[~ Inverted 2\,$\mu$m\,$\times$\,2\,$\mu$m\text{ traps separated by }2\,$\mu$m]{\includegraphics[trim = 0mm 0mm 0mm 0mm, clip=true, keepaspectratio=true, width=0.4\columnwidth, angle=0]{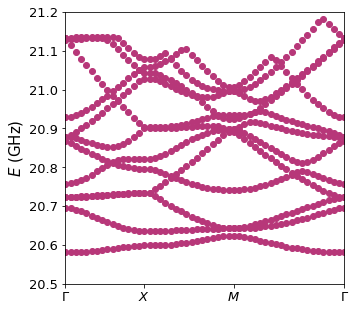}}
    \end{center}
    \vspace{-0.4 cm}
\caption{
{\bf Phonons in polariton-phonon hybrid lattices: comparison with inverted potential lattices.}
\textbf{(a)} In-plane dispersion of the acoustic modes of a square lattice of 2\,$\mu$m\,$\times$\,2\,$\mu$m squared traps, separated by 2\,$\mu$m. \textbf{(b)} Complementary lattice to (a) where the etched and non-etched regions are permuted.
}
\label{FigS8}
\end{figure}

The results of the calculations using the above described effective model are shown in Figs.~\ref{FigS7} and \ref{FigS8}. The acoustic phonon 2D band dispersion relation within the first Brillouin zone is shown in Fig.~\ref{FigS7}(a) for the case of a square array of 1\,$\mu$m\,$\times$\,1\,$\mu$m traps separated by $1\,\mu$m. The acoustic dispersion follows the usual path in $k$-space, i.e. $\Gamma\rightarrow X\rightarrow M\rightarrow\Gamma$, as indicated in panel (c) of this figure. Figure~\ref{FigS7}(b) corresponds to the acoustic phonon 2D band dispersion relation for the same square array of 1\,$\mu$m\,$\times$\,1\,$\mu$m traps, but with a distance between traps enlarged to $3.5\,\mu$m. 
As can be observed, only the lowest (fundamental) bands in (a) and (b) are energetically separated from the rest, resembling a tight-binding--like situation. The other higher energy bands are similar for both cases, resembling a free-electron--like physics. When analyzing the fundamental bands, it can be noted that the case with more separated traps ($3.5\,\mu$m barriers) displays a \textit{flatter band}, that corresponds near the $\Gamma$ point to more massive phonons [Fig.\,\ref{FigS7}(b)]. Otherwise, the less separated traps (1\,$\mu$m barriers) result in a broader band, and thus correspond to lighter phonons [Figs.\,\ref{FigS7}(a)]. To be noticed here is that a pure 3D phonon gap around $\sim 20.75$~GHz \textit{only} exists for the thicker barrier [panel (b)].

The different bands result from the coupling of the discrete energy levels that are confined in 0D within each trap. For example, the lower energy band in Fig.~2(d) of the main text arises mainly from the fully confined s-like ground state of the phonon trap, and thus displays a tight-binding--like parabolic dependence around the $\Gamma$ point, equivalent to the polariton case shown in panel (b) of that same figure. %The s-like character refers here to the in-plane spatial distribution of the mode within the trap. 
For this modelled structure, with small traps, the higher states are not confined and thus resemble those of almost free-electrons in a lattice. %As a result the second band in Fig.~1d has a preponderant p-like character, but it is also mixed with trap levels of higher order. 

A noticeable consequence is that the proposed technology can be used for phonon engineering in 2D based acoustic nanocavities, as has been previously proposed theoretically to obtain phonon molecules, band structures, and acoustic Bloch oscillations using simpler 1D layered media ~\cite{SM_Kimura2007b, SM_Bruchhausen2018}.\\

%%%%%%%%%%%%%%%%%%%%%%%%%%%%%%%%%%%%%%%%%%%%%
%%%%%%%%%%%%%%%%%%%%%%%%%%%%%%%%%%%%%%%%%%%%%
Additionally, in Fig.\ref{FigS8} two further examples of 2D dispersions for two trap arrays are shown. Panel (a) corresponds to a square array of 2\,$\mu$m\,$\times$\,2\,$\mu$m traps separated by 2\,$\mu$m, and \textbf{(b)} for the inverted complementary structure, i.e. the regions that were etched and non-etched are inverted. The path followed to draw the 2D dispersion is the same as the one shown previously in Fig.\ref{FigS7}(c). In the first array (panel (a)) the phononic ground state is localized in the trap, with low overlap between traps. Contrarily, in the second array (the inverted one, panel (b)), the ground state is a more dispersive band.

\subsection{Analogy between effective potentials for polaritons and phonons}

The cavity exciton-polariton confinement in the growth direction ($z$) of the samples is determined by the photonic component. As explained in Ref.\cite{SM_Fainstein2013}, for the Al$_{\textrm{x}}$Ga$_{1-\textrm{x}}$As--family based planar microcavity structures, due to a ``magical coincidence'' the optical and acoustic impedance mismatch is basically the same. Consequently such a structure confining photons, identically confines phonons along the growth direction. Moreover, the confined photons and phonons share the very same wavelength, and their frequency difference is simply determined by the light-to-sound speed ratio. 
An important difference for phonons however is the existence of longitudinal acoustic (LA) and transverse acoustic (TA) modes. For planar systems grown in the [001]-direction, the above considerations apply for both acoustic polarisations separately.
Because of the involved light-phonon coupling mechanism, in such planar structures only coupling to LA phonons becomes accessible. 

Conceptually, photon trapped states (consequently polariton states) as those treated in this work, will have a $z$-component that resembles that of the planar structure, and additional in-plane components that will be defined by the propagation dynamics and lateral boundary conditions. Within a simple picture the same can be applied to phonons. The ground states of both will have similar wave-functions (both in-plane symmetric, `$s$'-like), and both will have in-plane anti-symmetric ($p$-type) first excited states. Similar lateral effective potentials are expected, \textit{a priori}, given by the polariton or phonon frequency difference of the corresponding planar structure inside and outside the traps, an Ansatz that works well to describe our experimental findings.

The detailed character of the phonon modes in 0D traps, however, is more complex. When the in-plane symmetry is broken, as is the case of lateral micro-structuring for obtaining zero-dimensional (0D) trapped states, the acoustic vibrations are no longer pure LA and TA, but become mixed vibrational states, that -within a simple picture- are related by the Poisson ration. However, a detailed description of the corresponding mixed-vibrational eigen-states becomes non-trivial and numerical solutions need to be sought for \cite{SM_Anguiano2018b, SM_Lamberti2017}. Although confined phonon states with similar envelope functions and spatial dependence to the case of the confined photons will exist, the rigorous quantitative description of the acoustic system will be more complex \cite{SM_Lamberti2017}. See Supplementary Note 7B for an example of such simulations.

Another important contrast between the phonon and polariton system, comes from the excitonic part of the polariton. 
The frequency of confined phonons and bare photons monotonically increases for decreasing cavity spacer thickness (i.e., for increasing amount of etching). However, this is not so straightforward in the polariton regime. Due to the  exciton-photon strong coupling, the lower polariton level (from which the ground states of 0D traps and arrays are constituted) saturates at the exciton energy irrespective of the blue-shift of the bare photon states. Also the polariton effective mass depends on the cavity-exciton detuning. Consequently, the more excitonic the trap and barrier polaritons are, the weaker the analogy to the vibrational system is expected to be. In this work, the detuning is always kept negative for the trap states, i.e. the photonic components of the involved polariton states are dominant. But the energy of the barrier states is essentially saturated at the exciton energy. Qualitatively the analogy still holds, but one has to be careful when quantitative definitions are required.

In addition, the excitonic component of polaritons introduces another relevant difference when the particle occupation increased, due to the exciton-mediated polariton-reservoir and polariton-polariton interaction. 
For example, with increasing excitation power the confined polariton states blue-shift in an amount that is different for the non-etched well (more photonic) and etched barrier (more excitonic) regions of the effective polariton potential. These non-linearities do not exist for phonons.

\section*{Supplementary Note 5: Calculation of the effective mass of the involved acoustic mode}\label{sec: effective mass calculation}
\setcounter{subsection}{0}

%In order to obtain the effective mass ($m_{\textrm{eff}}$), a parameter that is usually used to describe the optomechanical coupling between the fundamental confined photonic (polaritonic) mode with the fundamental confined acoustic mode, we consider the involved potential energy. 
In order to obtain the effective mass ($m_{\textrm{eff}}$), a parameter that is usually used to describe the optomechanical coupling between the confined photonic (polaritonic) modes and a confined acoustic mode, we consider the involved potential energy.
The general phonon displacement $\vec{U}(\vec{r})$ of the acoustic mode can be parameterized as $\vec{U}(\vec{r})= u_0\,\vec{u}(\vec{r})$, where $\vec{u}(\vec{r})$ corresponds to the normalized displacement mode. The normalization of this mechanical mode is considered in such a way that at position $\vec{r}_0$, $|\vec{U}(\vec{r}_0)|^2\equiv 1$. $\vec{r}_0$ is defined as the reduction point and is chosen as the position where the modes displacement is maximal.\cite{SM_Baker2014} For our system, the reduction point is situated at the interfaces of the phonon cavity spacer. The potential energy of the parameterized mechanical oscillator must be equal to the actual potential energy of the confined phonon breathing mode:
\begin{eqnarray}
\mbox{$\frac{1}{2}$}\Omega_{\textrm{M}}^2\int d^3\vec{r}\, \rho(\vec{r})\,|\vec{U}(\vec{r})|^2 = \mbox{$\frac{1}{2}$}\,\Omega_{\textrm{M}}^2\,m_{\textrm{eff}}\,u_0^2\ .
\end{eqnarray}
Consequently, the effective mass is obtained as
\begin{eqnarray}
m_{\textrm{eff}}=\frac{\int d^3\vec{r}\, \rho(\vec{r})\,|\vec{U}(\vec{r})|^2}{u_0^2}\equiv \int d^3\vec{r}\, \rho(\vec{r})\,|\vec{u}(\vec{r})|^2\ ,
\end{eqnarray}
where $\rho(\vec{r})$ is the density distribution field of the structure. \\

The parameterized effective equation of motion for the cavity confined breathing mode, corresponding to a damped driven oscillator, will be given by \cite{SM_Metzger2008}
\begin{eqnarray}
m_{\textrm{eff}}\frac{d^2u}{dt^2}+m_{\textrm{eff}}\,\Gamma_{\textrm{M}}\,\frac{du}{dt}+m_{\textrm{eff}}\Omega_{\textrm{M}}^2\,u=F\ .
\end{eqnarray}
Here $m_{\textrm{eff}}$ is the effective mode's mass, and $\Gamma_{\textrm{M}}$ represents the mechanical damping. The driving force $F$ of the effective mechanical system involves the sum of different contributions, such as radiation pressure ($F_{\textrm{RP}}$) and electrostrictive ($F_{\textrm{ES}}$).

%%%%%%%%%%%%%%%%%%%%%%%%%%%%%%%%%%%%%%%%%%%
\section*{Supplementary Note 6: Effective polariton trap potential and higher-order optomechanical coupling mechanism}
%%%%%%%%%%%%%%%%%%%%%%%%%%%%%%%%%%%%%%%%%%%%%%
\setcounter{subsection}{0}
\subsection{The Gross-Pitaeskii equation with effective potential}
The effective confinement potential for the polaritons is estimated using an effective Gross-Pitaeskii equation that takes into account both the blue shift induced by the repulsive interactions with the exciton reservoir and the saturation of the Rabi splitting \cite{SM_Mangussi2021, SM_Reynoso2022}. The effective equation yields:
%%%%%%%%%%%%%%%%%%
\begin{equation}\label{Effective_LP_Equation}
 i \hbar \dfrac{\partial\psi}{\partial t} = \left[-\frac{\hbar^{2}\nabla^{2}}{2 m_{\textrm{LP}}}  + V_{\textrm{LP}}(\bm{r})+\frac{i\hbar}{2}\left(\frac{R\, P(\bm{r})}{\gamma_{\textrm{R}}+R |\psi|^2}-\kappa\right)\right] \psi\,,
\end{equation}
%%%%%%%%%%%%%%%%%%
where $\psi(\bm{r},t)$ is the complex field describing the lower polaritons and we have used the adiabatic approximation for the exciton reservoir \cite{SM_CarusottoRMP2013}. Equation (\ref{Effective_LP_Equation}) is a good approximation when describing the confined polaritonic levels $s$ and $p$. In addition, in our particular case, and for the purposes of describing only the energy of the polaritonic modes (not their occupation) and the effective potential, this equation can be further simplified by ignoring the last term \cite{SM_Mangussi2021, SM_Reynoso2022}. The effective potential is described as
%%%%%%%%%%%%%%%%%%%%%%%%%%%%
\begin{equation}\label{Effective_potential}
V_{\textrm{LP}}(\bm{r},n_{\textrm{R}})=\frac{1}{2}\left[ E_{\textrm{C}}+E_{\textrm{X}}-\sqrt{\Omega^{2}+\Delta^{2}}\right]\, ,
\end{equation}
%%%%%%%%%%%%%%%%%%
where $E_{\textrm{C}}(\bm{r})=\Delta_{0}+V_{\textrm{C}}(\bm{r})$ describes the photonic potential of the trap, $\Delta_{0}$ the bare detuning, $E_{\textrm{X}}(\bm{r},n_{\textrm{R}})=g_{\textrm{X}}n_{\textrm{R}}(\bm{r})$ the exciton energy ($g_{\textrm{X}}\approx 6\,\mu$eV$\mu$m$^2$) and  $\Delta(\bm{r},n_{\textrm{R}})=E_{\textrm{C}}(\bm{r})-E_{\textrm{X}}(\bm{r},n_{\textrm{R}})$ the effective detuning. Here we have ignored the contribution to $E_{\textrm{X}}$ from the repulsive interaction among the lower polaritons. This is fine in this case as the polaritons have a large photonic component.

The saturation of the Rabi splitting with increasing population of the reservoir is described  as
%%%%%%%%%%%%%%%%%%%
\begin{equation}
\label{Rabi}
\Omega(\bm{r},n_{\textrm{R}})=\frac{\Omega_{0}}{\sqrt{1+\frac{n_{\textrm{R}}(\bm{r})}{n_{\mathrm{Sat}}}}},
\end{equation}
%%%%%%%%%%%%%%%%%%%
where $\Omega_{0}$ is the Rabi splitting at zero carrier density and $n_{\mathrm{Sat}}\approx 3\times 10^3\mu$m$^{-2}$.
The effective mass is approximated as
%%%%%%%%%%%%%%%%%%
\begin{equation}
\label{m_LP}
\frac{1}{m_{\textrm{LP}}}=\frac{|X(\bm{r}_0,n_{\textrm{R}})|^2}{m_{\textrm{X}}}+\frac{|C(\bm{r}_0,n_{R})|^2}{m_{\textrm{C}}}\,, 
\end{equation}
%%%%%%%%%%%%%%%%%%
where $\bm{r}_0$  corresponds to the center of the pumped trap and the spatially dependent Hopfield coefficients are 
%%%%%%%%%%%%%%%%%%
\begin{eqnarray}
\nonumber
|X(\bm{r},n_{\textrm{R}})|^2 &=&\frac{1}{2}\left(1+\frac{\Delta(\bm{r},n_{\textrm{R}})}{\sqrt{\Omega(\bm{r},n_{\textrm{R}})^{2}+\Delta(\bm{r},n_{\textrm{R}})^{2}}}\right)\,, \\
|C(\bm{r},n_{\textrm{R}})|^2 &=& 1-|X(\bm{r},n_{\textrm{R}})|^2 \,.
\label{Hopfield}
\end{eqnarray}
%%%%%%%%%%%%%%%%%%
Note that we have explicitly taken into account the dependence of the parameters with the density of carriers in the reservoir.
The density of excitons in the reservoir as a function of the external pump power is estimated as   
%%%%%%%%%%%%%%%%%%
\begin{equation}
\label{n_R}
n_{\textrm{R}}(\bm{r}) = \frac{P(\bm{r}) \tau_{\textrm{R}} \alpha}{\hbar \omega_{\textrm{L}} \, 2 N_{\textrm{QW}}}\,,
\end{equation}
%%%%%%%%%%%%%%%%%%
where $P(\bm{r})$ is the pump power per unit area, $\tau_{\textrm{R}}$ is the effective lifetime of the exciton in the reservoir, $\alpha$ is the total effective absorption coefficient of the QWs, $\hbar \omega_{\textrm{L}}$ is the energy of the non resonant pumping laser, $N_{\textrm{QW}}$ is the number of quantum wells and the factor $2$ accounts for the dominant role of the triplet interactions.
We assume a Gaussian shape for the pump given by
%%%%%%%%%%%%%%%%%%%%%%
\begin{equation}
P(\bm{r})=\frac{P_{0}}{2 \pi \sigma_{\textrm{p}}^{2}} \exp\left[-\frac{(\bm{r}-\bm{r}_{\textrm{p}})^{2}}{2\sigma_{\textrm{p}}^{2}}\right]
\end{equation}
%%%%%%%%%%%%%%%%%%%%%
In the simulations we use an effective value for the standard deviation $\sigma_{\textrm{p}} \approx 3 \mu$m, and change the position of the spot ($\bm{r}_{\textrm{p}}$) to reproduce the particular experimental situation. 
The values for $\Omega_{0}$, $\Delta_{0}$ and the cavity parameters $m_{\textrm{C}}$ and $V_{\textrm{C}}(\bm{r})$ were obtained by fitting. The photonic cavity potential $V_{\textrm{C}}(\bm{r})$ was simulated following Refs. \cite{SM_Kuznetsov2018, SM_Mangussi2021}. The optimal results were obtained using $\Omega_{0}=6.0$meV and $E_{\textrm{C}}=\Delta_{0}=-10.5 \, \text{meV}$ and $E_{\textrm{C}}=\Delta_{0}+\Delta U_{\textrm{p}}=5.5 \, \text{meV}$ for the non etched and etched regions, respectively. Here $\Delta U_{\textrm{p}}=16$ meV is the potential barrier for photons, generated by the difference in the thickness of the cavity spacer between the two regions.\\

\begin{figure}[!!!ttt]
    \begin{center}
    \includegraphics[trim = 0mm 0mm 0mm 0mm, clip=true, keepaspectratio=true, width=0.6\columnwidth, angle=0]{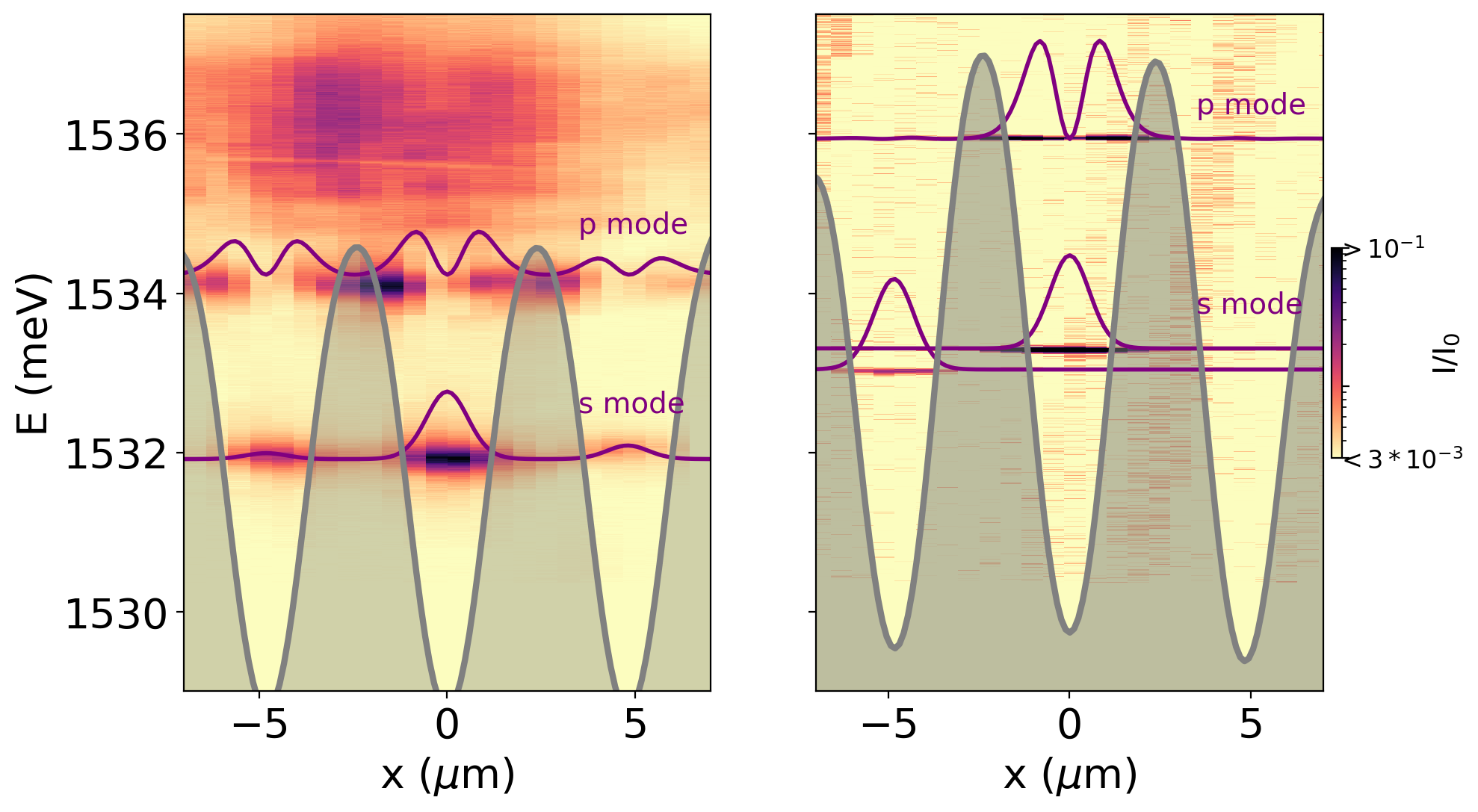}
    \end{center}
%    \vspace{-0.8 cm}
\caption{\textbf{Effective polariton trap potential.} Panel left (right) displays the spectral and spatial image corresponding to the low (high)-pump power condition. 
The effective trap potentials (shaded gray) that result from a Gross-Pitaevskii (GP) modelling of the quantum light fluid. Two situations are shown, corresponding to a power excitation condition below (left panel) and above (rig panel) the condensation threshold $P_{th}$.
The corresponding GP calculated confined polariton wavefunctions (thin purple curves) of the fundamental `s'-type and first excited `p'-type modes are also plotted and indicated.
The underlying density plot corresponds to the experimental measured photoluminescence at the corresponding excitation power. The data shown, was adapted from Ref.\cite{SM_Reynoso2022}.}
\label{Fig_PolaritonProfile}
\end{figure}

According to the model described above, in Fig.\,\ref{Fig_PolaritonProfile} we show the simulated effective trap polariton potential profile for two different non-resonant pump-power conditions, i.e. low- (left panel) and high-power (right panel) excitation. On each panel, the corresponding confined polariton wave-functions for each trap, derived from the Gross-Pitaevskii equation, are shown superimposed. The detuning between the pumped and neighbor trap ground states can be tuned through the excitonic-related repulsive interaction with the reservoir by varying the non-resonant pump-power. Notice that the polariton levels within each trap shift rigidly for increasing power.  As observed, in Fig.\,\ref{Fig_PolaritonProfile}, the `$s$'-type confined ground states are at near zero detuning at low power (left panel), while the detuning is evidently increased at high power (right panel). 

We would like to point out, that the overlap between the '$s$'-type confined ground states of the central trap and the neighboring trap is very small ($\sim$10$^{-4}$). These states turn out to be strongly confined within each trap, resulting in a negligible $s-s$ transition probability, and consequently a negligible inter-trap optomechanical coupling. 

On the contrary, as observed in Fig.\,\ref{Fig_PolaritonProfile}, the penetration into the barriers of the first excited (`$p$'-type) confined states is significantly higher than that of the `$s$' states. The extension and ``delocalization'' of these excited states is clearly evidenced. For these confined `$p$' states, the simulation shows that the overlap between neighbor traps' wave-functions is $\sim$0.4, implying a much larger transition probability
%transition probability that is seven orders of magnitude larger for the $p-p$  ($\sim0.16$) 
as compared to the $s$--$s$ case %($\sim10^{-8}$) 
\cite{SM_Reynoso2022}. 
The underlying density-plot corresponds to the system's photoluminescence, where the low- left panel) and high-power (right panel) situations are very well described by the model. \\

The negligible overlap between ground states of neighbor traps, and the delocalization of the first excited states, justifies a Hamiltonian model for the inter-trap optomechanical coupling that considers a second order process mediated through the $p$-states to couple the two fundamental $s$ modes.

%%%%%%%%%%%%%%%%%%%%%%%%%%%%%%%%%%%%%%%%%%%%%
%%%%%%%%%%%%%%%%%%%%%%%%%%%%%%%%%%%%%%%%%%%%%
%%%%%%%%%%%%%%%%%%%%%%%%%%%%%%%%%%%%%%%%
\subsection{Second order optomechanical coupling rate $g_2$}
%%%%%%%%%%%%%%%%%%%%%%%%%%%%%%%

The negligible inter-trap optomechanical coupling of the traps' ground states (see previous subsection), and the experimental evidence of the strong interconnection between the neighbour traps, which lead to the appearance of induced sidebands on the photoluminescence (PL) spectrum \cite{SM_Chafatinos2020} at phonon energy separations, drives us to introduce a simplified model with a \textit{second order inter-trap optomechanical coupling} mechanism, that indeed captures the main physical ingredients \cite{SM_Reynoso2022}. 

The model takes into account the two fundamental polaritonic modes of two neighboring cavities, a single polariton excited state that is shared by \textit{both} traps, and an on-site phonon-mediated coupling between ground and excited states.
The Hamiltonian then has two contributions, $H=H_0+H_\mathrm{OM}$. Here 
%%%%%%%%%%%%%%%%%%%%%%%%%
\begin{equation}
H_0=\sum_{j=1}^3 \hbar\omega_j\,\hat{a}_j^\dagger\hat{a}_j^{}+\sum_{n} \hbar\Omega_{ n}\,\hat{b}_{n}^\dagger\hat{b}_{ n}^{ }\, ,
\end{equation}
%%%%%%%%%%%%%%%%%%%%%%%%%
describes the decoupled polariton and phonon modes: i)   $\hat{a}_j^{\dagger}$ ($\hat{a}_j$) creates (annihilates) a polariton in the $j$-mode with energy $\hbar\omega_j$, where  $j=3$ refers to the excited mode; ii) $\hat{b}_{n}^{\dagger}$  ($\hat{b}_{n}$) creates (annihilates) a $p$-phonon in the ${n}$-mode with energy $\hbar\Omega_{n}$.  The index $n$ labels the fundamental and the overtone mechanical modes so that, for example,  $\Omega_1=3\Omega_0 \sim 2\pi \times 60$GHz (for simplicity, we take $\Omega_0= 2\pi \times 20$GHz). The linear optomechanical coupling reads
%%%%%%%%%%%%%%%%%%%%%%%%%%%%%%
\begin{eqnarray}
H_\mathrm{OM}&=&\sum_{j=1}^2\sum_n \hbar g_{jn}(\hat{a}_j^\dagger \hat{a}_3^{}+\hat{a}_3^\dagger \hat{a}_j^{})(\hat{b}_{n}^\dagger+\hat{b}_{n}^{})
\,.
\end{eqnarray}
%%%%%%%%%%%%%%%%%%%%%%%%%%%%%%

The optomechanical features in the PL spectrum appear when the modes $\hat{a}_1$ and $\hat{a}_2$ are tuned to a particular energy difference (related to the phonon frequencies). This calls for a description where these two modes play the more important role. Since the excited mode is well-separated from the fundamental modes in comparison with the phonon energy, $\Delta_j=\omega_3-\omega_j\gg\Omega_n$, it is thus reasonable to assume that one can describe the dynamics with an effective reduced Hamiltonian. This can be done by means of a suitable canonical transformation \cite{SM_Reynoso2022}.
It can be readily seen that,
to leading order in $g_{jn}/\Delta_j$ and $\Omega_n/\Delta_j$,  and retaining only those terms involving the phonon operators, 
%%%%%%%%%%%%%%%%%%%%%%%%%%%%%%%%%%%%%%%
\begin{eqnarray}
\label{H'}
H'&=&\sum_{j=1}^3 \hbar\omega_j\,\hat{a}_j^\dagger\hat{a}_j^{}+\sum_{n} \hbar\Omega_{ n}\,\hat{b}_{n}^\dagger\hat{b}_{ n}^{ }\\
\nonumber
&&+\sum_{j=1}^2\sum_{n,m}\frac{\hbar g_{jn}g_{jm}}{\Delta}(\hat{a}_3^\dagger\hat{a}_3^{}-\hat{a}_j^\dagger\hat{a}_j^{})(\hat{b}_{n}^\dagger+\hat{b}_{n}^{})(\hat{b}_{m}^\dagger+\hat{b}_{m}^{})\\
\nonumber
&&-\sum_{n,m}\frac{\hbar g_{1n}g_{2m}}{\Delta}(\hat{a}_1^\dagger\hat{a}_2^{}+\hat{a}_2^\dagger\hat{a}_1^{})(\hat{b}_{n}^\dagger+\hat{b}_{n}^{})(\hat{b}_{m}^\dagger+\hat{b}_{m}^{})\,.
\end{eqnarray}
%%%%%%%%%%%%%%%%%%%%%%%%%%%%%%%%%%%%%%
Here, for simplicity, we took $\Delta_j\pm\Omega_n\sim\Delta_j\equiv\Delta$. The second line in $H'$ reflects a coupling between the phonon displacement and the polariton mode occupations. This leads to a renormalization of the phonon energies induced by the polaritons in both the ground and excited states. When a mechanical coherent state sets in, it also leads to a modification of the polariton energies depending on the phonon occupation. The last line in $H'$, in turns,  makes explicit that there is an effective \textit{quadratic} phonon coupling between the two fundamental polariton modes of the neighboring traps, of order $g_2=g_0^2/\Delta$ \cite{SM_Reynoso2022}.

\subsection{Estimation of the inter-trap single-photon electrostrictive coupling rate $g_2$}

The linear on-site optomechanical coupling factor $g_0$, which is governed by the processes that couple two polariton levels within the \textit{same} trap, is estimated considering the effective exciton-mediated optomechanical coupling reported for similar polariton traps as those considered in this work \cite{SM_Kuznetsov2021}. 

The on-site coupling in  Ref.~\cite{SM_Kuznetsov2021} results mainly from a deformation-potential interaction modulated by intense electrically generated bulk acoustic waves (BAWs). A value of $g^\mathrm{eff}_\mathrm{om}/2\pi \sim 50$~THz/nm was obtained, which accounts for the change of polariton energy per unit of acoustic displacement \cite{SM_Kuznetsov2021}. 

This later parameter is related to the actual on-site optomechanical coupling constant ($g^\mathrm{eff}_0$) considering the displacement induced by the zero point fluctuations ($x_\mathrm{zpf}$) by $g^\mathrm{eff}_0=g^\mathrm{eff}_\mathrm{om}\,x_\mathrm{zpf}$\,\cite{SM_RMP}.
For a similar structure of $\sim2\,\mu$m lateral size, this value has been estimated to be roughly $x_\mathrm{zpf}\sim 0.5$\,fm~\cite{SM_Villafane2018}. Consequently, $g^\mathrm{eff}_{0}/2\pi \sim 50$~MHz for this system, which represents a  very large value compared to other reported optomechanical systems that only account for an optical radiation back-action mechanism based on radiation pressure interaction\,\cite{SM_RMP}. The Hopfield coefficient for this cavity polariton trap system was of the order of $|X|^2\sim 0.7$, and the reported structure had only one embedded QW~\cite{SM_Kuznetsov2021}. 

The structure investigated in this work, has four cavity embedded QWs instead of one, proportionally increasing the corresponding interaction of the involved fields. For high excitation powers used for exciting the traps analyzed here, the excitonic Hopfield coefficient is estimated to be around $|X|^2\sim 0.05$ \cite{SM_Mangussi2021}, i.e. the involved polariton states have large photonic components. Considering these two differences, the $g_0$ would be roughly a factor of $4$ times larger and a factor $0.05/0.7$ smaller with respect to the above obtained $g^\mathrm{eff}_{0}$. Therefore, the on-site optomechanical coupling factor for the present work, resulting from a deformation-potential interaction, can be estimated to be $g_{0}/2\pi\sim 14\,\mathrm{MHz}$.\\ 

Based on this estimated on-site (intra-trap) optomechanical coupling factor, considering the model discussed in the previous subsection, and accounting that the energy distance between the excited state and the ground state is $\Delta_j\approx 2$\,meV ($\sim$484\,GHz) (see e.g. Fig.\ref{Fig_PolaritonProfile}), we can estimate an inter-trap single-photon coupling rate of $g_2=g_0^2/\Delta\sim 400$\,Hz. This inter-site second order coupling was shown in Ref.\cite{SM_Reynoso2022} to lead to the development of mechanical self-oscillation due to an optomechanical parametric process in the resonant condition when polaritons condense. That is, if the energy separation between neighbor traps is 2$\Omega$, in Ref.\cite{SM_Chafatinos2020}, the magnitude of the polariton energy modulation induced by this coherent self-substained mechanical wave was estimated to be a fraction $\sim$0.6 of the mechanical energy $\hbar\Omega$.

%%%%%%%%%%%%%%%%%%%%%%%%%%%%%%%%%%%%%%%%%%%%%%%%%%%%%%%%%%%%%%%%%%%%%%%%%%%%%%%%%%%%%%%%%%%%%%%%%%%%%%

\section*{Supplementary Note 7: Experimental results of polariton properties}
\setcounter{subsection}{0}
%%%%%%%%%%%%%%%%%%%%%%%%%%%%%%%%%%%%%%%%%%%%%
\begin{figure}[!!!ttt]
    \centering
    \includegraphics[width=0.6\columnwidth]{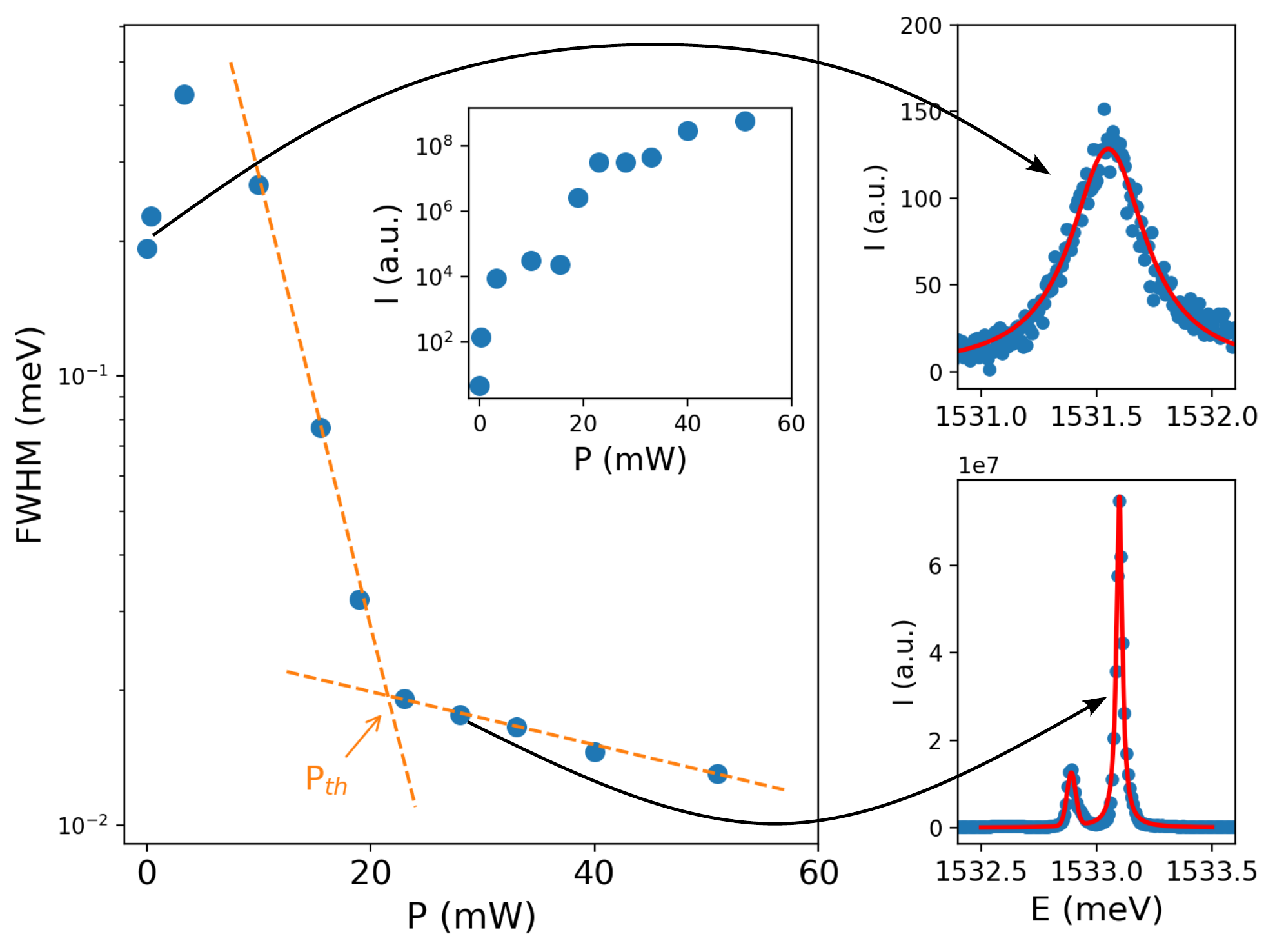}
    \caption{\textbf{Determination of the condensation threshold power $P_{\textrm{th}}$}. Left panel: the pump power dependence of the full width at half maximum (FWHM) of PL spectra of the polaritonic fundamental mode of the pumped trap (in logarithmic scale). The condensation threshold power $P_{\textrm{th}}$ is identified by the intersection of the change in slope of the FWHM as a function of power (orange dashed lines). The inset panel shows the intensity of fundamental pumped mode (in logarithmic scale). The right panels show the measured PL spectra (blue dots) for two pump-power situations, indicated correspondingly by the black-arrows. The top(bottom) right panel depict the situation before(after) condensation. For the latter two fundamental modes can be identified: the one of the central pumped trap (at $\sim$1533.1 meV), and a neighbor trap to lower energy.}
    \label{fig:Pth}
\end{figure}

\subsection{Condensation threshold}
The pump power threshold of the polariton condensation $P_{\textrm{th}}$ was established by analysing the photoluminescence (PL) of the fundamental polaritonic mode of the pumped central trap. For determining $P_{\textrm{th}}$, the full width at half maximum (FWHM) was extracted as a function of excitation power, as shown in Fig. \ref{fig:Pth} (left panel). For $P\ll P_{\textrm{th}}$, with increasing pump power, the polaritonic mode broadens due to an increase in polariton decoherence, consequence of the polaritonic self-interaction noise introduced by polariton-polariton and polariton-reservoir interactions. Afterwards, the mode's width drops off steeply evidencing the transition to the condensed phase \cite{SM_Kasprzak}. For $P\sim P_{\textrm{th}}$, the PL line-width changes its evolving slope (as indicated in Fig.\,\ref{fig:Pth}). $P_{\textrm{th}}$ is determined as the power for which this change is noted (see orange dashed lines). The inset of the left panel shows the PL intensity as function of excitation power. A change in the evolution of this magnitude is observed near $P\sim P_{\textrm{th}}$.
In the right panels of Fig.\,\ref{fig:Pth}, typical spectra of the fundamental modes' PL, before (top-right) and after (bottom-right) condensation, are shown. The increase of five orders of magnitude in intensity signals the transition to the condensed phase (see inset).

\subsection{Mechanical modulation of polariton mode: Appearance of sidebands}
%%%%%%%%%%%%%%%%%%%%%%%%%%%%%%%%%%%%%%%%%%%%%
\begin{figure}[!!!ttt]
    \begin{center}
    \subfigure[]{\includegraphics[width=0.4\columnwidth]{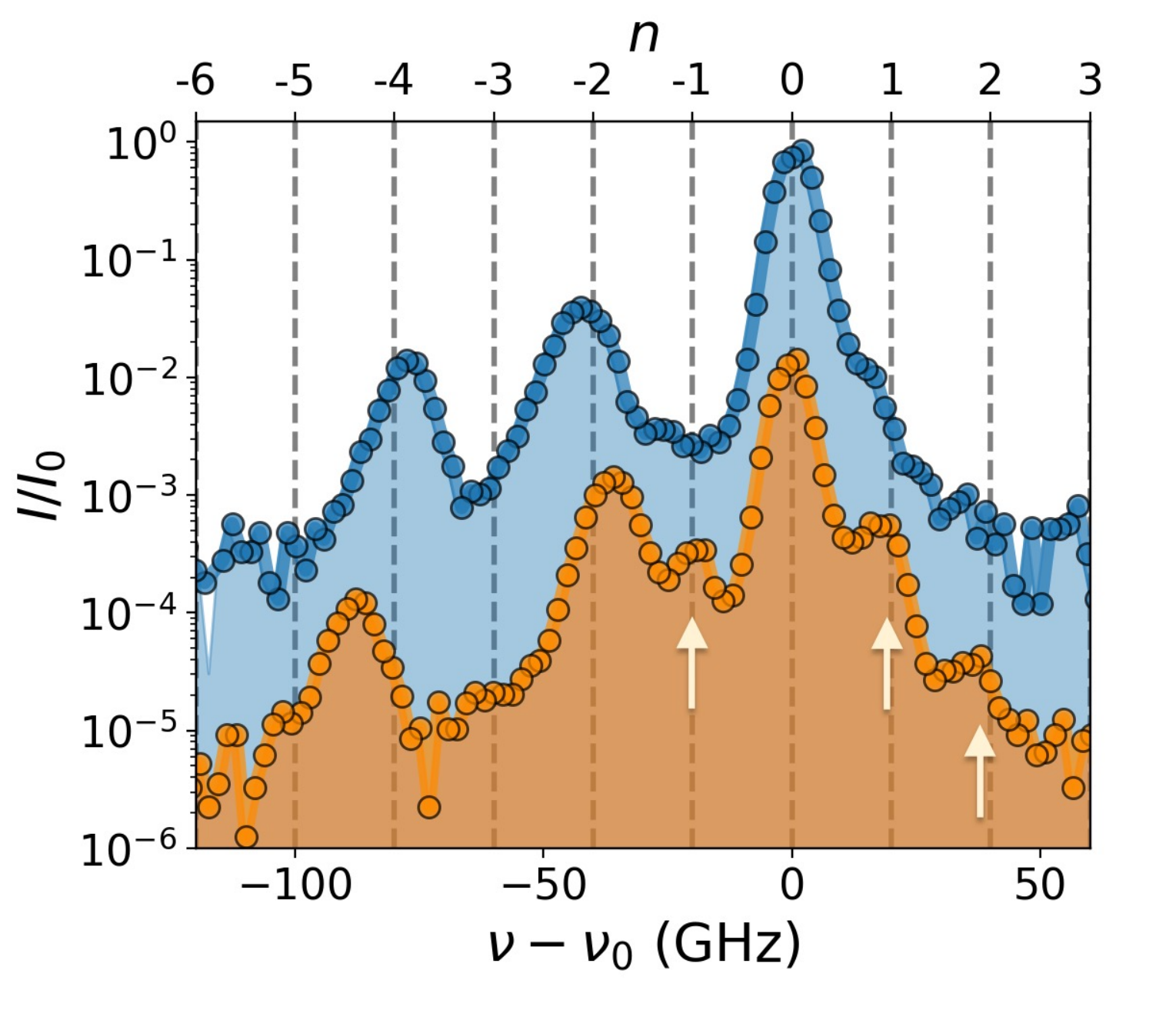}}
    \subfigure[]{\includegraphics[width=0.4\columnwidth]{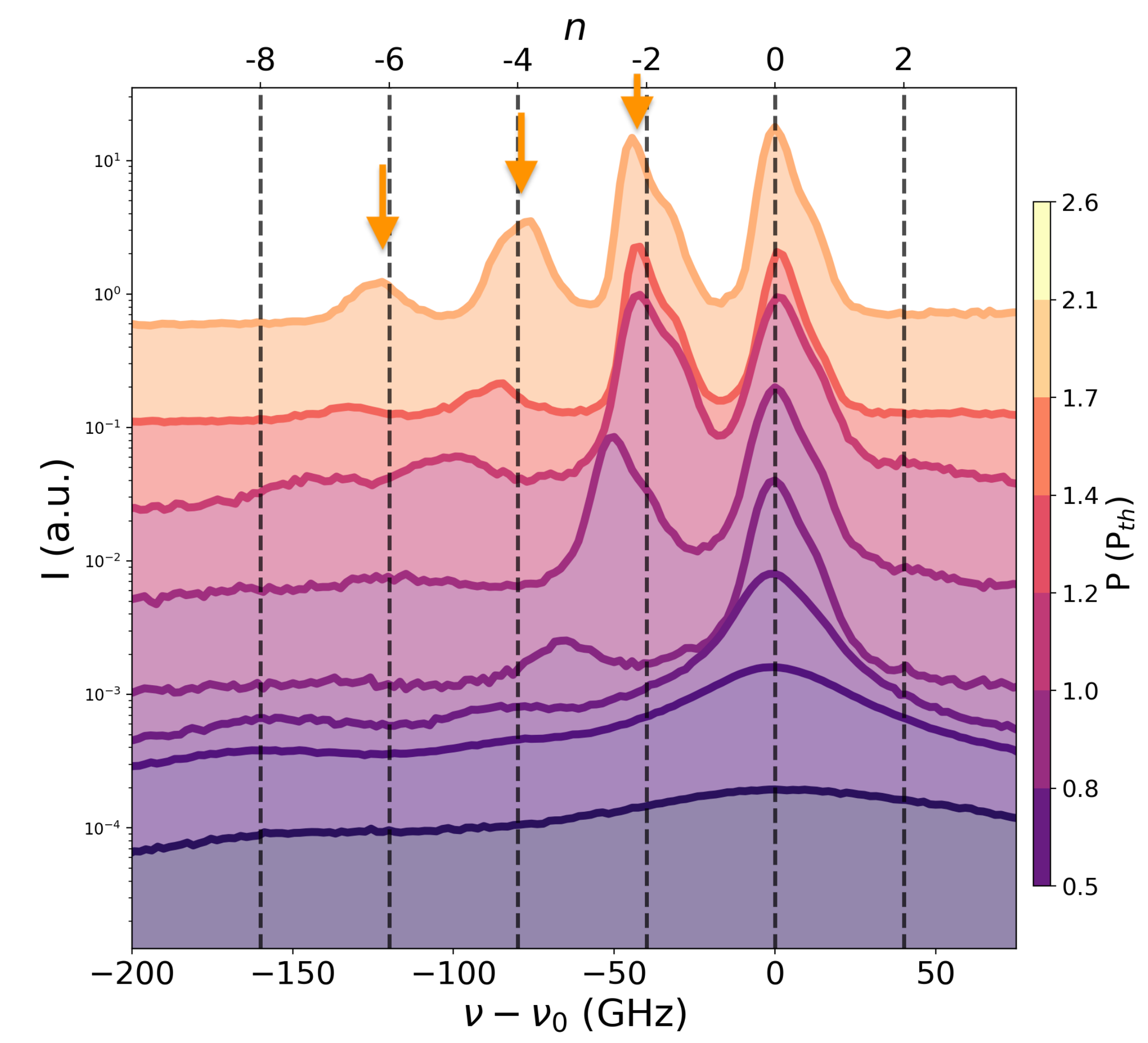}}
   	\end{center}
    \vspace{-0.4 cm}
    \caption{\textbf{Appearance of sidebands by optomechanical interaction.} \textbf{(a)} PL emission with (orange spectrum) and without (blue spectrum) side-bands (indicated with white arrows at multiples of $\nu_m\sim20$GHz). Here the frequency scale is given relative to the fundamental mode's energy of the central pumped trap. The fundamental mode of the nearest neighbour trap asynchronously locks at $\nu-\nu_0\sim -2\nu_m=-40$GHz. A third fundamental mode appears to lower relative frequencies for both cases, corresponding to an additional neighbour trap.
    Furthermore, for the blue spectrum, this third model locks at $\nu-\nu_0\sim -4\nu_m=-80$GHz. This data has been adapted from Ref. \cite{SM_Chafatinos2020}.
    \textbf{(b)} Spectrum cascade of modes of an array of 1.6$\mu$m$\times$1.6$\mu$m square traps separated by 3.2$\mu$m. Above of pump power condensation threshold, four clear modes appear (indicated with orange arrows), which asynchronously locked at multiples of $2\times\nu_m$ for the highest powers \textit{without} sidebands.
    }   
    \label{fig:Sidebands}
\end{figure}
%%%%%%%%%%%%%%%%%%%%%%%%%%%%%%%%%%%%%%%%%%%%%

As we have reported in Ref.\cite{SM_Chafatinos2020}, an important characteristics that is observed when the two involved polariton states (the fundamental modes of the central and neighbour traps in this case) satisfy a proper resonance condition $\delta\nu=\nu-\nu_0=-2\times n \times \nu_m$ (with $n>0$ a integer number and $\nu_m$ the fundamental mechanical mode), is the appearance of strong side-bands. This phenomena indicates that the system entered a condition of optomechanical parametric oscillation\cite{SM_Reynoso2022}. In Fig.\,\ref{fig:Sidebands}(a), the orange spectrum corresponds to a near-resonance situation $\delta\nu\approx-2\nu_m$. Here, clear side-bands are observed (white arrows) for the central trap at $\nu-\nu_0=0$. However, this is not the general case. 

The blue spectrum in Fig.\,\ref{fig:Sidebands}(a), corresponds to the situation were locking resonance condition is satisfied, $\delta\nu=-2\nu_m$. In this case, the asynchronous locking is also observed for the third trap's fundamental mode at $\delta\nu=-4\nu_m$, but no evidence of side-bands is found. Moreover, another example of asynchronous locking is shown in Figure \ref{fig:Sidebands}(b) for an array of $1.6\mu$m$\times$1.6$\mu$m square traps, separated by $3.2\mu$m. From bottom to top the evolution of the PL of the traps' modes is plotted for increasing power. It is observed that, above the condensation threshold, four trap modes appear. At the highest pump power, all modes \textit{lock asynchronously} at a separation of $2\times\nu_m$ with each other. However, side-bands were not observed.

\subsection{Further examples of asynchronous locking of the polariton modes}
\begin{figure}[!!!ttt]
    \begin{center}
   	\includegraphics[trim = 0mm 0mm 0mm 0mm, clip=true, keepaspectratio=true, width=0.5\columnwidth, angle=0]{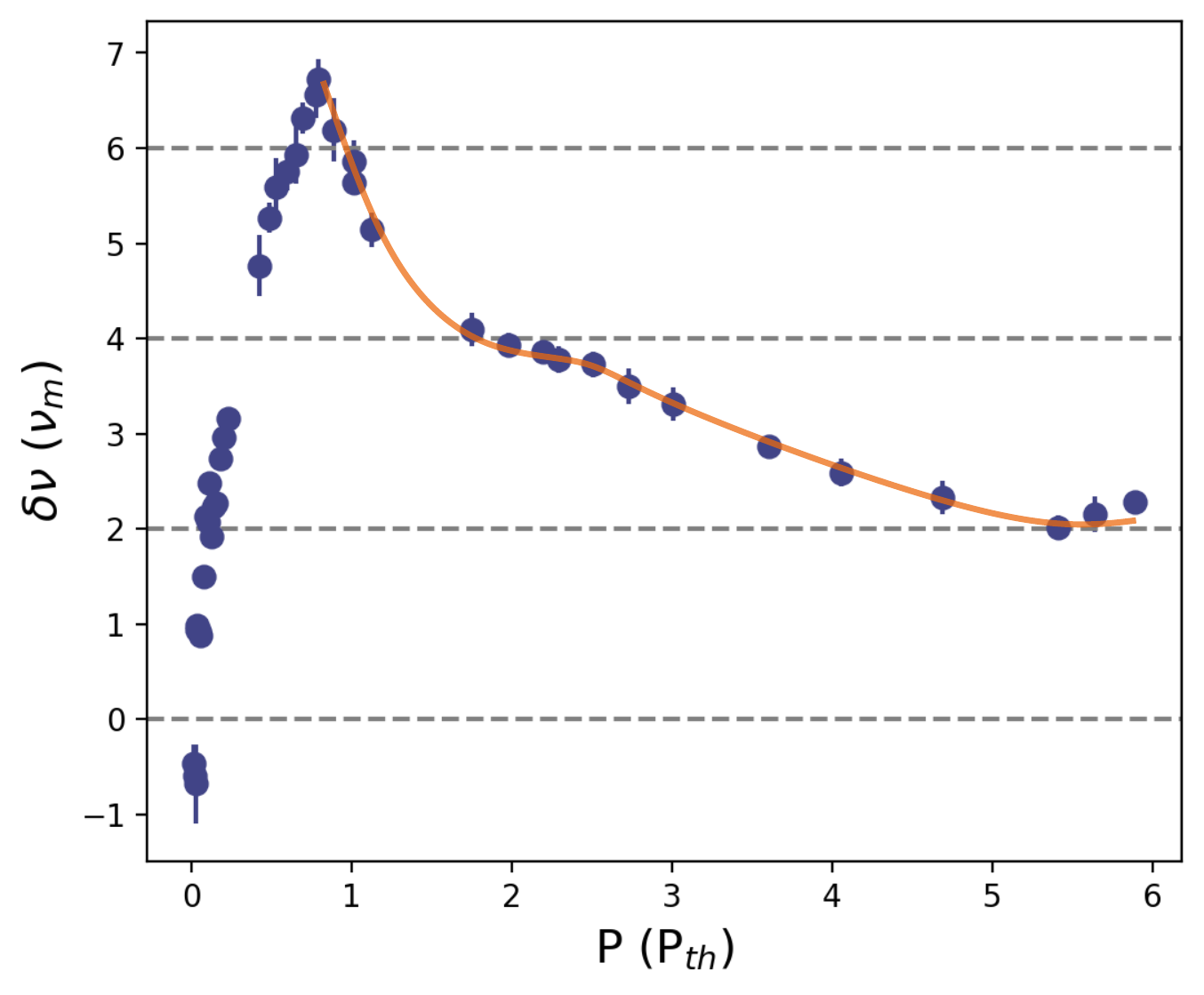}  
    \end{center}
    \vspace{-0.4 cm}
\caption{\textbf{Partial and total locking.} Frequency detuning (given in terms of fundamental mechanical frequency $\nu_m$) between the pumped and the neighbor trap as a function of pump power (given in terms of $P_{\textrm{th}}$, the condensation threshold power). The red full line is a guide to the eye. The data corresponds to Ref.\cite{SM_Reynoso2022}.}
\label{fig:locking}
\end{figure}

%%%%%%%%%%%%%%%%%%%%%%%%%%%%%%%%%%%%%%%%%%%%%
Figure \ref{fig:locking} shows the frequency detuning (given in terms of the fundamental mechanical frequency $\nu_m$) between the pumped and the neighbor trap as a function of pump power (given in terms of $P_{\textrm{th}}$, the condensation threshold power
The horizontal dashed lines indicate the matching of this detuning with even multiples of $\nu_m$. %Below $P_{\textrm{th}}$, the figure shows how the energy of both fundamental modes depart from each other.  
The detuning between modes above $P_{\textrm{th}}$ diminishes for increasing excitation power. Notably, when $\delta\nu$ reaches a value near $4\nu_m$ the evolution gets into a plateau (partial locking). When further increasing the excitation power the detuning continues to decrease, and finally locks at a separation of $2\nu_m$ for the highest powers. The data corresponds to Ref.\cite{SM_Chafatinos2020}. 

\begin{figure}[!!!ttt]
    \begin{center}
   	\includegraphics[trim = 0mm 0mm 0mm 0mm, clip=true, keepaspectratio=true, width=0.6\columnwidth, angle=0]{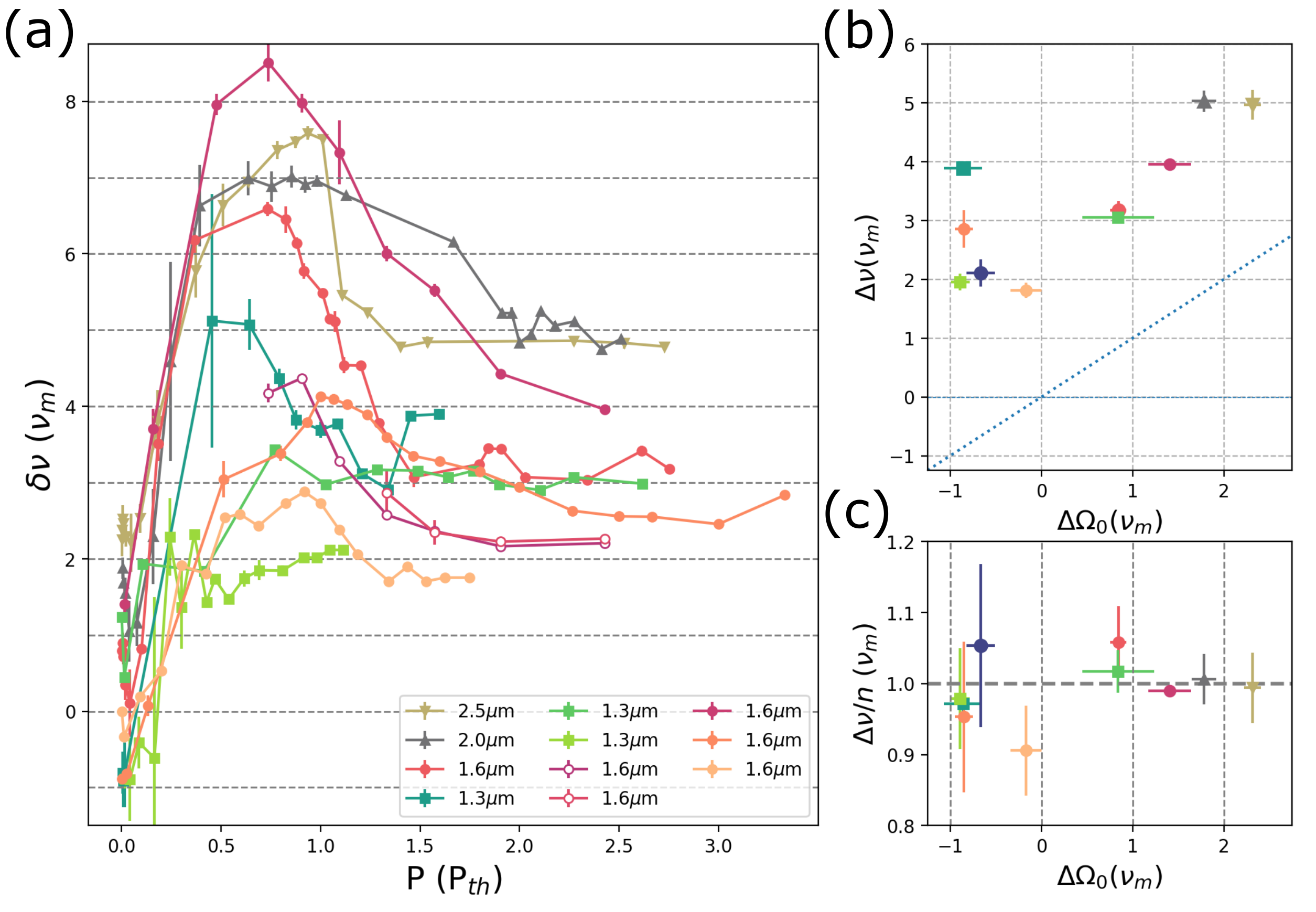}
    \end{center}
    \vspace{-0.4 cm}
\caption{\textbf{Additional experiments that evidence asynchronous locking.} \textbf{(a)} Dependence of the inter-trap energy detuning as a function of excitation power, performed on arrays with different square trap sizes and inter-trap separation: 2.5$\mu$m/5.0$\mu$m, 2.0$\mu$m/4.0$\mu$m, 1.6$\mu$m/3.2$\mu$m, 1.3$\mu$m/2.6$\mu$m (square size / separation size). \textbf{(b)} Data presented as in Fig. 4 of Ref. \cite{SM_Baas2008}, where $\Delta\Omega_0$ is the bare frequency detuning (obtained at small pump powers (P$\ll P_{\textrm{th}}$) and $\Delta\nu$ is the dressed frequency detuning (obtained above threshold, at powers for which asynchronous locking occurs). The blue dotted line corresponds to a non-synchronized phase $\Delta\nu = \Delta\Omega_0$. \textbf{(c)} Same result as panel (b) but with $\Delta\nu\rightarrow\Delta\nu/n$.}
\label{Fig_Locking_further examples}
\end{figure}
%%%%%%%%%%%%%%%%%%%%%%%%%%%%%%%%%%%%%%%%%%%%%
Following the above discussion, we now address different cases where the asynchronous locking of the polariton modes is observed. In Figure \ref{Fig_Locking_further examples}(a), the results of the evolution of the detuning corresponding to the fundamentals modes for increasing pump power are shown for a variety of different samples and trap sizes as indicated. In order to identify the existence of locking, in Fig.\,\ref{Fig_Locking_further examples}(b), the dressed detunings $\Delta\nu$ (obtained at high powers where the modes are locked) are presented following Fig.\,4 of A. Baas, \textit{et al.} \cite{SM_Baas2008}. Here $\Delta\Omega_0$ corresponds to the bare frequency detuning (obtained well below the threshold power ($P\ll P_{\textrm{th}}$). Bare and dressed detunings are given in units of the fundamental mechanical mode $\nu_m$. The blue-dashed line with unity slope ($\Delta\nu\equiv \Delta\Omega_0$) indicates the non-synchronized situation ($\Delta\Omega_0=\Delta\nu$). A fully synchronized phase, in turn, would appear as $\Delta\nu=0$. Note that neither of these two cases are observed: all experiments fall above the dashed line, with the dressed detunings corresponding closely to integer numbers of the confined phonon frequency.
In fact, if the same plot is normalized by the corresponding multiple $n$ of $\nu_m$, i.e. replacing $\Delta\nu\rightarrow\Delta\nu/n$, it is emphasized that all the curves converge to the same value, as shown in Fig.\,\ref{Fig_Locking_further examples}(c). Indeed all the points fall near the unity value given by the horizontal dashed line. 
%The data shown in Fig.\ref{fig:locking} is also included in the panel (b) and (c) of Fig.\,\ref{Fig_Locking_further examples} indicated by the blue full circles. 
One point worth mentioning is that we have not to date observed asynchronous locking at $1\times\nu_m$.\\

A final remark regarding the crystalline direction in which experiments have been performed: All but one of the experiments analyzed in the main text and this Supplementary Notes were performed collecting the emitted light with the traps aligned along the X crystalline direction, i.e. the [1 -1 0], as defined in Fig.1(a) of the main text. The only experiment shown with the traps aligned along the Y direction ([1 1 0]) is shown in Fig\,4 of the main text. We do not observe fundamental differences when collecting in either of the two directions.

%%%%%%%%%%%%%%%%%%%%%%%%%%%%%%%%%%%%%%%%%%%%%%%%%%%%%%%%%%%%%%%%%%%%%%%%%%%%%%%%%%%%%%%%%%%%%%%%%%%%%%
\section*{Supplementary Note 8: Model for the synchronization of the modes}

%%%%%%%%%%%%%%%%%%%%%%%%%%%%%%%%%%%%%%%%%%%%%%%%%%%
\setcounter{subsection}{0}
\subsection{Model without a mechanical wave ($J=\mathrm{constant}$)}
%%%%%%%%%%%%%%%%%%%%%%%%%%%%%%%%%%%%%%%%%%%%%%%%%%
Following Wouters \cite{SM_Wouters2008} we describe two coupled polariton modes ($j=1,2$)  and the corresponding reservoir's densities as (we take $\hbar=1$)
%%%%%%%%%%%%%%%
\begin{eqnarray}
\nonumber
i \dot{\psi}_j&=&(\varepsilon_j+U_j |\psi_j|^2+U_j^{\textrm{R}} n_j)\psi_j-J\psi_{3-j}+\nonumber\\ && +\frac{i}{2}(R\, n_j-\gamma)\psi_j\,, \nonumber\\
\dot{n}_j&=& P_j-\gamma_{\textrm{R}}\,n_j-R|\psi_j|^2 n_j\,.
\label{GP}
\end{eqnarray}
%%%%%%%%%%%%%%%%%%
Here $\varepsilon_j$ is the bare energy of the $j$-mode, $U_j$ and $U_j^{\textrm{R}}$ are the polariton-polariton and polariton-reservoir interaction couplings, $J$ describes a direct hopping term between modes, $\gamma$ the polariton decay and $R$ the stimulated loading from the reservoir. The dynamic of the later is controlled by the pump power $P_j$, the excitonic decay rate $\gamma_{\textrm{R}}$ and the stimulated decay to the condensate. We notice here that since we are considering two separated modes (on different tramps) the direct overlap between them is assumed small and so the equation for the local value of the reservoir density only depends on the amplitude of the corresponding polariton mode. A more involved model, including a cross term (as would be the case for a single tramp) could be easily included. In that case, synchronisation due to the competition of the reservoir-mediated population of the modes is also possible \cite{SM_Eastham2008}.

We seek for a synchronized solution of the form $\psi_j=\sqrt{\rho_j}e^{-i\omega t\pm \theta/2}$ (here the $+$ sign corresponds to $j=1$) and $\dot{n}_j=0$. Introducing this Ansatz in Eqs. \eqref{GP} we get the following set of algebraic equations
%%%%%%%%%%%%%%%%%%%%%%%%%%%%%%%%%%%%
\begin{eqnarray}
\nonumber
\frac{1}{\alpha}\frac{\left(\xi _1^0-\xi_1\right)}{\xi _1+1}&=&-2  J_g \sin(\theta )\\
\nonumber
\alpha\frac{ \left(\xi _2^0-\xi _2\right)}{\xi_2+1}&=&2 J_g \sin (\theta )\\
\nonumber
\alpha  J \cos (\theta )+\omega&=&\frac{U_1^{\textrm{R}} n_0\left(\xi _1^0+1\right)}{\xi _1+1}+\xi _1 \rho_0 U_1\\
\nonumber
\frac{J\cos(\theta )}{\alpha }+\omega&=&\frac{U_2^{\textrm{R}} n_0 \left(\xi _2^0+1\right)}{\xi _2+1}+\Delta\varepsilon +\xi _2 \rho _0 U_2\\
\label{constrains}
\end{eqnarray}
%%%%%%%%%%%%%%%%%%%
where $\alpha=\sqrt{\rho_2/\rho_1}$, $\Delta\varepsilon=\varepsilon_2-\varepsilon_1$ and we have introduced some dimensionless  parameters so that:  $J_g=J/\gamma$, $\xi_j=\rho_j/\rho_0$, $\rho_0=\gamma_{\textrm{R}}/R$, $P_j=(1+\xi_j^0)P_\mathrm{th}$, $P_\mathrm{th}=\gamma \gamma_{\textrm{R}}/R$, $n_0=\gamma/R$. We also made use of the stationary solution for the reservoir, $n_j=P_j/(\gamma_{\textrm{R}}+R\rho_j)=n_0 \left(\xi_j^0+1\right)/(\xi_j+1)$.

Equations \eqref{constrains} need to be solved for $\xi_j$, $\omega$ and $\theta$. Notice that the first two allow for the determination of $\xi_j(\theta)$, independently of the interactions (for instance $J=0$ implies $\xi_j=\xi_j^0$).  
While it is possible to find a general solution for $\xi_j(\theta)$, it involves a quartic polynomial with generic roots and, in practice, it is in general difficult to determine the one with physical meaning ($\xi_j,\alpha>0$).

However, an analytical condition for $\theta$ can be found if one assumes that $R \rho_j\gg \gamma_{\textrm{R}}\rightarrow \xi_j\gg 1$, i.e. $P_j\gg P_\mathrm{th}$, deep into the condensed regime \cite{SM_Wouters2008}.  In that case, $\xi_j+1\rightarrow \xi_j$ in the above equations (and for consistency the same applies to $\xi_j^0$). Hence, from the first two equation in Eq. \eqref{constrains} one gets
%%%%%%%%%%%%%%%%%%%%%5
\begin{eqnarray}
\xi_1(\theta)&=&\Theta(-\theta) g(\theta)+\Theta(\theta) \frac{(\xi _{1}^0)^2}{(1+4J_g^2\sin^2(\theta) )g(\theta)}\\
\xi_2(\theta)&=&\frac{\left(\xi_1(\theta)-\xi_1^0\right) \xi_2^0}{4 J_g^2 \xi_1(\theta) \sin^2(\theta)+\xi_1(\theta)-\xi_1^0}
\end{eqnarray}
%%%%%%%%%%%%%%%%%
with
%\begin{widetext}
%%%%%%%%%%%%%%%%%%%%%%
\begin{equation}
g(\theta)=\frac{(\xi _{1}^0)^2}{2 J_g \sin ^2(\theta ) \left(\sqrt{\xi _1^0 \xi _2^0 \csc ^2(\theta )+J_g^2
   \left(\xi _1^0+\xi _2^0\right){}^2}+J_g \left(\xi _1^0+\xi _2^0\right)\right)+\xi_1^0} ,
\end{equation}
%%%%%%%%%%%%%%%%%%%%%%%%%%%
and $\Theta(x)$ the step function.
The final equation that determines $\theta$ is then
\begin{eqnarray}
\nonumber
\Delta\varepsilon &=&f(\theta)\\
\nonumber
&=& J\cos(\theta)\left(\frac{1}{\alpha(\theta)}-\alpha(\theta)\right)
%\\
%\nonumber
%&&
+\frac{U_1^{\textrm{R}} n_0\xi _1^0}{\xi_1(\theta)}+ U_1\rho_0 \xi_1(\theta)-\frac{U_2^{\textrm{R}} n_0\xi_2^0}{\xi_2(\theta)}-U_2\rho_0 \xi_2(\theta)\,.\\
\end{eqnarray}
%\end{widetext}
Once $\theta$ is determined, the locking frequency is given by
\begin{equation}
\omega=\frac{\bar{\varepsilon}_1+\bar{\varepsilon}_2}{2}-\frac{J\cos(\theta)}{2}\left(\alpha+\frac{1}{\alpha}\right)\,,
\end{equation}
and since
\begin{equation}
    \bar{\varepsilon}_2-\bar{\varepsilon}_1=J \cos(\theta)\, \left(\frac{1}{\alpha}-\alpha\right)\,,
\end{equation}
we get
\begin{equation}
\omega=\bar{\varepsilon}_1-\alpha\, J\cos(\theta)\,.     
\end{equation}
Here $\bar{\varepsilon}_j=\varepsilon_j+U_j \rho_0 \xi_j+U_j^{\textrm{R}}n_0 \frac{\xi_j^0}{\xi _j}$ is the dressed energy. Notice that we allowed here for different pump powers on each trap, $\xi_1^0\neq\xi_2^0$. The resulting phase diagram for the synchronization is identical to the one reported in Ref. \cite{SM_Wouters2008}, except that $\Delta\varepsilon$ is not longer symmetric around zero when  $\xi_1^0\neq\xi_2^0$.

%%%%%%%%%%%%%%%%%%%%%%%%%%%%%%%%%%%%%%%%%%%%%%%%%%%%%%%%%
\subsection{Model with a $\Omega$-mechanical wave in the RWA ($J(t)$)}
%%%%%%%%%%%%%%%%%%%%%%%%%%%%%%%%%%%%%%%%%%%%%%%%%%%%%%%%%%
Here we consider the case when the coupling constant $J$ is time dependent, being modulated by a coherent mechanical wave of frequency $\Omega$. In the simplest case, making use of the rotating wave approximation (RWA) and assuming $\varepsilon_1>\varepsilon_2$, we have 
%%%%%%%%%%%%%%%%%%%%%%%%%%%%%%%
\begin{eqnarray}
\nonumber
i \dot{\psi_1}&=&(\varepsilon_1+U_1 |\psi_1|^2+U_1^{\textrm{R}}\, n_1)\psi_1-\left(Je^{i\Omega t}\right)^*\psi_{2}+ \nonumber \\ && +\frac{i}{2}(R\, n_1-\gamma)\psi_1 \nonumber\\
i \dot{\psi_2}&=&(\varepsilon_2+U_2 |\psi_2|^2+U_2^{\textrm{R}}\, n_2)\psi_2-Je^{i\Omega t}\psi_{1}+\nonumber \\ && +\frac{i}{2}(R\, n_2-\gamma)\psi_2 
\label{GP_ph}
\end{eqnarray}
%%%%%%%%%%%%%%%%%%%%%
so that the only difference with the analysis of the previous sections is that we now propose $\psi_1=\sqrt{\rho_1}e^{-i\omega t+\theta/2}$ and $\psi_2=\sqrt{\rho_2}e^{-i(\omega-\Omega) t-\theta/2}$ as the solution for the asynchronized state. The resulting equations are the same as before except that $\varepsilon_2\rightarrow \varepsilon_2+\Omega$. Hence, we have for instance that 
%%%%%%%%%%%%%%%%%%%%%%%%%%%
\begin{equation}
    \bar{\varepsilon}_2-\bar{\varepsilon}_1=-\Omega+J \cos(\theta)\, \left(\frac{1}{\alpha}-\alpha\right)\,.
\end{equation}
%%%%%%%%%%%%%%%%%%%%%

\subsection{Full model and simulations}
%%%%%%%%%%%%%%%%%%%%%%%%%%%%%%%%%%%%%%%%%%%%%
\begin{figure*}[!!!ttt]
    \begin{center}
    \includegraphics[trim = 0mm 0mm 0mm 0mm, clip=true, keepaspectratio=true, width=0.9\columnwidth, angle=0]{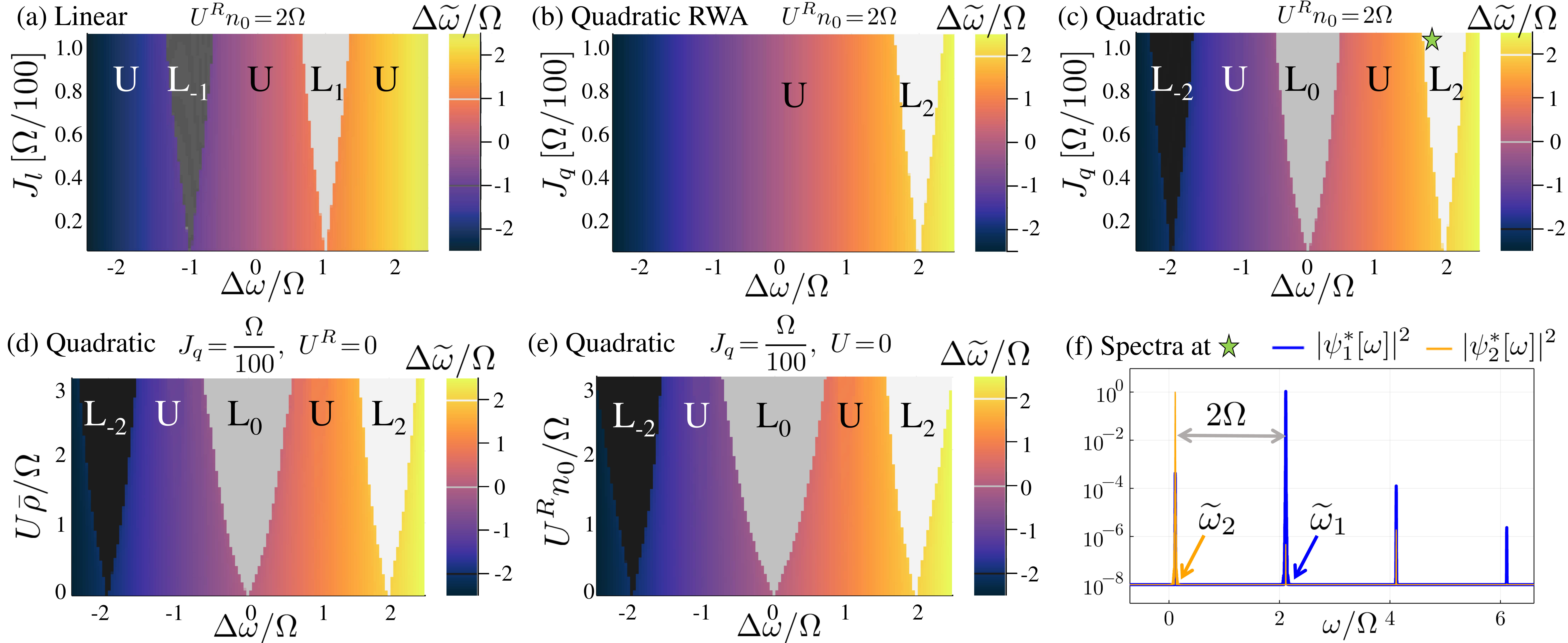}
    \end{center}
    \vspace{-0.5 cm}
\caption{\textbf{Interaction-enhanced intermode frequency locking of the steady state subject to a phonon-induced $J(t)$.} Color-maps of the computed dressed detuning $\Delta\widetilde{\omega}$, as a function of the bared detuning,  $\Delta\omega$ and some of the physical parameters of the model  for $\kappa=0.2\Omega$,  $\bar{\rho}=10^5$ and $J(t)$ corresponding to the linear case  $J_l (e^{i  \Omega t}+e^{-i  \Omega t})$ (a), the quadractic-RWA  $J_q e^{-i 2 \Omega t}$ (b), or full quadratic $J_q (e^{i 2 \Omega t}+e^{-i 2 \Omega t}+2)$ (c). In panels (a-c)  we take $U=0$ and $U^{\textrm{R}} n_0=2\Omega$ and change the phonon coupling strengths, $J_l$ or $J_q$. The  dressed frequencies of the two modes get \emph{locked} in the regions labeled as $\mathrm{L}_n$ having constant $\Delta\widetilde{\omega}=n\Omega$: These are regions centered at values of the bare detuning $\Delta{\omega}=n\Omega$. At the remaining regions, labelled as $\mathrm{U}$, the obtained $\Delta\widetilde{\omega}$ is \emph{unlocked}. 
%Remarkably, the interaction $U^R$ substantially enlarges the bare-detuning width of the $\mathrm{L}_n$  tongues even for $J_{l/q}$  being less than one percent of the phonon frequency $\Omega$.  
Panels (d) and (e) show, for the full quadratic case with $J_q=0.01\Omega$,  that both $U$ and $U^{\textrm{R}}$ enhance the possibility of locking even for   $J_q$ of the order of  one percent of the phonon frequency $\Omega$.
%We note that at larger values of $J_{q/l}$ or the interactions (not shown) other fractional $\Delta\widetilde{\omega}= \frac{n}{m}\Omega$ locked regions, having smaller area in parameter space, can develop. 
(f)  polariton power spectra of the steady state (normalized with $\bar\rho$) for the locked condition shown in panel (c).  The main spectral peaks at $\widetilde{\omega}_j$ present $2\Omega$-separated sidebands having spectral weight that are around four orders of magnitude, or more, smaller than the main peak.}
\label{fig:simsLocking}
\end{figure*}

%\begin{figure}[!!!ttt]
%    \begin{center}
%    \includegraphics[trim = 0mm 0mm 0mm 0mm, clip=true, keepaspectratio=true, width=0.95\columnwidth, angle=0]{FigS15.pdf}
%    \end{center}
%    \vspace{-0.5 cm}
%\caption{\textbf{Fourier transform of the steady state for the quadratic case, i.e., $J(t)=J_q(t)$.} (a) Color-maps of $\log_{10}|\psi_j[\omega]|$ (with $\psi_j[\omega]$ the Fourier transform of the $\psi_j(t)$: the steady state for mode $j$) as a function of $J_q$. We took the parameters as given in Fig.\ref{fig:simsLocking}(c) with the bared detuning fixed away from resonance at $\Delta\omega=-1.8\Omega$. Above a critical value of $J_q$ the system enters locked region $\mathrm{L}_{-2}$, i.e., $\Delta\widetilde{\omega}=-2\Omega$. (b) Cut of $\log_{10}|\psi_1[\omega]|$ for $J_q=0.01\Omega$. The main spectral peak at $-\widetilde{\omega}_1$ has a height of about $\sqrt{\bar\rho}$. It presents $2\Omega$-separated sidebands having spectral weight that are two orders of magnitude or more below the main peak.}
%\label{fig:simsLocking2}
%\end{figure}

%%%%%%%%%%%%%%%%%%%%%%%%%%%%%%%%%%%%%%%%%%%%%
In this section we present numerical simulations obtaining the stationary solutions of the equation of motion of Eqs.\eqref{GP_ph}. We include situations beyond the RW approximation by replacing $J e^{i\Omega t}$ in Eqs.\eqref{GP_ph} by $J(t)$ describing optomechanical coupling being quadratic  ($J(t)=J_q (e^{i 2 \Omega t}+e^{-i 2 \Omega t}+2)$ with $J_q=g^{(2)} n_b$), or linear ($J(t)=J_l (e^{i\Omega t}+e^{-i\Omega t})$ with $J_l=g^{(1)} \sqrt{n_b}$) in the phonon displacement. 
This follows from assuming the presence of a coherent population of $n_b$ phonons, i.e., $b(t)+b(t)^* =2\sqrt{n_b} \cos(\Omega t)$, and an OM interaction having a hopping term between the two polariton modes with a prefactor containing the $n$-order power of the phonon displacement, namely, $-\hbar g^{(n)} (\hat{b}+\hat{b}^\dagger)^n$. 

As in the last section, we work well above the condensation threshold where $R|\psi_j|^2\gg\gamma_{\textrm{R}}$. Then $n_j= n_0 \bar{\rho} /|\psi_j|^2$ with equal power for both modes (here, for the sake of simplicity, we introduced $\bar{\rho}=P/\gamma=\rho_0\, (P/P_\mathrm{th})\sim\rho_0\xi^0$). As both modes are driven to $\bar{\rho}$ we replace $(R n_j-\gamma)$ by $\gamma(\bar{\rho}/|\psi_j|^2 -1)$ in Eqs.\eqref{GP_ph}. We choose identical interaction strengths: $U_1=U_2=U$ and $U_1^{\textrm{R}}=U_2^{\textrm{R}}=U^{\textrm{R}}$. Therefore the reservoir-polariton and the polariton-polariton interaction induce a blue shift for mode $j$ of $U^{\textrm{R}} n_j = U^{\textrm{R}} n_0 (\bar{\rho} /|\psi_j|^2)$ and $U |\psi_j|^2 = U\bar{\rho}\, (|\psi_j|^2/\bar{\rho})$, respectively. In the results below we vary the quantities $U^R n_0$ and $U\bar{\rho}$,  corresponding to the expected blue shifts for $|\psi_j|^2=\bar{\rho}$. 

Our goal is to study the tendency of the two modes to get locked, i.e., presenting frequencies separated by fixed amounts related to the phonon frequency. This requires the appearance of large areas in terms of the bare detuning, $\Delta\omega\equiv(\varepsilon_1-\varepsilon_2)/\hbar$, in which the dressed detuning is locked.  In order to find the dressed detuning in a operative way from the simulated steady state we define  $\Delta\widetilde\omega\equiv\widetilde\omega_1-\widetilde\omega_2$, with the dressed frequencies $\widetilde\omega_j$ given by the frequencies having the greatest spectral weight in $\psi^*_j[\omega]$, i.e., the Fourier transform of each mode amplitude, $\psi^*_j(t)$.

In Fig.\ref{fig:simsLocking} we present and describe the results of our simulations showing how the interactions enhance the area of the locked regions. In panels (a-c) we set $U=0$ and simulate the effect of $U^{\textrm{R}} n_0$ being of the order of $\Omega$. This choice is due to the dominant photon-like character of the measured polaritons making the reservoir-polariton coupling $U^{\textrm{R}} n_0$ larger than the polariton-polariton coupling $U\bar{\rho}$; this is consistent with our Gross-Pitaeskii equation simulations \cite{SM_Reynoso2022, SM_Mangussi2021}. However, we note that as is expected from the analytical theory of the previous section \cite{SM_Wouters2008}, similar enhancement of the locked regions is observed for $U^{\textrm{R}}=0$ and $U\neq 0$ (see Fig.\ref{fig:simsLocking}(d)). Importantly, panels (a-e) show that even for $J_q$ or $J_l$ of one percent of $\Omega$ the locked regions acquires a significant portion of the parameter space. We can also see, from comparing Fig.\ref{fig:simsLocking}(b) with Fig.\ref{fig:simsLocking}(c), that at the simulated moderate values of the parameters $J_q$, $U\bar{\rho}$, and $U^{\textrm{R}} n_0$ the RWA --and thus the theory of Ref.\cite{SM_Wouters2008}-- is enough to describe the limits of the $\mathrm{L}_2$ locked area fairly well, i.e., the region having dressed detuning $\Delta\widetilde\omega=2\Omega$. For larger values of the relevant parameters the RWA breaks down as other locked situations overlap, including cases with $\Delta\widetilde\omega=\frac{n}{m}\Omega$ (not shown). 
We note that  the linear coupling case  shown in Fig.\ref{fig:simsLocking}(a) presents its dominant locked regions  at $\Delta\widetilde\omega=\pm\Omega$, something the it was not observed in our experimental data. Finally, it is worth pointing out that inside the locked region the power spectrum of the modes (Fig.\ref{fig:simsLocking}(f)) shows sidebands that are several orders of magnitude smaller than the main peak and hence extremely hard to resolve experimentally.

\paragraph*{\\ \textbf{Correspondence} and requests for materials should be addressed to A.F.}

\end{document}